\documentclass[aps,prb,amsfonts,array,13pt]{revtex4}
\usepackage{epsfig}
\def\be{\begin{equation}}
\def\ee{\end{equation}}
\def\bea{\begin{eqnarray}}
\def\eea{\end{eqnarray}}
\def\z{\mathbf}
\newcommand{\BM}[1]{\mbox{\boldmath$#1$}}

\begin{document}

\title{Slow two-level systems in point contacts}
\author{A. Halbritter,$^1$ L. Borda,$^{2,3}$ A. Zawadowski$^{2,4}$}
\address{$^1$Electron Transport Research Group of the Hungarian Academy of
Sciences and Department of Physics,\\
Budapest University of Technology and Economics, 1111 Budapest,
Hungary}
\address{$^2$Department of Theoretical Physics and Research Group of the
Hungarian Academy of Sciences,\\
Budapest University of Technology and Economics, 1111 Budapest,
Hungary}
\address{$^3$Sektion Physik and Center for Nanoscience, LMU M\"unchen, Theresienstrasse 37, 80333 M\"unchen, Germany}
\address{$^4$Research Institute for Solid State Physics and Optics,
Hungarian Academy of Sciences,\\
1525 Budapest, Hungary}

\begin{abstract}
A great variety of experiments, like heat release measurements,
acoustic measurements, and transport measurements on mesoscopic
samples have proved that two level systems (TLSs) have a crucial
role in the low temperature thermal and electric properties of
disordered systems. This paper is aimed at reviewing the role of
slow TLSs in point contacts. First the theory of point contacts is
summarized, concentrating on the discussion of different point
contact models, and on the different regimes of electron flow in
the contact, mainly focusing on the ballistic and diffusive limit.
The Boltzmann equation is solved in both regimes, and the position
dependence of the electrical potential is determined. Then the
scattering processes in point contacts are investigated,
particularly concentrating on the scattering on slow TLSs. If the
the electron assisted transitions between the two states are
negligible the electron-two level system interaction can be
treated with a simplified Hamiltonian. The scattering on such slow
TLSs causes nonlinearity in the current-voltage characteristics of
the point contact, which can be determined using Fermi's golden
role. These calculations are presented showing both the
contribution of elastic and inelastic scattering, and including
the dependence on the position of the TLS, and on the effect of
high frequency irradiation. These results are used to discuss the
differences between these slow TLSs and the fast centers which may
be described by the two channel Kondo model. The available
experimental results are analyzed, distinguishing between the
effects due to the different types of TLSs.
\end{abstract}

\maketitle

\newpage
\draft \tableofcontents
\newpage

\section{INTRODUCTION}
\label{sec:intro} In the last decades point contact spectroscopy
has been proved to be a very powerful method to study
electron-phonon interaction in metals (for reviews see
\onlinecite{YS},\onlinecite{Jansen1980} and
\onlinecite{Duit1989}). In the recent years it has been realized
that in mesoscopic metallic systems the electron scattering by
some dynamical defects may play a major role in understanding the
energy relaxation and dephasing processes of the electrons at low
temperature (for reviews see \onlinecite{Pecs2000} and
\onlinecite{orsi2001}). Such defects can also be studied by point
contact spectroscopy provided that they are situated in the
contact region. These defects can either be magnetic impurities or
structural defects where an atom or a group of atoms have a
metastable, almost double degenerate groundstate forming a
two-level system (TLS). The TLS spectroscopy is very similar to
the phonon spectroscopy, but as the TLSs cannot propagate like
phonons, they are not in thermal equilibrium with the bath. The
TLS can be formed as defects in the bulk region of crystalline
materials, or at dislocations, at surfaces due to impurities and
also in amorphous materials. In most of the cases the internal
structure of the TLSs are unknown and they can behave very
differently.

The dynamics of magnetic impurities shows the Kondo
effect\cite{kondo1988} where by lowering the temperature a Kondo
resonance is developed at the range of the characteristic Kondo
temperature, $T_K$ and approaching zero temperature a magnetic
singlet is formed by the impurity spin and the electrons. In that
case near zero temperature the magnetic singlet acts just like a
static impurity and the dynamical processes are frozen out.

The subject of slow defects can be described by the conventional
theory of TLSs and its application to the point contact is the
present review. There are, however, many experimental facts which
cannot be described in that way, and show resemblance to the
magnetic Kondo anomalies even if their magnetic origin is very
unlikely. We make an effort to point at those experiments by
comparing the observed behavior with the predictions of the slow
TLS model. Several of these are believed to be due to dynamical
defects, different from slow TLS. It must be emphasized that there
is no theory which is generally accepted and can be applied.
Earlier it has been proposed that they are fast TLSs, which can be
described by the so called two-channel Kondo
model.\cite{VZa,VZb,VZc} That proposal has recently been
criticized and
debated.\cite{Aleiner2001a,Aleiner2001b,orsi2001,Borda2003} These
scatterers now appear rather as ``phantoms'' and the present
review does not cover that issue, we, however, present the earlier
interpretation just to help in reading the earlier experimental
publications.
\begin{itemize}
\item[(i)]\underline{slow TLS}. In that case an atom or a group of
atoms can have a transition between the two states of lowest
energies by direct tunneling or at higher temperature by thermal
excitation over the barrier. In the former case those systems are
also called as tunneling centers (for review see \cite{Black}).
The transition rate can be on the scale of seconds and even on
longer scales. In that case e.g. the specific heat may depend on
the speed of the measurement as to reach thermal equilibrium takes
longer time. The resistivity of such mesoscopic system may jump in
time between two or several values on the above time scale, as in
small mesoscopic systems the resistivity depends on the position
of a single atom even if the two possible atomic positions are
much smaller than an atomic distance. The study of point contacts
containing TLSs in the contact region turned out to be very
powerful as the TLS may result in zero bias anomalies shown by the
$I(V)$ characteristics. The electron passing through the point
contact can interact dynamically with the TLS or it can be
elastically scattered, which scattering depends on the actual
position of the atom forming the TLS. The inelastic scattering may
result in a back-scattering of the electron in which case the
electron does not proceed from one electrode to the other when the
applied voltage is larger than the splitting of the TLS and the
process shows up as a resistivity minimum at zero bias. In the
case of elastic scattering the transmission rate may depend on the
position of a single atom and the anomalies can have either sign.
The characteristic relaxation times from the TLS causing slow
telegraph fluctuation of resistance to those being responsible for
zero bias anomalies in the $I(V)$ curve varies over several orders
of magnitude, but for all these so-called slow TLSs the average
transition time is much longer than the electron-TLS interaction
time.

\item[(ii)]\underline{fast centers.} The fast TLSs are strongly
debated whether they can show Kondo
anomaly.\cite{Aleiner2001b,orsi2001,Borda2003} Here we present the
previous ideas just to make contact with the extended literature,
and the present status of the problem will be discussed at the end
of the paper. If a conduction electron experiences several
switching occasions one after the other we speak about fast TLS.
These fast TLSs may show Kondo anomaly as the TLS's two states can
be described by a quasi-spin replacing the real impurity spin of
the magnetic Kondo problem and the scattered electron can have
different spherical momenta e.g. $s$- and $p$-states playing
similar role to the real spin of the conduction electron. The
spin-flip process of the magnetic Kondo problem is replaced in the
latter case by electron assisted transition between the two states
of the TLS. That trasnition was is originally believed to be
electron assisted tunneling, which is unlikely according to the
present theories,\cite{Aleiner2001a,Aleiner2001b} however an
electron assisted transition without tunneling is still
possible.\cite{Borda2003} In point contacts containing magnetic
impurities the Kondo resonance scattering contributes to the
back-scattering rate, thus the Kondo impurities result in zero
bias peak in the dynamical resistivity, which has a width
characterized by the Kondo temperature, $T_K$. The sign of the
peak is just opposite to those caused by slow TLSs. I.K. Yanson
and his
co-workers\cite{YS,Lysykh1980,Omelyanchouk1980,Naidyuk1982,Yanson2001}
and also other groups\cite{Jansen1980,Ralph1983} studied the zero
bias anomalies due to intentionally placed magnetic impurities.
Considering the Kondo effect there is a significant difference
between the magnetic case and the TLS. In the latter case the
conduction electrons, additionally to their angular momenta
occurring in the Hamiltonian and the coupling, have a further
degree of freedom, the real spin, which makes the electronic sea
double degenerate although it does not appear in the coupling
constants. Referring to that extra degeneracy the phenomenon is
called two-channel Kondo problem (2-CK) in contrast to the
magnetic one-channel case (1-CK) (for earlier review see
\onlinecite{CZ}). That extra degeneracy prevents the system to
have a singlet ground state at low temperature, it has finite
entropy and, therefore, the Fermi liquid behavior does not show up
as far as a low energy cutoff (e.g. the energy difference between
the two lowest states of the TLS, known as splitting) is reached
as the temperature is lowered.

The condition for the formation of a 2-CK ground state is that the
energy splitting is negligible compared to the Kondo temperature,
which is generally not expected. The zero bias anomalies have been
believed to be powerful tools to make difference between 1-CK and
2-CK problems (for a review see \onlinecite{Ralph2,vonDelft1999}).
For example annealing or electromigration could result in
disappearance or modification of the zero bias anomalies in the
case of structural defects while such effects are not expected for
magnetic impurities.\cite{Upadhyay1997} Similar issues also appear
in the electron dephasing time which has been intensively debated
recently\cite{Zawadowski1999,Aleiner2001a,Aleiner2001b,orsi2001,Imry1999}.
Another version of the two-channel Kondo effect was proposed to
explain the zero bias anomalies in URu$_2$Si$_2$ point
contacts\cite{Rodrigo1997} where two degenerate localized electron
orbitals of the U atom play the role of the TLS which interact
with the conduction electrons.\cite{Cox1987,CZ}
\end{itemize}

The issue of the 2-CK problem was responsible for the great
interest in the zero bias anomalies due to TLS and, therefore, it
has a great importance to distinguish between anomalies due to
slow TLSs and fast scatterers.

Even if a considerable amount of the discussions are carried out
considering fast TLSs there is no doubt that in many cases the
slow ones are dominant e.g. where the zero bias anomaly in the
point contact spectrum has a positive
sign.\cite{Akimenko1993,anomaly,Keijsers1998} Working on the
issues listed above we learned that the literature on the theory
of slow TLSs in point contacts is very much scattered over many
papers and journals, the predictions should be collected from a
large number of papers, especially, published in Russian and
Ukrainian physical journals some of them not available at smaller
libraries. That experience encouraged us to collect and at several
occasions to complete those results. The present review is written
in a self-contained manner, thus no further reading is necessary
to follow the theory.

We want to emphasize that the authors did not consider as a task
to give the historically proper complete list of references but
they picked up those which are the most appropriate ones to get
further information.

\section{Two level systems in solids}
\label{TLSinsolids}

\subsection{The model of two level systems}
\label{tls}

\begin{figure}[t]
\centering
\includegraphics[width=9truecm]{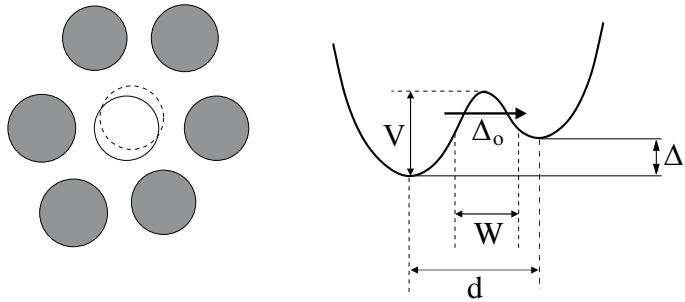}
\caption{\it A two-level system is modelled as a double well
potential with an energy difference $\Delta$ between the two
positions, and a tunneling probability $\Delta_0$ for crossing the
barrier between the two metastable states. $W$ and $d$ denote the
width of the barrier and the distance between the minima,
respectively.} \label{TLS}
\end{figure}

The model of two level tunneling systems was at first introduced
to explain the unexpected low-temperature behavior of specific
heat and thermal conductivity in disordered solids which cannot be
described by phonon excitations.\cite{Zeller} The  model based on
two level systems, which was able to explain the linear
temperature dependence of specific heat of amorphous dielectrics
at low temperatures was suggested by Anderson, Halperin and
Varma\cite{Anderson} and Phillips.\cite{Phillips} This so-called
tunneling model assumes that in an amorphous state some atoms or
groups of atoms may switch between two, energetically nearly
equivalent configurations. This situation is modeled as a double
well potential with two stable states differing by energy
$\Delta$, and a potential barrier between them with a tunneling
rate $\Delta_0$ (see Fig.~\ref{TLS}).
The tunneling probability can be approximated in terms of the
Gamow parameter ($\lambda$): $\Delta_0 \sim e^{-\lambda}$, where
$\lambda$ is defined by the characteristic height, $V$ and width,
$W$ of the potential barrier, and the mass, $M$ of the tunneling
particle:
\begin{equation}
\lambda=\sqrt{\frac{2MV}{\hbar^2}}W.
\end{equation}
Furthermore it is assumed that the TLSs in a solid are uniformly
distributed in terms of the parameters $\Delta$ and $\lambda$,
that is the density of states $P$ for the first excited states of
the TLSs is constant:
\begin{equation}
P(\Delta,\lambda){\rm d}\Delta {\rm d}\lambda=P_0{\rm d}\Delta
{\rm d}\lambda.
\end{equation}
Experimental studies do indeed support this
assumption.\cite{Hunklinger,Golding1973,Black}

The two positions of the system are described by a quasispin, and
the Hamiltonian of the TLS is expressed in terms of the
quasispin's Pauli-operators: \be H_{\rm TLS}={1\over
2}\sum\limits_{\alpha ,\beta} \left({\Delta b^+_{\alpha}\sigma
^z_{\alpha\beta}b_{\beta}+ \Delta_0b^+_{\alpha}\sigma
^x_{\alpha\beta}b_{\beta}}\right), \label{h_tls} \ee where the
indexes $\alpha$ and $\beta$ correspond to the two states of the
TLS ($|1\rangle$ and $|2\rangle$), and $b^+_{\alpha}$ and
$b_{\alpha}$ are the creation and annihilation operators for the
state $\alpha$. This Hamiltonian can be diagonalized by writing
the eigenstates as:\bea
|E_+\rangle &=&\mu |1\rangle+\nu |2\rangle\nonumber\\
|E_-\rangle &=&\nu |1\rangle-\mu |2\rangle.\eea The eigenenergies
are:\be E_{\pm}=\pm\frac{1}{2}\sqrt{\Delta^2+\Delta_0^2},\ee and
thus the energy splitting between the two eigenstates is: \be
E=\sqrt{\Delta^2+\Delta_0^2}.\ee The coefficients $\mu$ and $\nu$
can be expressed as: \bea
\mu=\frac{E+\Delta}{\sqrt{(E+\Delta)^2+\Delta_0^2}}\nonumber\\
\nu=\frac{\Delta_0}{\sqrt{(E+\Delta)^2+\Delta_0^2}}.\label{munu}\eea
For a highly asymmetric TLS, where $\Delta\gg\Delta_0$ the energy
eigenstates are the two positions of the TLS, that is $\mu=1,0$
and $\nu=0,1$. In the opposite case, where the tunneling energy is
much larger than the splitting ($\Delta_0\gg\Delta$) the energy
eigenstates are delocalized between the two positions, and
$\mu=\nu=\frac{1}{\sqrt{2}}$.

\subsection{Scattering on two level systems}

In a disordered material two level systems are not isolated
objects, but they interact with their neighborhood. For instance a
strain induced by an elastic wave may change the parameters of the
two level systems. It is assumed that only the asymmetry parameter
$\Delta$ is influenced by an elastic wave, thus the interaction
between two level systems and phonons can be treated as a
perturbation in $\Delta$:
\begin{equation}
\delta \Delta=\gamma 2\delta u_{ik},
\end{equation}
where $\gamma$ is the coupling constant, and $\delta u_{ik}$ is
the appropriate local elastic strain tensor.

In metallic samples the coupling to conduction electrons has to be
taken into account as well. The general form of the Hamiltonian
describing the interaction between the
 electrons and the TLS is:\cite{Z,ZV}
\be H_{\rm
el-TLS}=\sum\limits_{{\z{p}}_1,{\z{p}}_2,\\\alpha,\beta,\sigma}
b^+_{\beta}a^+_{{\z{p}}_2\sigma}V^{\beta\alpha}_{{\z{p}}_1{\z{p}}_2}a_{{\z{p}}_1\sigma}b_{\alpha},
\label{gen_int_tls} \ee where $a^+$ and $a$ are the electron
creation and annihilation operators, $\sigma$ is the spin index of
the electron and $V$ is the interaction matrix element, which can
be decomposed in terms of Pauli operators:\cite{ZV} \be
V^{\beta\alpha}_{{\z{p}}_1{\z{p}}_2}=\sum\limits_{i=x,y,z}
V^i_{{\z{p}}_1{\z{p}}_2}\sigma^i_{\beta\alpha}+V^0_{{\z{p}}_1{\z{p}}_2}\delta_{\beta\alpha}.
\label{V_pauli} \ee The last term stands for potential scattering
on the average positioned TLS, the term with $\sigma^z$ describes
the difference between scattering amplitudes for the two positions
of the TLS, while $\sigma^x$ and $\sigma^y$ correspond to the
electron assisted transitions between the two states of the TLS.

\subsubsection{Slow two level systems}

The interaction Hamiltonian is further simplified if the
electron-TLS interaction time is much shorter than the transition
time of the TLS, thus the TLS stays in a given position during the
electron scattering. For these so-called slow two level systems
the electron assisted transitions described by $V^x$ and $V^y$ can
be neglected, and the Hamiltonian is: \be H_{\rm
el-TLS}=\sum\limits_{{\z{p}}_1,{\z{p}}_2,\sigma}
V^z_{{\z{p}}_1{\z{p}}_2}a^+_{{\z{p}}_2\sigma}a_{{\z{p}}_1\sigma}\cdot\sum\limits_{\alpha,\beta}
\sigma^z_{\beta\alpha}b^+_{\beta}b_{\alpha}+\sum\limits_{{\z{p}}_1,{\z{p}}_2,\sigma}V^0_{{\z{p}}_1{\z{p}}_2}a^+_{{\z{p}}_2\sigma}a_{{\z{p}}_1\sigma}
\cdot\sum\limits_{\alpha,\beta}\delta_{\beta\alpha}b^+_{\beta}b_{\alpha},
\label{int_slowtls}\ee the interaction matrix elements between the
delocalized energy eigenstates are: \be \langle E_-|H_{\rm
el-TLS}|E_+\rangle
=\sum\limits_{{\z{p}}_1,{\z{p}}_2,\sigma}\left[2\mu\nu
V^z_{{\z{p}}_2{\z{p}}_1}\right]a_{{\z{p}}_2\sigma}^\dagger
a_{{\z{p}}_1\sigma},\label{matrix_elements_inel}\ee which gives
the amplitude of the electron's inelastic scattering on the TLS,
which is associated with an energy change, $E$. The matrix
elements corresponding to the elastic processes are written as:
\bea \langle E_+|H_{\rm el-TLS}|E_+\rangle
&=&\sum\limits_{{\z{p}}_1,{\z{p}}_2,\sigma}\left[V^0_{{\z{p}}_2{\z{p}}_1}+(\mu^2-\nu^2)V^z_{{\z{p}}_2{\z{p}}_1}\right]a_{{\z{p}}_2\sigma}^\dagger a_{{\z{p}}_1\sigma}\nonumber\\
\langle E_-|H_{\rm el-TLS}|E_-\rangle
&=&\sum\limits_{{\z{p}}_1,{\z{p}}_2,\sigma}\left[V^0_{{\z{p}}_2{\z{p}}_1}-(\mu^2-\nu^2)V^z_{{\z{p}}_2{\z{p}}_1}\right]a_{{\z{p}}_2\sigma}^\dagger
a_{{\z{p}}_1\sigma}.\label{matrix_elements_elastic}\eea In both
cases the energy of the scattered electron is conserved, however
the scattering cross section is different if the TLS is in the
$|E_+\rangle$ or $|E_-\rangle$ state.

By simple argumentation the interaction strength, $V^z$ can be
estimated as: $V^z\sim(p_{F}d)V$,\cite{Black1979,VZa} as the
difference between the electron scattering amplitudes in the two
positions must be proportional to the distance of those positions
on the scale of the Fermi wavelength, and the strength of the
scattering, $V$. $V^0$ is very model dependent, e.g.\ the
tunneling atom can be different from the most of the atoms.

\subsubsection{Fast two level systems}
\label{fastTLS}

The electron assisted transition processes are only important in
the case of fastly relaxing two level fluctuators. In this case
the Hamiltonian given by Eq.~(\ref{gen_int_tls}) can be scaled to
the two channel Kondo problem (2CK) with the Hamilton operator:
\be H_{\rm int}=\BM{\sigma}_{\rm
TLS}V\sum\limits_{p_1,p_2,\alpha,\beta,\sigma}
a^+_{p_1\alpha\sigma}\z{S}_{\alpha\beta}a_{p_2\beta\sigma},
\label{2CK} \ee where the the momentum dependence of
$a_{\z{p},\sigma}$ is decomposed into appropriately choosen
spherical waves with indices $\alpha,\beta$, and $p=|\z{p}|$ (For
details see \onlinecite{VZa,VZb}). $\z{S}$ is a pseudo-spin
corresponding to the spherical indices of the electrons, whereas
$\BM{\sigma}_{\rm TLS}$ is a pseudo-spin corresponding to the
states of the TLS, and $V$ is the interaction strength.  In this
model the electron spin is an extra degree of freedom compared to
the single channel magnetic impurity Kondo problem, and this
quantum number is conserved during the interaction and the
coupling is independent of that. In contrast to the single-channel
Kondo problem, which has a non-degenerated, $S=0$ ground state
because of the spin compensation cloud formed by the conduction
electrons, the ground state of the two channel Kondo problem has a
non-zero entropy. In the two channel Kondo model the conduction
electrons overscreen the impurity spin by forming a non-trivial
spin state. The low temperature regime of the two channel Kondo
model was found to have non-Fermi-liquid behavior.

This paper is devoted to review the literature of the slow TLSs
and point contacts. The Kondo regime of the TLSs is out of present
consideration, but one can find an up-to-date review of this field
in the work of Cox and Zawadowski.\cite{CZ}

\subsubsection{Scattering rate beyond the Born approximation}
\label{Born}

In the previous part of this Section the scattering rates due to
TLSs are calculated in Born approximation. Considering only the
screening interaction $V^z$ and ignoring the assisted tunneling
rates ($V^x, V^y$) Kondo\cite{kondo1984} has found power law
correction to the tunneling rate which is due to the building up
of a screening cloud or Friedel oscillation around the changed
position of the tunneling atom. That is closely related to the
Anderson's orthogonality
catastrophe\cite{anderson1967,kondo1976,kondo1988} and the X-ray
absorption problem (see e.g.\ \onlinecite{nozieres1969}). That
phenomenon is associated with creating a large number of
electron-hole pairs as the position of the tunneling atom is
changed. The renormalized dimensionless coupling constant (usually
denoted by $K$) can be expressed in terms of the phase shift
$\delta$ describing the scattering of the electrons by the atom in
the s-wave channel (see e.g.\ \onlinecite{yamada1985,kondo1988})
and the separation distance, $d$ between the two positions of the
TLS
\begin{equation}
K=2\left[ {1\over\pi}\arctan {{\sqrt{1-a^2}\tan\delta}\over
{(1+a^2\tan^2\delta)^{1/2}}} \right]^2, \label{K}
\end{equation}
where $a=j_0(k_Fd)$ with the spherical Bessel function, $j_0$ and
$K<1/2$ for only $s$-wave scattering. In the weak coupling limit
$\delta\ll1$ and $k_Fd\ll1$ that expression is proportional to
$(\varrho_0V^z)^2$. In many other publications the notation
$\alpha$ is used for $K$ (see e.g.\ \onlinecite{Costi1996}).

The behavior of a TLS coupled to electronic heat bath (Ohmic
region) by the coupling $K$ has been studied in great detail (see
e.g.\ \onlinecite{Leggett1987} and \onlinecite{Weiss}). In general
there are two different regions regarding the coupling strength
$K$. Here the asymmetry parameter $\Delta$ is disregarded and
$T=0$ is taken.
\begin{itemize}
\item[(i)] \underline{$0<K<{1\over 2}$ damped coherent
oscillation}. At $K=0$ the atom has a periodic motion between the
two positions. As $K$ is increased the oscillation is more damped
and the renormalized tunneling frequency $\Delta_{ren}$ is given
by
$$
\Delta_{ren}=\Delta_0\left({\Delta_0\over{\omega_c}}\right)^{K\over{1-K}},
$$
where $\omega_c$ is the high frequency cutoff, which has first
believed to be the electronic bandwidth (see e.g.\
\onlinecite{CZ}, but according to the adiabatic renormalization it
is rather the energy scale of the lower excitations in the
potential well where several excited levels are
considered.\cite{Kaganbook,kondo1988} \item[(ii)]
\underline{${1\over 2}<K<1$ incoherent relaxation}. The motion is
incoherent and the atom is spending longer times in a position
before tunneling as $K$ is increased.
\end{itemize}
Many experimental facts indicate (see next subsection) that the
TLSs are in the first region ($K<1/2$) and even can be in the weak
coupling limit ($K\ll 1$).

The main issue is how the quasielastic and inelastic scattering
rates are modified by the coupling $K$. That problem was
investigated by H. Grabert, S. Linkwitz, S. Dattagupta, U.
Weiss\cite{Grabert1986} considering the neutron scattering by TLS.
Those results are very instructive for the present case, there
are, however, essential differences:
\begin{itemize}
\item[(i)] Korringa type of relaxations are very important in the
relaxation of the TLS coupled to the ohmic electronic heat bath.
If energy $\omega$ is transferred to the TLS the spectrum of the
TLS is broadened by a relaxation rate proportional to $\omega$. In
the case of point contacts that energy may arise from the energy
of the electron passing through the point contact. At finite bias,
$V$ the system is, however, out of equilibrium and in the
stationary state of the TLS the averaged energy of the TLS is
proportional to $V$ also. Thus, the nonequlibrium situation should
be treated by using e.g. the Keldysh technique.\cite{keldysh1985}
\item[(ii)] The energy transfer is not determined by the applied
voltage, in contrast to the neutron scattering experiment where
the energy of the incoming and outgoing neutrons are
simultaneously measured. In the present case by small change of
the voltage the energy of the extra electrons passing through the
point contact are given ($\sim V$) but the energy of the scattered
electrons is not determined, thus the energy of the outgoing
electron shows some distribution. That results in an additional
smearing in the spectrum.
\end{itemize}

In the following we give a brief summary of the neutron scattering
case.\cite{Grabert1986} Very accurate results are obtained for low
temperature and energies except the range $\omega\ll
\Delta_{ren}$. That is certainly the interesting range considering
especially for the long range tails of the characteristics which
are crucial for the interpretations of the
experiments.\cite{Keijsers1996,Zarand1998} The calculation was
carried out by expanding in the tunneling events and summing up.

More comprehensive studies were performed by using numerical
renormalization group\cite{Costi1996,Costi1998} for the entire
energy range of energy and even including the asymmetry, $\Delta$
and later even for nonequilibrium.\cite{Costi1997}.

The previous results\cite{Grabert1986} have a very good fit, by
the formula
$$
S(\omega)\sim{A\over{\exp (kT)
-1}}{\omega\over{(\omega-\omega_0)^2 +b\omega^2}}
$$
where $a$, $B$, and $\omega_0$ are fitted parameters. That clearly
shows a double peak structure with energies $\pm\omega_0$. The
lines are broadened by a Korringa type energy relaxation rate
proportional to $\omega$. In the nearly Lorentzian lineshape no
direct role of coupling $K$ was discovered.

Those results are strongly indicating that in weak coupling region
that would be very hard to discover any sign of anomalous
relaxation rate in the tunneling characteristics. That situation
is further complicated by the averaging over the distribution of
the energy of the scattered electrons and the out of equilibrium
situation. We find, that the experimental studies of the tunneling
characteristics are not appropriate tool to look for the
interaction effects, except that $\Delta_0$ is renormalized. That
situation is very different from the cases of one and two channel
Kondo impurities, where the characteristics are determined by the
Kondo resonances.

\subsection{Experimental investigation of two level systems}

There are several experimental methods to investigate the kinetics
of TLSs. The first two methods to be discussed are studying the
TLSs by their interaction with phonons. In heat release
measurements the time dependence of heat release or specific heat
is measured: after a sudden cool down of the sample due to the
energy relaxation of the TLSs it takes a long period of time to
reach thermal equilibrium. The second group of measurements
investigates the propagation of sound in amorphous materials
measuring sound attenuation and sound
velocity.\cite{Hunklinger1976}

Heat release measurements basically focus on amorphous dielectrics
and superconductors, as in metals the contribution of conduction
electrons to the heat capacity usually dominates that of TLSs,
thus the estimation of $P_0$ is more difficult. According to the
heat release measurements of Kol\'a\v{c} et al.\cite{Kolac} in
amorphous metal Fe$_{80}$B$_{14}$Si$_6$ the density of the TLSs is
about $P_0\approx 1.5\times10^{44}J^{-1}m^{-3}$.

Though in acoustic measurements basically the phonons are
investigated, if the electron - TLS coupling is strong, the
results may significantly differ from that on amorphous
dielectrics.\cite{Golding1978} A simple description for ultrasound
absorption by two level systems is as follows. The occupation
numbers for the upper and the lower states of the TLS are $n_{+}$
and $n_{-}=1-n_{+}$, respectively. The probability for ultrasound
absorption is $\alpha Pn_{-}$, where $P$ is the ultrasound
intensity, and $\alpha$ is the absorption coefficient. The
probability of relaxation is $n_{+}/\tau$, where $\tau$ is the
relaxation time of the TLS. In balance these two quantities are
equal, thus one obtains:
\begin{equation}
n_{-}=\frac{1}{\tau}\frac{1}{\alpha P+\frac{1}{\tau}}.
\end{equation}
The absorbed energy is:
\begin{equation}
E_{abs}=\hbar\omega\alpha Pn_{-}=\hbar\omega\frac{\alpha
P\frac{1}{\tau}}{\alpha
P+\frac{1}{\tau}}.
\end{equation}
It is easy to see, that at large enough intensities
($P\to\infty$), the energy absorption saturates to a value
inversely proportional to the relaxation time ($E_{abs}\to
const./\tau$). In insulators and semiconductors the available
maximal ultrasound power is enough to drive the absorption into
saturation at $T=1-4$K, but in metals such saturation is only
obtained at temperatures as low as $T=10$mK. It implies that the
relaxation times in metallic samples are much shorter than in
disordered insulators. The interpretation for this experimental
observation is, that the electron-TLS scattering processes are
dominating in the relaxation of TLSs. Assuming a standard
Korringa-like electron - TLS relaxation, where an electron-hole
pair is created, at low enough temperatures one obtains:
\begin{equation}
\frac{1}{\tau}\sim KP_0 kT,
\end{equation}
that is, the ultrasound absorption measurements give the
opportunity to make an estimation for the electron - TLS coupling
parameter ($K$), and the density of the TLSs ($P_0$). In fact the
acoustic properties of disordered metals do not show universal
behavior as in the case of insulating glasses. There are several
theoretical approaches (the mentioned standard Korringa-like
relaxation,\cite{Black} the strong coupling theory of Vlad\'ar and
Zawadowski,\cite{VZa,VZb} the electron-polaron effect considered
by Kagan and Prokof'ev\cite{Kagan,kondo1988}) that may explain the
aspects of the relaxation of TLSs due to conduction
electrons.\cite{bez2000} Still, taking into account the simple
electron-TLS coupling given by Eq.~(\ref{gen_int_tls}), the
acoustic measurements can give estimations for the density of two
level systems ($P_0$) or the electron-TLS coupling constant ($K$).
In case of superconductors the Korringa relaxation is suppressed
by the presence of the superconducting gap, but the gap disappears
if a high enough magnetic field is applied on the sample,
switching on the TLS-electron interaction.\cite{Coppersmith}
According to the measurements of Esquinazi et al.\cite{Esquinazi2}
in the normal and superconducting state of $Pd_{30}Zr_{70}$
compared with the theory of Kagan and Prokof'ev\cite{Kagan} the
coupling parameter is approximately $K\approx0.4$. The acoustic
measurements of Coppersmith and Golding\cite{Coppersmith2} on the
normal conducting amorphous metal
Pd$_{0.775}$Si$_{0.165}$Cu$_{0.06}$ estimated the coupling
constant as $K\approx0.2$. For the density of the two level
systems a lot of results are available in the literature, and it
can be stated generally, that it is the same within a factor of
less than 10 over a large variety of disordered material, being
metallic glasses or dielectric amorphous material. (For TLSs with
energy splitting less than 1 Kelvin it is approximately 1-10ppm.)
A detailed review of heat release and acoustic measurements in
disordered material can be found in
\onlinecite{Esquinazi,Esquinazi2}, and for earlier data see
\onlinecite{VZc}.

Electron-TLS interaction can be studied directly by measuring
electric transport in disordered metals. Point contact
spectroscopy offers the possibility to investigate the properties
of even a single two level system centered in the vicinity of the
contact. The most spectacular sign of two level systems in point
contacts is the so-called telegraph noise: the resistance of the
contact is fluctuating between two (or more) discrete values on
the timescale of seconds. One can estimate the average lifetimes
in the excited ($\tau_e$) and ground state ($\tau_g$) by recording
hundreds of transitions, and fitting the resulting histograms of
lifetimes ($P_e(t)$ and $P_g(t)$) with exponential decay functions
($P_e(t)\sim e^{-t/\tau_e}$ and $P_g(t)\sim e^{-t/\tau_g}$,
respectively). The two life times are related to each other by the
detailed balance, $\tau_e/\tau_g=e^{-\Delta/k_bT}$, where $\Delta$
is the asymmetry parameter. If the TLS jumps to the other state,
the electron screening cloud also needs to rearrange, which makes
the jumps of the TLS slower. (The building up of the electronic
screening cloud is related to a process which is similar to the
X-ray absorption in metallic systems\cite{nozieres1969,kondo1988},
and which is also called as electron-polaron
effect.\cite{Kagan,Kaganbook}) From the theoretical point of view
this slow-down of the TLS motion is treated as a renormalization
of the tunneling amplitude:\cite{kondo1984}
\begin{equation}
\Delta_{ren}=\Delta_0\left(\frac{\Delta_0}{\omega_c}\right)^{\frac{K}{1-K}},
\end{equation}
where $\omega_c$ is the bath cutoff frequency for which originally
the electron bandwidth had been taken,\cite{kondo1984} but later
it was proposed that it is replaced by a typical energy of the
next higher excitation in the potential
well.\cite{Kagan,kondo1988} Golding et al.\cite{Golding} studied
the two level fluctuation in mesoscopic disordered Bi samples.
They argued that their samples were in the strong coupling limit,
where the coupling of the TLS to conduction electrons is very
strong compared to the tunneling matrix element, thus the latter
can be treated as a small perturbation. In this limit the
tunneling of the TLS is incoherent, because of the dephasing due
to the fast oscillations of the electron bath. Here a theoretical
scaling function can be set up:
$T^{1-2K}\gamma_f=f_{K,\Delta_0}(k_BT/\Delta)$, which does indeed
agree with the experimental result. (The energy splitting of the
the TLS can be tuned by changing the electron density, which can
be reached by applying weak magnetic field.\cite{Zimmerman}) In
the limit $k_BT/\Delta\gg1$ the scaling function is constant, i.e.
$\gamma_f\sim T^{2K-1}$, that is the coupling parameter can be
determined from a single fit, giving $K\approx 0.24$ for the
\label{couplingconst} particular TLS measured.

The effect of fastly relaxing TLS in point contacts cannot be
resolved as a telegraph fluctuation of the resistance, it causes
an anomalous behavior in the voltage dependence of the
differential resistance around zero bias, a so-called zero bias
anomaly (ZBA), which will be a subject of detailed discussion in
this review.

\section{Point contacts}
\label{sec:models}

Metallic systems, where two macroscopic electrodes are connected
via a contact with small cross section are called ``point
contacts'' (PC) in general, regardless of the actual size of the
contact area. It can be generally stated, that the resistance of a
point contact is mostly determined by the narrow neighborhood of
the junction; therefore, a PC acts as a ``microscope'' magnifying
all kinds of phenomena occurring in the small contact region.

\begin{figure}[t]
\includegraphics[width=0.245\textwidth]{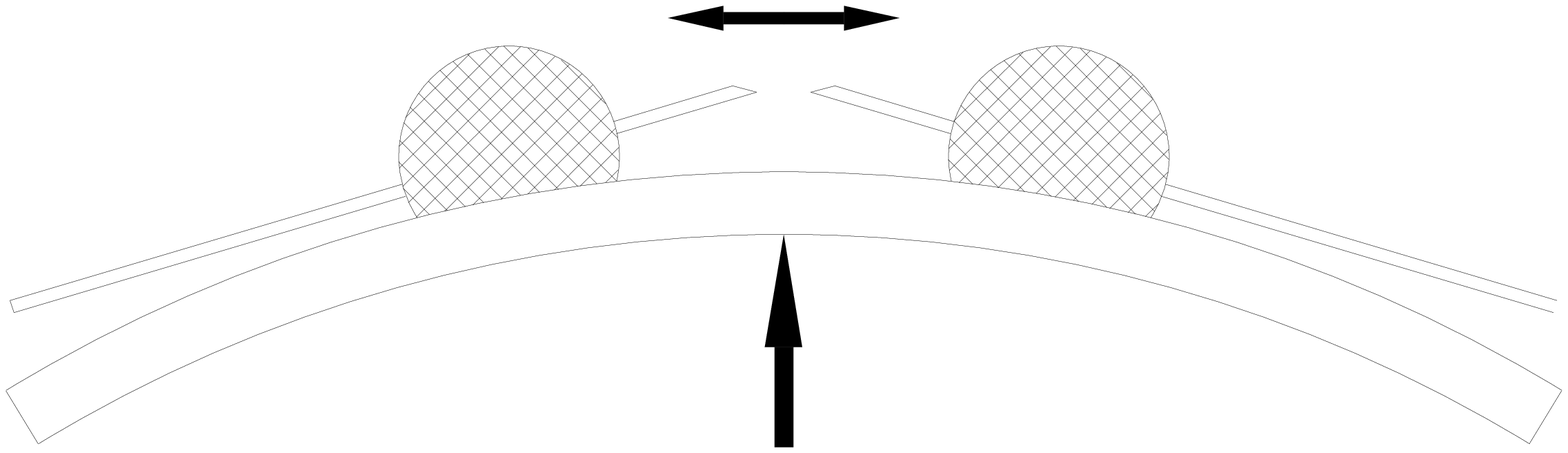}
\includegraphics[width=0.245\textwidth]{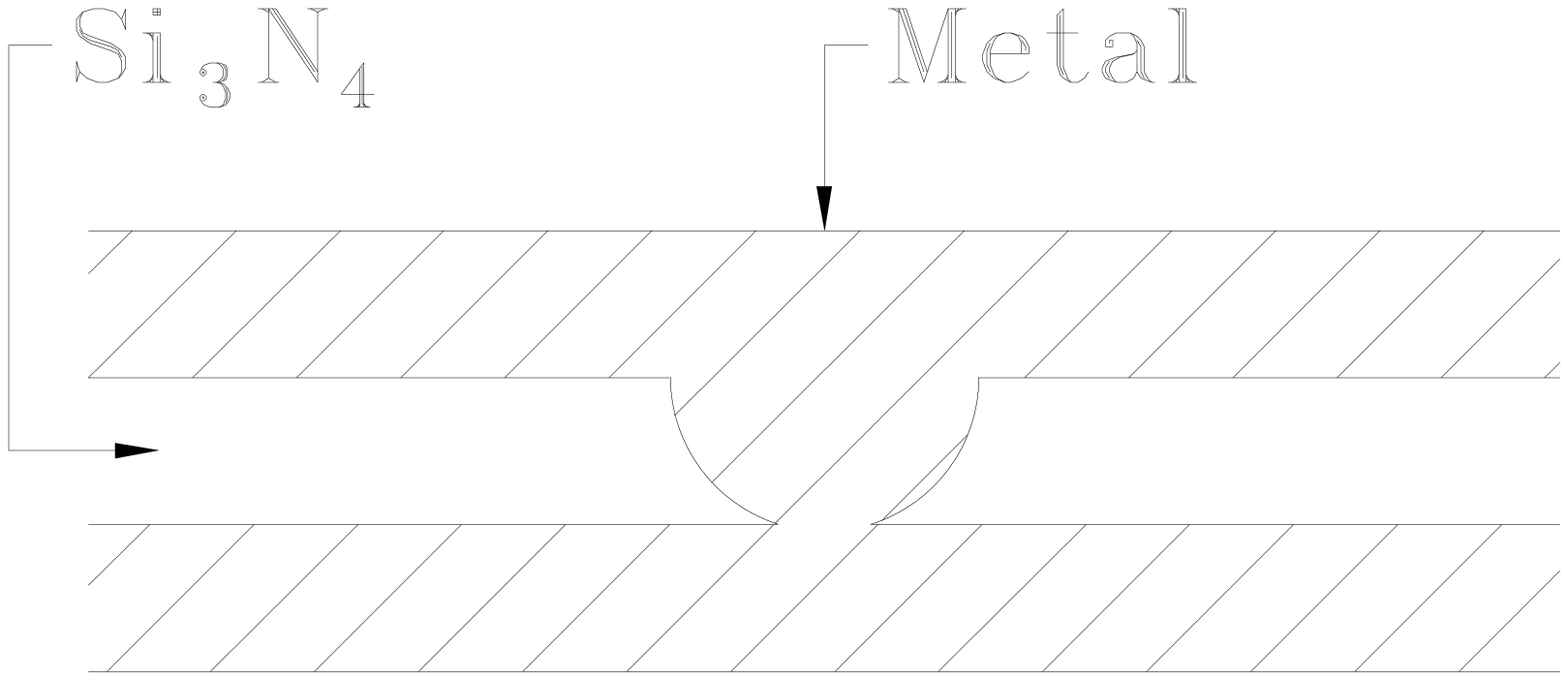}
\includegraphics[width=0.245\textwidth]{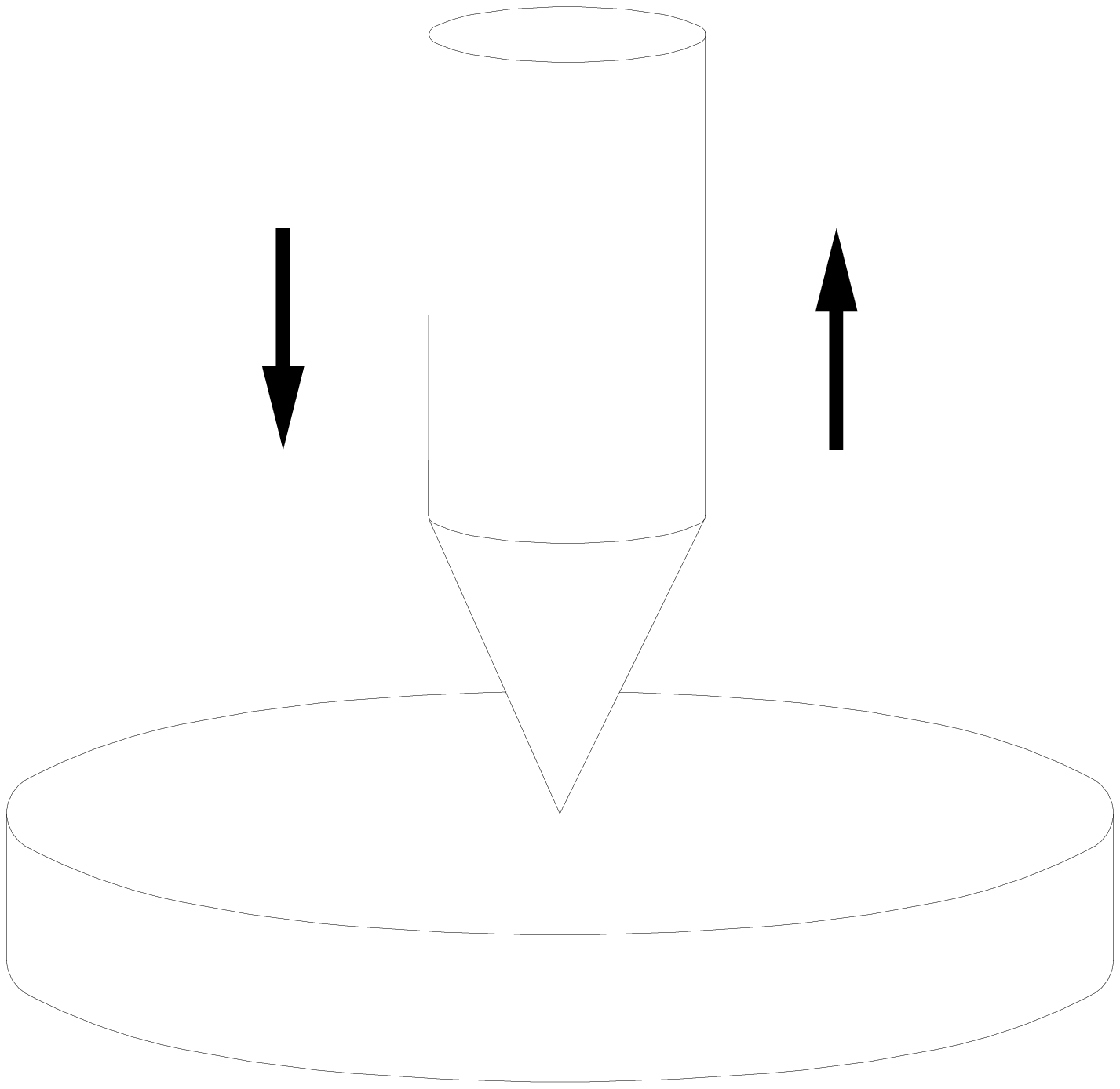}
\includegraphics[width=0.245\textwidth]{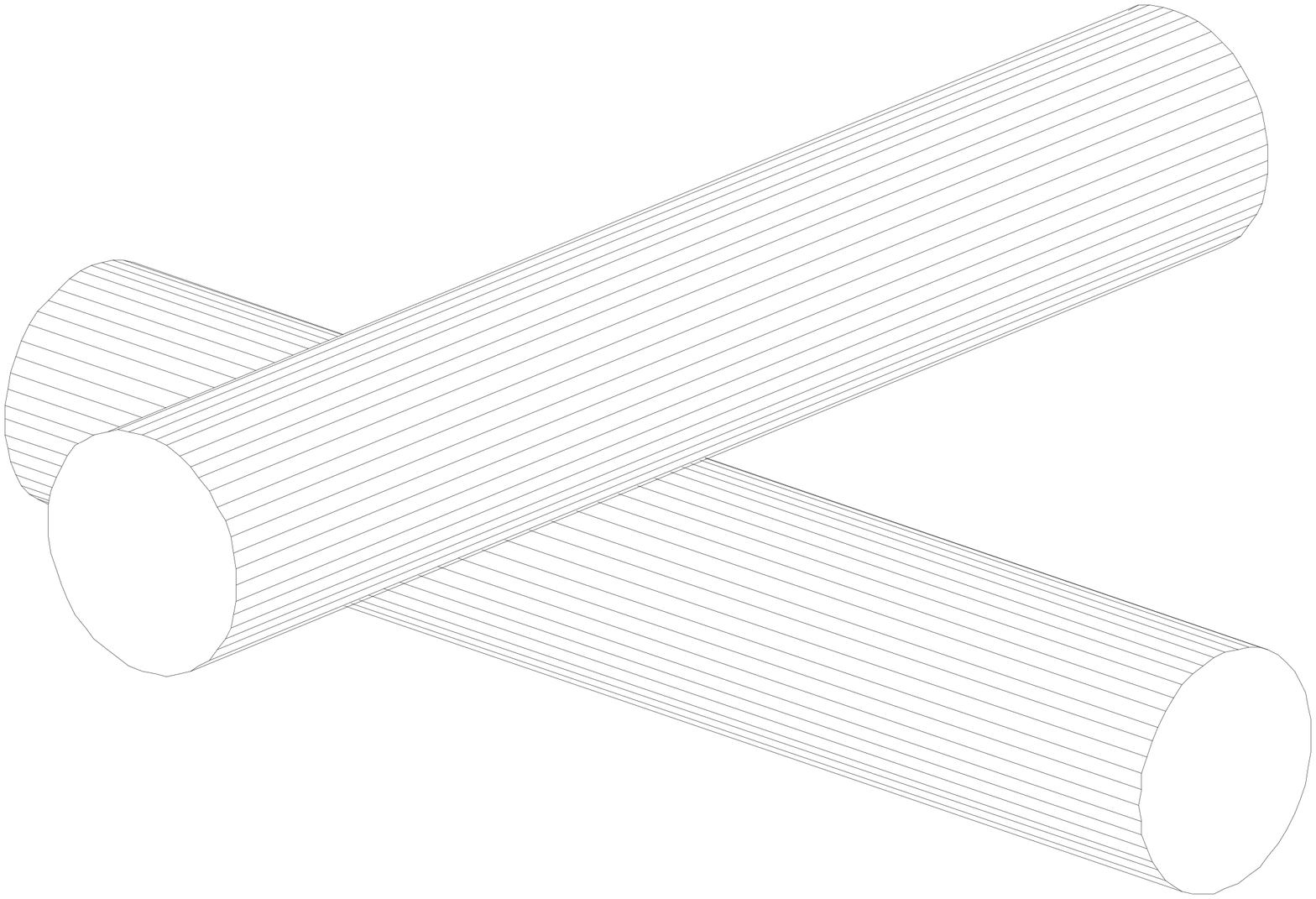}\\
\flushleft \hspace{0.11\textwidth}(a) \hspace{0.22\textwidth} (b)
\hspace{0.22\textwidth} (c) \hspace{0.22\textwidth} (d)
\caption{\it The four main techniques for establishing point
contacts. (a) Mechanically controllable break junction, (b)
nanoconstriction, (c) spear-anvil geometry, (d) touching two
wires.} \label{4methods}
\end{figure}

Several methods have been worked out to produce extremely small
contacts between two conducting leads. Figure~\ref{4methods}
presents the four most important experimental techniques. The
first one (Fig.~\ref{4methods}a; \onlinecite{MCBJ}) is referred to
as mechanically controllable break junction technique (MCBJ). Here
the sample (practically a piece of metallic wire) is fixed on the
top of a flexible beam, and a small notch is established between
the anchoring points. The contact is created in situ at low
temperature by breaking the sample with bending the beam, thus one
obtains clear and adjustable junction on atomic length scale. The
second method (Fig.~\ref{4methods}b; \onlinecite{Ralph}) uses
nanolithography to establish a small hole in a silicon nitride
insulating membrane by etching. If the etching is stopped just
when the hole breaks through, the diameter at the bottom edge
remains extremely small ($\approx$ 3nm), well below the usual
resolution of lithography ($\approx$ 40nm). In the next step metal
is evaporated on both sides, creating a high-quality point contact
device. This method can provide extremely stable and clear
contacts on atomic length scale, however the contact diameter
cannot be varied during the measurement. The third drawing
(Fig.~\ref{4methods}c) shows a similar arrangement to a
scanning-tunneling microscope, a so-called spear-anvil geometry: a
vertically movable, sharply tapered needle is pressed onto a flat
surface. Finally, Fig.~\ref{4methods}d shows a simple technique,
where the edges of two wires are brought into contact. In
arrangements (a), (c) and (d) usually a differential screw
mechanism is used to adjust the contact supplemented with a
piezo-crystal for fine tuning.

The first application of point contacts was carried out by Igor
Yanson\cite{Yanson1} to investigate electron-phonon scattering in
nanojunctions. He found, that the point contact spectrum, obtained
as the second derivative of current with respect to the bias
voltage ($d^2I/dV^2$) contains structure due to the
electron-phonon interaction described by the Eliashberg function
$\alpha^2(\omega)F(\omega)$.\cite{Khotkevich1995} This simple
method for measuring electron-phonon interaction spectrum became a
popular application of PC spectroscopy, however, it can be used to
probe other electron scattering processes as well, like
electron-TLS interaction, which is the central topic of this
paper.

\begin{figure}[t!]
\centering
\includegraphics[height=4.5cm]{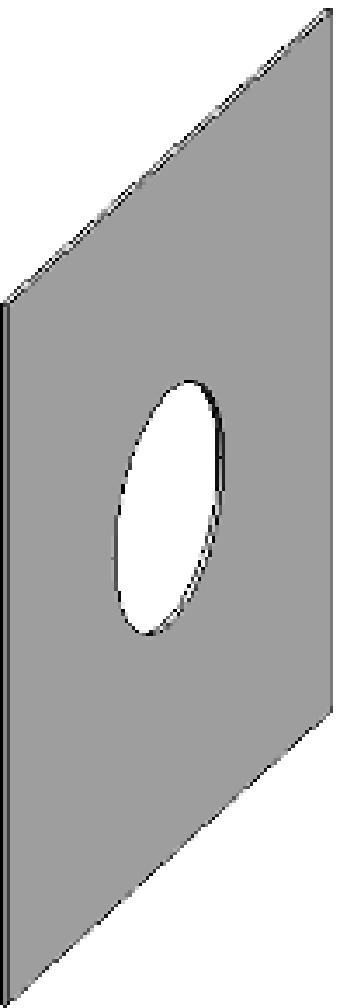}\hspace{2cm}
\includegraphics[height=4.5cm]{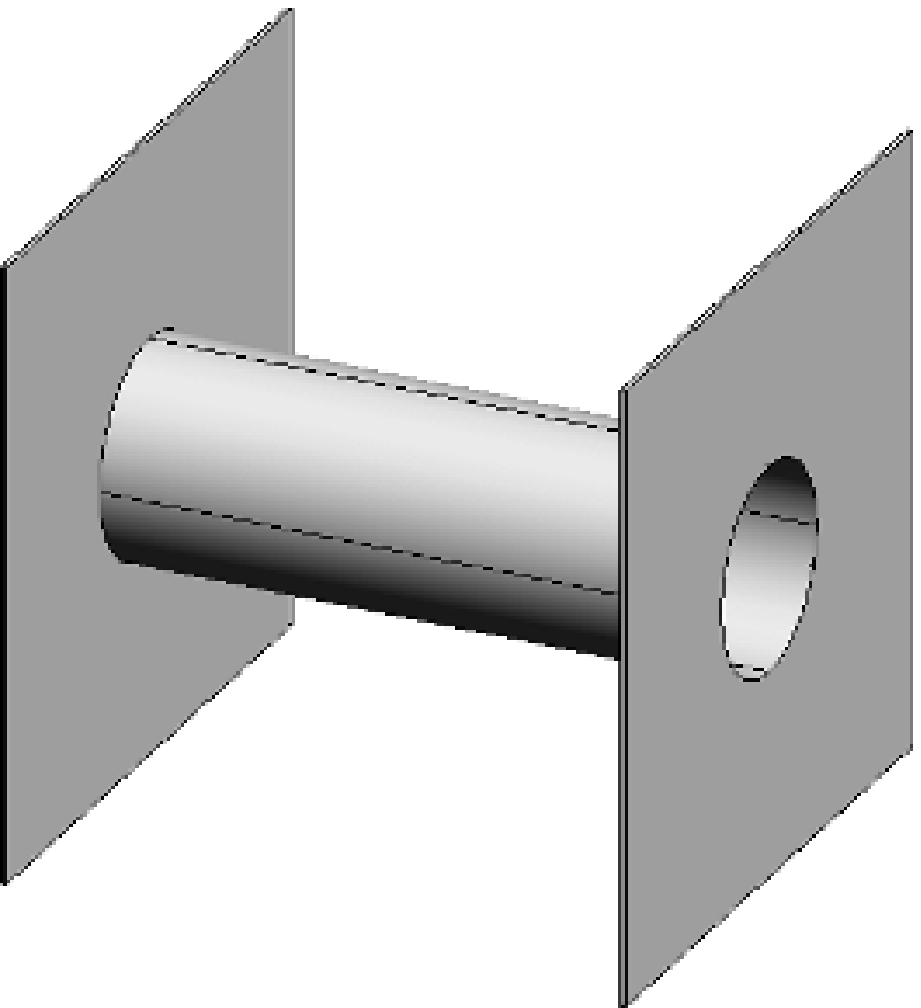}\hspace{2cm}
\includegraphics[height=4.5cm]{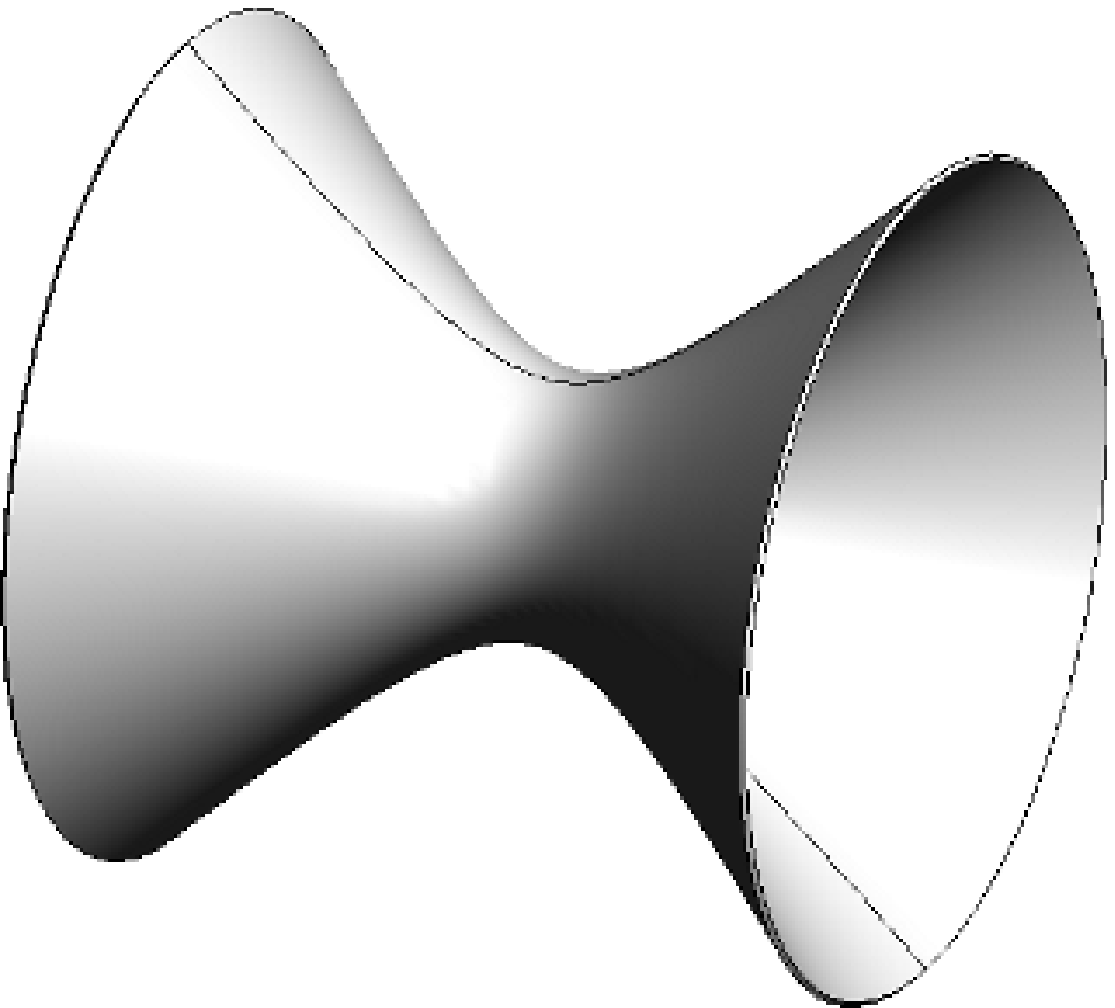}\\
\flushleft \hspace{3cm} (a) \hspace{3.5cm} (b) \hspace{6cm} (c)
\caption{\it Three point contact models. (a) An orifice with
diameter ``$d$'' in an insulating screen between two metallic
half-spaces. (b) Two bulk regions are connected with a long,
narrow conducting neck. (c) Rotational hyperboloid.}
\label{PCmodels.fig}
\end{figure}

Theoretically, point-contacts are considered as two bulk
electrodes connected through a narrow constriction. The simplest,
and most commonly used PC model is presented in
Fig.~\ref{PCmodels.fig}a. This so-called opening-like point
contact is an orifice with diameter $d$ in an infinite isolating
plane between the two electrodes. Another extreme limit is the
channel-like PC: a long, narrow neck between the bulk regions with
the length being much larger than the diameter, $L\gg d$
(Fig.~\ref{PCmodels.fig}b). The crossover between the both cases
can be obtained by considering the point-contacts as rotational
hyperboloids with different opening angles
(Fig.~\ref{PCmodels.fig}c). In most of the cases the shape of the
PC does not influence the character of physical processes in the
constriction radically, and the main parameter is the ratio of the
contact diameter ($d$) and other characteristic length scales in
the system. Three fundamental length scales are the mean free
paths connected to different scattering processes ($l$); the Fermi
wavelength of electrons ($\lambda_{\rm F}$); and the atomic
diameter ($d_{\rm at}$).

If the Fermi wavelength or the atomic diameter becomes comparable
to the contact size, we speak about \emph{quantum point contacts}
and \emph{atomic-sized point contacts}, respectively. These
systems are reviewed in detail in \onlinecite{Agrait2003}. In this
paper we consider contacts, that are neither \emph{atomic} nor
\emph{quantum}, but that are small enough compared to certain mean
free paths, and the calculations are basically performed for an
opening type geometry.

Concerning the mean free paths, we have to make difference between
the elastic and inelastic scatterings. Under usual experimental
conditions the elastic mean free path ($l_{\rm el}$) is smaller
than the inelastic one ($l_{\rm in}$). Here the inelastic mean
free path is the \emph{length of the path} an electron travels
between two inelastic scatterings ($l_{\rm in=}v_F\tau_{\rm in}$).
Mostly the important parameter is not $l_{\rm in}$, but the
inelastic diffusive length, that is the average \emph{distance} an
electron can diffuse between two inelastic scatterings: $L_{\rm
in=}\sqrt{l_{\rm in}l_{\rm el}}$. If the contact diameter is much
smaller than any of the mean free paths $d\ll l_{\rm el}, l_{\rm
in}$, we speak about a \emph{ballistic} contact. In this case the
electron travels through the constriction without any scattering
(except for the reflection on the walls). On the other hand, if
$d\gg l_{\rm el}$, the electron makes a diffusive motion in the
contact, and accordingly we speak about the \emph{diffusive}
regime. At contact diameters exceeding the inelastic diffusive
length ($d\gg L_{\rm in}$) the excess energy of the electrons is
dissipated inside the constriction, which causes a considerable
Joule heating in the contact. This limit is called \emph{thermal}
regime. In the following subsections these different limits are
discussed. In many cases the system is characterized by the
Knudsen number, $K=d/l_{\rm el}$, which was first introduced for
the problem of the gas outflowing from a tank through a
hole,\cite{Knudsen1934} but in the recent decades it has been used
to characterize point contacts as well.

\subsection{Diffusive regime}
\label{maxwell}

First the diffusive contacts are treated, for which the electric
potential ($\Phi$) can be determined by classical equations. (For
a general discussion see \onlinecite{RH}) If the mean free path of
the electrons is much shorter than the dimension of the contact
($l_{\rm el}\ll d$) then the current density, $\z{j}$ is given by
Ohm's law in terms of the electric field, $\z{E}$ or the electric
potential, $\Phi$:
\begin{equation}\label{ohms}
\z{j}=\sigma\z{E}=-\sigma\BM{\nabla}\Phi,
\end{equation}
where $\sigma$ is the conductivity of the metal. Furthermore due
to the charge neutrality in metals the continuity equation holds:
\begin{equation}\label{continuity}
\nabla\z{j}=0.
\end{equation}
If the conductivity is considered to be constant, these two
equations result in the Laplace equation for the electric
potential:
\begin{equation}\label{laplace}
\triangle\Phi=0.
\end{equation}

In this phenomenological approach the scattering processes are
included in the conductance, which is inversely proportional to
the mean free path, $\sigma\sim l^{-1}$. This treatment does not
distinguish between elastic and inelastic electron scattering. In
the further treatment of the structures in the dynamical
conductance (Sec.~\ref{scatteringinPC}) the inelastic scatterings
play crucial role, therefore, instead of the phenomenological
theory the kinetic equations must be used. Here the solution of
the Laplace equation is presented in order to determine the linear
resistance of a diffusive contact, and to determine the position
dependence of the electric potential.

Furthermore, we note that the Laplace equation is only applicable
if $\sigma$ is constant is space. This assumption is certainly not
valid if the temperature of the contact neighborhood is
inhomogeneous, which happens in the thermal regime. This situation
is treated in the next subsection. Here we assume that the
inelastic diffusive length is much larger than the dimension of
the contact, thus the temperature of the contact region is
constant.

\begin{figure}
\centering \epsfxsize=8.0cm \epsfbox{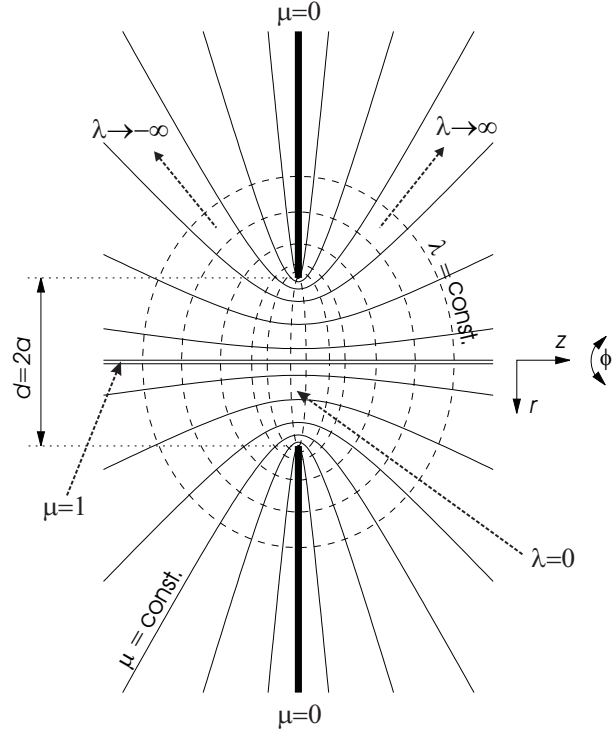}
\vskip0.7truecm \caption{\it The adequate elliptical coordinate
system, in which the Laplace equation is solved for an opening
like point contact.}\label{elliptic_system}
\end{figure}

\begin{figure}
\epsfxsize=0.6\textwidth \centering \epsfbox{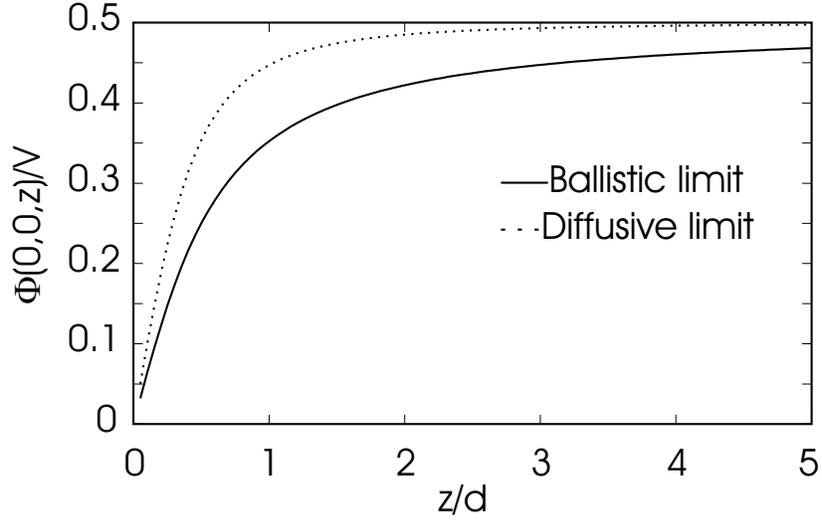}
\vskip0.7truecm \caption{\it The potential as a function of the
distance measured from the center of the opening type contact
along the $z$ axis, in the ballistic and the diffusive limit.}
\label{potent_fig}
\end{figure}

The Laplace equation should be solved with the boundary
conditions: \bea \Phi(z\rightarrow\pm\infty)\rightarrow\mp{V\over
2}. \label{bound} \eea This problem was first solved by
Maxwell.\cite{JCM} Due to the geometry of the problem the Laplace
equation is most easily solved in an elliptical coordinate system
demonstrated in Fig.~\ref{elliptic_system} (see e.g.\
\onlinecite{Morse1953}, \onlinecite{Simo}).
The transformation between a traditional cylindrical coordinate
system $(r,z,\varphi)$ and the elliptical coordinates
$(\lambda,\mu,\varphi)$ is determined by the following equations
of the ellipses and hyperbolas: \bea
{r^2\over{a^2\left(1+\lambda^2\right)}}+{z^2\over{a^2\lambda^2}}&=&1
\qquad {\rm and}
\nonumber\\
{r^2\over{a^2\left(1-\mu^2\right)}}-{z^2\over{a^2\mu^2}}&=&1,
\label{ell_hyp} \eea where $a=d/2$ is the radius of the contact;
$-\infty\leq\lambda\leq\infty$; and $0\leq\mu\leq 1$. Using
Eq.~(\ref{ell_hyp}) the old coordinates can be expressed in terms
of the new ones: \be
r=a\sqrt{\left(1+\lambda^2\right)\left(1-\mu^2\right)},\qquad
\;z=a\lambda\mu . \ee In this coordinate system the orifice is
given by the $\lambda\to 0$ surface, while the isolating layer is
defined by $\mu=0$. The next task is to determine the components
of the metric tensor in the new coordinates: \bea
{\rm d}s^2&=&{\rm d}r^2+{\rm d}z^2+r^2{\rm d}\varphi^2\nonumber\\
\nonumber \eea or, using $\lambda$ and $\mu$, \bea {\rm
d}s^2&=&a^2{{\mu^2+\lambda^2}\over{1+\lambda^2}}{\rm d}\lambda^2+
a^2{{\mu^2+\lambda^2}\over{1-\mu^2}}{\rm d}\mu^2+
a^2\left(1+\lambda^2\right)\left(1-\mu^2\right){\rm d}\varphi^2.
\eea The prefactors are the components of the metric tensor in the
new (also orthogonal) coordinate system: \bea
g_{\lambda\lambda}=a^2{{\mu^2+\lambda^2}\over{1+\lambda^2}}, \quad
g_{\mu\mu}=a^2{{\mu^2+\lambda^2}\over{1-\mu^2}}, \quad
g_{\varphi\varphi}=a^2{\left(1+\lambda^2\right)\left(1-\mu^2\right)}.
\eea Substituting these into the form of the Laplace operator
expressed by the general orthogonal coordinates: \be
\triangle\Phi= {1\over\sqrt{g_{11}g_{22}g_{33}}} \left[
{\partial\over{\partial x_1}}\left( \sqrt{g_{22}g_{33}\over
g_{11}}{{\partial\Phi}\over{\partial x_1}}\right)+
{\partial\over{\partial x_2}}\left( \sqrt{g_{11}g_{33}\over
g_{22}}{{\partial\Phi}\over{\partial x_2}}\right)+
{\partial\over{\partial x_3}}\left( \sqrt{g_{11}g_{22}\over
g_{33}}{{\partial\Phi}\over{\partial x_3}}\right) \right], \ee one
can get easily \be \triangle\Phi= {1\over {a^2\left(\lambda^2
+\mu^2\right)}} \left[ {\partial\over{\partial \lambda}}\left(
\left(1+\lambda^2\right){{\partial\Phi}\over{\partial\lambda}}\right)+
{\partial\over{\partial \mu}}\left(
\left(1-\mu^2\right){{\partial\Phi}\over{\partial\mu}}\right)+
{\partial\over{\partial \varphi}}\left( {{\left(\lambda^2
+\mu^2\right)}\over{\left(1+\lambda^2\right)\left(1-\mu^2\right)}}{{\partial\Phi}\over{\partial\varphi}}\right)
\right]. \ee Taking the geometry of the problem into account it is
obvious that \be {\partial\Phi\over{\partial\varphi}}=0. \ee The
boundary condition Eq.~(\ref{bound}) can only be satisfied if at
$\lambda\to\pm\infty$ the potential is independent of $\mu$, that
is ${\partial\Phi\over{\partial\mu}}=0$. As an Ansatz we extend
this condition to arbitrary $\lambda$ values which means that the
equipotential surfaces are the $\lambda={\rm constant}$ surfaces.
As these surfaces are orthogonal to the $z=0$ plane it is obvious
that electric field has no normal component at the insulating
layer. After this consideration the Laplace equation takes the
following simple form \be {{\rm d}\over{{\rm d}\lambda}}\left[
\left(1+\lambda^2\right){{{\rm d}\Phi}\over{{\rm d}\lambda}}
\right]=0. \label{laplace_apert} \ee The solution of this equation
with the boundary conditions is: \be
\Phi(\lambda)=-{V\over\pi}\arctan (\lambda). \label{phi_lambda}
\ee Along the $z$-axis ($\mu=1$), the potential is changing as \be
\Phi(z)=-{V\over\pi}\arctan{z\over a}, \label{eq31} \ee which is
plotted in Fig.~\ref{potent_fig}.

Inside the opening ($\lambda=0$) the electric field has only $z$
component: \be E_z=-{\partial\Phi\over{\partial
z}}=-{1\over{a\mu}}{\partial\Phi\over{\partial\lambda}}\bigg|_{\lambda=0}={V\over
{a\pi}}{1\over{\sqrt{1-({r\over a})^2}}}.\label{Ez}\ee It should
be noted that the field is the largest at the edge of the opening
($r\to a$) and most of the current flows through that region. The
total current flowing through the contact is:\be
I=\sigma\int\limits_0^a 2\pi r{\rm d}r\;E_z(r)=2a\sigma V,\ee and
thus the resistance of a diffusive point contact is: \be
R={1\over{\sigma d}}.\ee

This calculation was performed for an opening type contact. In
case of a channel type geometry with $L\gg d$ the potential drops
in the channel, and the resistance is easily written using Ohm's
law:
\begin{equation}\label{diffchannel}
R=\frac{4L}{\sigma d^2\pi}.
\end{equation}

\subsection{Thermal regime} \label{thermal} If the dimension of the
contact is larger than the inelastic mean free path, then the
electron's excess energy is dissipated in the contact region, and
Joule heating takes place. In this case the contact center can be
considerably overheated compared to the bath temperature, and thus
the conductivity is also position dependent: $\sigma(T(\z{r}))$.
This phenomenon can be treated by considering the equations both
for the electrical and the heat conduction. The electrical and
thermal current densities are:
\begin{equation}\label{currdensthermal}
\z{j}=-\sigma\BM{\nabla}\Phi,\quad\z{q}=\kappa\BM{\nabla}T,
\end{equation}
where $\kappa$ is the heat conductivity and $T$ is the position
dependent temperature. The continuity equations for the electrical
and thermal current are determined by the charge neutrality and
the Joule heating respectively:
\begin{equation}\label{thermcont}
\nabla\z{j}=0,\quad\nabla\z{q}=-\z{j}\BM{\nabla}\Phi.
\end{equation}
Furthermore, we assume the validity of the the Wiedemann-Franz
law: \be {\kappa\over T}=\mathcal{L}\sigma, \label{wiedemann} \ee
where $\mathcal{L}=(\pi k_B)^2/3e^2$ is the Lorentz number.
Combining these equations we get:
\begin{eqnarray}\label{combined}
\nabla(\sigma\BM{\nabla}\Phi)&=&0\\
\frac{\mathcal{L}}{2}\nabla(\sigma\BM{\nabla}T^2)+\sigma(\BM{\nabla}\Phi)^2&=&0.
\end{eqnarray}
From these equations it can be easily seen that the temperature is
generally related to the electric potential as:
\begin{equation}\label{therm1}
T^2(\z{r})={\rm const.}-\frac{\Phi^2(\z{r})}{\mathcal L}.
\end{equation}
The constant term is determined by the boundary condition:
\begin{equation}
    T(z\to\pm\infty)\to T_{\rm bath},
\end{equation}
thus:
\begin{equation}\label{therm2}
T^2(\z{r})=T^2_{\rm bath}+\frac{V^2}{4\mathcal
L}-\frac{\Phi^2(\z{r})}{\mathcal L}.
\end{equation}
In an opening type contact the potential is zero at the contact
surface, so the local temperature is written as:
\begin{equation}\label{therm3}
T^2_{\rm PC}=T^2_{\rm bath}+\frac{V^2}{4\mathcal L},
\end{equation}
According to this relation a bias voltage of $100$\,mV already
heats up a contact from cryogenic temperatures ($\sim 4$\,K) to
room temperature. This feature has crucial importance in
experimental studies. If the measurements are performed on a large
contact or at high bias voltages, it can easily happen that the
$I(V)$ curve presents the temperature dependence of the
conductance instead of the spectroscopic features,\cite{VYK} which
will be discussed in Sec.~\ref{scatteringinPC}.

\subsection{Ballistic regime} \label{ballistic} If
the linear dimension of the contact is much smaller than the
minimum of the mean free paths the electrons are travelling along
straight trajectories, and in the first approximation no
scatterings are taken into consideration. This system is described
by the equations of semiclassical dynamics, so the space and
momentum dependent distribution function, $ f_{\z{p}}(\z{r})$ is
to be determined beside the electric potential,
$\Phi(\textbf{r})$. In this limit the electrons ``remember'' which
side of the contact they are coming from, thus the distribution
function at position $\z{r}$ can be expressed as a sum of the
terms corresponding to electrons coming from the left or the right
side of the contact.\cite{KY} The distribution function can be
determined by solving the homogeneous Boltzmann equation which
contains no collision integral:\cite{KSO}
\begin{equation}
\z{v}_{\z{p}}{\partial f_{\z{p}}(\z{r})\over{\partial \z{r}}}
-e\z{E}{\partial f_{\z{p}}(\z{r})\over{\partial \z{p}}}=0,
\label{boltzmann}
\end{equation}
where $\z{E}$ is the electronic field, $\z{E}= -\frac{\partial
\Phi(\z{r})}{\partial \z{r}}$, and $\z{v}_{\z{p}}$ is the velocity
of the electron with momentum $\z{p}$. Note that the electronic
charge $e$ is {\it positive}, so the charge of an electron is
$-e$.  Far from the constriction the boundary condition for the
distribution function is that it has to tend to the equilibrium
distribution function.
\begin{equation}
\lim_{|\z{r}|\rightarrow \infty}f_{\z{p}}(\z{r})=f_0(\varepsilon
_{\z{p}}), \label{limit1}
\end{equation}
where
$f_0(\varepsilon_{\z{p}})=(e^{\varepsilon_{\z{p}}-\mu\over{kT}}+1)^{-1}$.

\begin{figure}
\centering \epsfxsize=0.25\textwidth \epsfbox{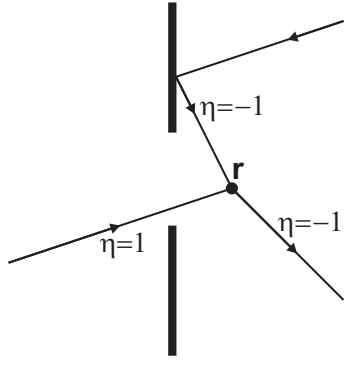}
\vskip0.7truecm \caption{\it Different electron trajectories
arriving at point $\z{r}$. The values of $\eta(\z{p},\z{r})=\pm1$
are also indicated.} \label{eta.fig}
\end{figure}

\begin{figure}
\centering \epsfxsize=0.7\textwidth
\epsfbox{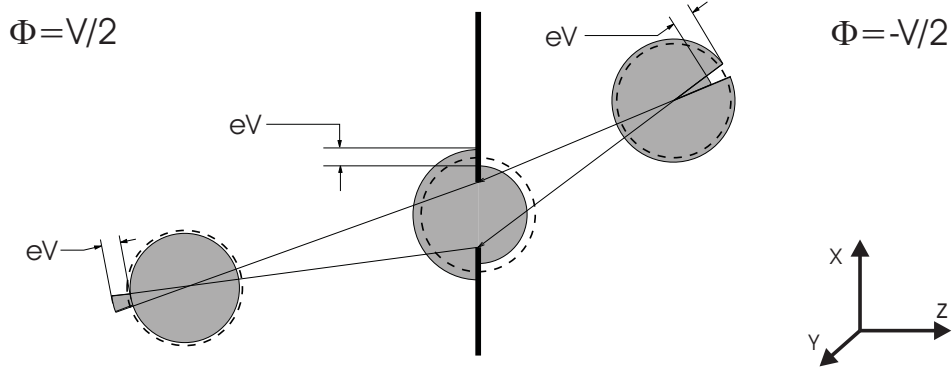} \vskip0.7truecm \caption{\it
Visualization of the momentum distribution function at three
different points in the case of ballistic limit. The shadowed
areas represent the occupied states at function of the direction
of the momentum, $\z{p}$. The dashed circle represents the
equilibrium Fermi surface in case of completely isotropic
distribution. The volume of the shadowed areas are independent of
the position $\z{r}$ (charge neutrality). } \label{distr_ball}
\end{figure}

If we apply $V$ voltage on the sample, the boundary condition for the
potential $\Phi$ takes the form
\begin{equation}
\Phi (z\rightarrow \pm \infty )=\mp {V\over 2}. \label{limit2}
\end{equation}
The solution of the Boltzmann equation can be found by using the
trajectory method.\cite{OKS} (The trajectories of the electrons
are considered to be straight because we are interested in the
limit $e|V|/\varepsilon_{\rm F}\ll 1$ and the current we are
searching for, depends linearly on the electrical field. It can be
shown that the bending of trajectory only contributes to the
current in the higher order of the field.) In this way
\begin{equation}
f_{\z{p}}(\z{r})=f_0\left(\varepsilon _{\z{p}}
+e\int\limits_{-\infty}^{\z{r}} \z{E}{\rm d}\z{l}\right)
\label{trajectory}
\end{equation}
holds combined with the charge neutrality condition
\begin{equation}
2e\int {\rm d}^3 \z{p}
\left[{f_{\z{p}}(\z{r})-f_0(\varepsilon_{\z{p}})}\right]=0,
\label{charge}
\end{equation}
where $\int\limits_{-\infty}^{\z{r}}$ denotes the integration over
the trajectory of the electron coming from the distant reservoir
to the contact region at point $\z{r}$. The distribution function
of an electron satisfying the conditions (\ref{charge}),
(\ref{trajectory}) and (\ref{limit1}) takes the form in the linear
order of the field
\begin{equation}
f_{\z{p}}(\z{r})=f_0\left({\varepsilon_{\z{p}}-e\Phi (\z{r})
+{eV\over 2}\eta (\z{p},\z{r})}\right), \label{distrfunc}
\end{equation}
which is constant along the trajectory as the energy
$\varepsilon_{\z{p}}-e\Phi(\z{r})$ is conserved.
The expression $eV\eta (\z{p},\z{r})/2$ describes the different
value of the chemical potential in the right and left hand sides
of the constriction ($\mu =\mu|_{V=0}\pm eV/2$):
\begin{equation}
\eta(\z{p},\z{r})=\cases{+1 & for electrons arriving at $\z{r}$
from the left reservoir \cr -1 & for electrons arriving at $\z{r}$
from the right reservoir} \label{eta}
\end{equation}
(For an illustration see Fig.~\ref{eta.fig}.) That can be also
expressed in terms of the solid angle, $\Omega (\z{r})$ at which
the opening of the contact is seen from point $\z{r}$:
\begin{equation}
\eta(\z{p},\z{r})=\cases{$sign$(z) &, if $-\z{p}\in \Omega
(\z{r})$ \cr $-sign$(z) &, if  $-\z{p}\not\in \Omega (\z{r})$}.
\label{eta1}
\end{equation}
A simple visualization of the distribution function is presented
in Fig.~\ref{distr_ball}.

The potential can be derived by substituting the form
(\ref{distrfunc}) of the distribution function into the neutrality
condition (\ref{charge}):\cite{OKS}
\be 2e\int {\rm d}^3\z{p}\left[{
f_0\left({\varepsilon_{\z{p}}-e\Phi(\z{r})+{eV\over
2}\eta(\z{p},\z{r}) }\right)-f_0(\varepsilon_{\z{p}})}\right]=0,
\ee where the integral for equienergetic surfaces is easily
expressed by the solid angle $\Omega(\z{r})$, and just an energy
integral remains: \be \int {\rm d}\varepsilon\left[{
{\Omega(\z{r})\over
4\pi}f_0\left({\varepsilon-e\Phi(\z{r})\pm{eV\over
2}}\right)+\left({1-{\Omega(\z{r})\over 4\pi}}\right)
f_0\left({\varepsilon-e\Phi(\z{r})\mp{eV\over
2}}\right)-f_0(\varepsilon)}\right]=0, \label{neutr2} \ee
where the upper and lower signs correspond to the cases $z>0$ and
$z<0$, respectively. The equation can be solved easily by using
the identity \be \int(f_0(\varepsilon+a)-f_0(\varepsilon)){\rm
d}\varepsilon=-a\label{identity}\ee for an arbitrary energy shift
$a$. We can get the potential in the whole space as:\cite{OKS}
\begin{equation}
\Phi(\z{r})=-{V\over 2}\left[{1-{{\Omega(\z{r})}\over
{2\pi}}}\right] \mathrm{sign}(z). \label{potent_op}
\end{equation}
Here $\Omega(\z{r})$ is the solid angle at which the contact is
seen from the position $\z{r}$. The potential depending on the
distance measured from the contact along the $z$ axis is shown in
Fig.~\ref{potent_fig} (solid curve).

The current density, $\z{j}$ is determined by the distribution
function as follows:
\bea \z{j}(\z{r})=-2e\int \frac{{\rm d}^3\z{p}}{(2\pi)^3}
{\z{v}_{\z{p}}f_{\z{p}}(\z{r})}. \label{currdens} \eea
The current through the contact is calculated by integrating the
$z$-component of the current density over the area of the contact,
$A$:
\begin{equation}
I=\int\limits_A {\rm d}A\;j_z(z=0), \label{curr1}
\end{equation}
At the contact surface half of the electrons go to the left with a
distribution function $f_0(\varepsilon-eV/2)$ and half of the
electrons goes to the right with a distribution function
$f_0(\varepsilon+eV/2)$, thus the current is written as \be
I=-e\int\limits_A {\rm
d}A\int\limits_{\varepsilon_{\z{p}}=\varepsilon}\frac{{\rm
d}S_{\varepsilon}}{(2\pi)^3}\int {\rm d}\varepsilon
\frac{|v_z|}{\hbar|\z{v_{\z{p}}}|}\left(f_0(\varepsilon+\frac{eV}{2})-f_0(\varepsilon-\frac{eV}{2})\right),
\ee where the integral $\int {\rm d}S_{\varepsilon}$ is taken over
an equienergetic surface in the $\z{p}$-space. At low enough
temperature and voltage ($k_{\rm B}T\ll \varepsilon _{\rm F}$ and
$eV\ll \varepsilon _{\rm F}$) the expression $\int
({f_0(\varepsilon +{eV\over 2})- f_0(\varepsilon -{eV\over
2})}){\rm d}\varepsilon$ equals $eV$ according to the identity
(\ref{identity}), so the current is written as: \be
I=\frac{e^2AS_{\rm F}}{\hbar(2\pi)^3}\langle
\cos(\vartheta)\rangle_{\rm FS}\cdot V,\label{sharvin}\ee where
$A$ is the area of the contact, $S_{\rm F}$ is the area of the
Fermi surface, $\cos\vartheta =v_z/v_{\rm F}$ and
$\langle{...}\rangle_{\rm FS}=\int {\rm d}S_{\rm F}(...)/S_{\rm
F}$ denotes the average taken over the Fermi surface. The
conductance of a ballistic point-contact given by the formula
(\ref{sharvin}) is known as the Sharvin conductance.\cite{Sh} For
free electron gas the Sharvin conductance is simplified as: \be
G_{\rm S}=\frac{2e^2}{h}\left(\frac{k_{\rm F} d}{4} \right)^2.
\label{sharvinconductance}\ee

\subsection{Intermediate case between the ballistic and diffusive
limit} \label{inter}

In the intermediate region between the diffusive and ballistic
regime an interpolating formula can be set up by numerically
solving the Boltzmann equation for arbitrary ratio of the contact
diameter and mean free path, $l$:\cite{Interp}
\begin{equation}
R=l/d\cdot\frac{16}{3\pi \sigma d}+\Gamma(l/d)\frac{1}{\sigma d},
\label{interpol.eq}
\end{equation}
where $\Gamma(l/d)$ is a numerically determinable monotonous
function, $\Gamma (0)=1$; $\Gamma(\infty)=0.694$. Note that the
first term is exactly the Sharvin resistance by putting the Drude
resistivity into the formula, $\rho=mv_F/le^2n$, thus it is
actually independent of $l$. This formula provides the possibility
to estimate the contact diameter from the contact resistance.

\section{Scattering on slow TLSs in point contacts.}
\label{scatteringinPC}

The voltage applied on a point contact results in a nonequilibrium
distribution of the conduction electrons in the contact region. An
electron coming from the right reservoir has an energy larger by
$eV$ than those coming from the left reservoir. This energy can be
released through inelastic scattering processes, which can happen
in such a way that an electron that has already crossed the
contact is scattered back through the opening. This so-called
backscattering reduces the current, thus the energy dependence of
the scattering probability can be traced by measuring the
nonlinearity of the current as the bias voltage is varied. This
phenomenon was first used by Igor Yanson\cite{Yanson1} (for a
review see \onlinecite{YS}) to study the phonon spectra and the
electron-phonon interaction and since that it is widely applied.

In this section we discuss the nonlinearities in the current
voltage characteristics due to the scattering on slow TLSs, which
show strong similarities to localized phonons. There is, however,
an essential difference between the two cases, while the phonons
contribute to the voltage region up to $30$\,meV, the TLSs
manifest themselves below or even well below $1$\,meV due to their
characteristic energies. The main contribution to the $I(V)$
characteristics comes from the close region of the contact,
therefore a microscopic process like a transition of one atom from
one position to another in the contact region occurs as a
significant measurable change in the current. Considering the TLSs
the main advantage of the point contact spectroscopy is that in
case of small enough contacts one can investigate  just one or few
scattering centers.

The current through the contact can be derived by solving the classical
stationary Boltzmann equation
\begin{equation}
\z{v}_\z{p}{{\partial
f_\z{p}(\z{r})}\over{\partial\z{r}}}-e\z{E}{{\partial
f_\z{p}(\z{r})}\over{\partial\z{p}}}=I(\z{p},\z{r}),
\label{boltzmann_new}
\end{equation}
where $I(\z{p},\z{r})$ is the collision integral for the electron
with momentum $\z{p}$ at position $\z{r}$, which is assumed to be
local in real space.

In the case of bulk phonons the collision integral is
\begin{eqnarray}
I_{\rm ph}(\z{p},\z{r})&=&\sum\limits_\alpha\int{\rm
d}^3\z{q}\;W_{\z{q}}^\alpha
\Bigg\{\left[f_{{\z{p}}+{\z{q}}}(1-f_{\z{p}})(N_{\z{q}}^\alpha+1)-f_{\z{p}}(1-f_{{\z{p}}+{\z{q}}})N_{\z{q}}^\alpha\right]
\delta(\varepsilon_{{\z{p}}+{\z{q}}}-\varepsilon_{\z{p}}-\omega_{\z{q}}^\alpha)+
\nonumber\\
&{ }&
\left[f_{{\z{p}}-{\z{q}}}(1-f_{\z{p}})N_{\z{q}}^\alpha-f_{\z{p}}(1-f_{{\z{p}}-{\z{q}}})(N_{\z{q}}^\alpha+1)\right]
\delta(\varepsilon_{{\z{p}}-{\z{q}}}-\varepsilon_{\z{p}}-\omega_{\z{q}}^\alpha)\Bigg\},
\label{eq70}
\end{eqnarray}
where $W_{\z{q}}$ is proportional to the squared matrix element of
the electron-phonon coupling, $N_{\z{q}}^\alpha$ is the phonon
occupation number for momentum ${\z{q}}$ and energy
$\omega_{\z{q}}$ in phonon branch $\alpha$. There is another term
of the collision integral which is due to the elastic (impurity)
scattering but that is not given here. The collision integral
vanishes far from the point contact because well inside the
electrodes thermal equilibrium exists. The situation is further
simplified when the following assumptions are valid: (i) the
electron-phonon interaction is homogeneous in the real space; (ii)
the phonon distribution corresponds to thermal equilibrium. In
that case the information on the contact geometry and also the
elastic scattering due to impurities in the dirty limit can be
expressed by a geometrical factor $K(\z{p},\z{p}')$ in the
expression of the current which depends only on the momenta
$\z{p}$ and $\z{p}'$ of the incoming and outgoing electrons. That
factor is frequently called as the $K$-factor (see
e.g.~\onlinecite{KSS,Ashraf1982}). In the case of phonons the
$K$-factor for a ballistic opening like contact is written as:
\begin{equation}\label{Kfactorball}
K(\z{p},\z{p}')=\frac{|\hat{p}_z\hat{p}_z'|}{|\hat{p}_z'\hat{\z{p}}-\hat{p}_z\hat{\z{p}}'|}\Theta(-\hat{p}_z\hat{p}_z'),
\end{equation}
where $\hat{\z{p}}$ $\hat{\z{p}}'$ are the unit vector parallel
with $\z{p}$ and $\z{p}'$, while $\hat{p}_z$ and $\hat{p}_z'$ are
the $z$ components of these unit vectors, where the $z$ direction
is the axis of the contact. The $K$-factor can be also calculated
for a diffusive contact:
\begin{equation}\label{Kfactordiff}
K(\z{p},\z{p}')=\frac{9\pi}{32}\frac{l_{\rm
el}}{d}(\hat{p}_z^2-\hat{p}_z'^2),
\end{equation}
where $l_{\rm el}$ is the elastic mean free path, and $d$ is the
contact diameter.

At zero temperature the logarithmic derivative of the resistance
can be expressed by a function
$G(\omega)=\tilde{\alpha}^2(\omega)F(\omega)$, where $F(\omega)$
is the density of states of phonons and $\alpha(\omega)$ is the
modified electron-phonon coupling which contains the $K$-factor.
Without that factor it is just the electron-phonon coupling
strength and in that case $G(\omega)$ is the Eliashberg function
known in the theory of superconductivity. The main result is
\begin{equation}
{1\over R}{dR\over{dV}}\sim G({eV\over\hbar}).
\label{phonon_G}
\end{equation}
In case of single crystals $G(\omega)$ depends also on the
orientation of the crystal, which makes it possible to study the
anisotropy of the phonon spectrum.\cite{YS} Measuring ${1\over
R}{dR\over{dV}}$ the phonon spectra were determined for many
different cases and the structures in them were identified as the
details of the spectra.\cite{Khotkevich1995} Those structures,
however, are superimposed on a continuous background.

Until now, it is assumed that the phonons are in thermal
equilibrium, which is the correct assumption as far as the phonons
at the contact region arrive from a large distance without any
collision where the thermal bulk distribution is realized. That
is, however, not the case when in the contact region the phonon
mean free path is comparable with the contact size. Then the
phonons generated by the non-equilibrium electrons in the contact
region cannot escape thus the phonons are also out of equilibrium.
The measure of non-equilibrium is limited by the energy relaxation
time of the phonons $\tau_{\rm ph}$. The non-equilibrium localized
phonons contribute to the background
resistance.\cite{vanGelder1978,vanGelder1980,Jansen1980} That can
be studied by measuring the amplitude of the background as a
function of the frequency of the applied
voltage\cite{Kulik1985,Yanson1985,Balkashin1992} and at high
frequency ($\omega\tau_{\rm ph}\gg 1$) that is decreasing. These
phenomena will be studied in the context of scattering on TLSs as
well.

\subsection{Scattering on TLSs in a ballistic contact}

In the following the Boltzmann equation will be solved considering
only the electron-TLS interaction. It will be assumed that the
density of the TLSs is low thus an electron scatters only once in
the orifice region and the contribution of the different TLSs are
additive. First the ballistic limit is treated, where both the
elastic and inelastic mean free paths are larger than the size of
the contact.

The collision term related to a single TLS is very similar to the
case of phonons, the momentum is, however, not conserved.
Furthermore, the TLS has only two states with population $n_+$ and
$n_-$ ($n_++n_-=1$) corresponding to the TLS energy levels $E_+$
and $E_-$ (see Sec.~\ref{TLSinsolids}). The situation is different
from those cases of phonons, where the phonon mean free path is
large compared to the size of orifice, $l_{\rm ph} \gg d$ and the
phonon distribution is the bulk one which is in thermal
equilibrium with the contact regions. In the case of TLSs the
interaction is localized in space, thus the TLSs in the contact
region are decoupled from the bulk region and they can be
considerably out of thermal equilibrium due to the voltage drop at
the contact. Therefore, the situation is similar to the special
case of localized phonons with $l_{\rm ph}\leq d$ shortly
discussed in the introductory remarks of this section.

The collision term similar to Eq.~(\ref{eq70}) can be written as
the sum of inelastic and elastic terms as $I_{{\z{R}},\rm
TLS}(\z{p},\z{r})=I_{{\z{R}},\rm
TLS,in}(\z{p},\z{r})+I_{{\z{R}},\rm TLS,el}(\z{p},\z{r})$. In the
inelastic case the TLS jumps to its other state due to the
scattering, and thus the energy of the electron changes by
$E=E_+-E_-$. The strength of the interaction is characterized by
$W_{\z{p}\z{p}'}$, which is obtained from the interaction matrix
elements (Eq.~\ref{matrix_elements_inel}) by Fermi's golden rule
as
\begin{equation}\label{scatteringstrength}
W_{\z{p}\z{p}'}=\frac{2\pi}{\hbar}\left|\langle \z{p},E_+|H_{\rm
e-TLS}|\z{p}',E_-\rangle\right|^2=\frac{2\pi}{\hbar}\left|\langle
\z{p},E_-|H_{\rm
e-TLS}|\z{p},'E_+\rangle\right|^2=\frac{2\pi}{\hbar}\left|2\mu\nu
V^z_{\z{p}\z{p}'}\right|^2.
\end{equation}
The interaction strength is symmetric in the incoming and outgoing
momentum, and in the neighborhood of the Fermi surface it depends
only on the unit vectors $\hat{\z{p}}$ and $\hat{\z{p}}'$, i.e.\
$W_{\z{p}\z{p}'}=W_{\hat{\z{p}}\hat{\z{p}}'}=W_{\hat{\z{p}}'\hat{\z{p}}}$.
The collision integral for the inelastic scattering is written as:
\begin{eqnarray}
I_{{\z{R}},\rm TLS,in}(\z{p},\z{r})&\simeq&\delta(\z{r}-\z{R})
\int{{\rm d}^3\z{p}'\over{(2\pi\hbar)^3}}W_{\z{p}\z{p}'}\times\nonumber\\
&{ }&\Bigg\{
f_{{\z{p}}'}(1-f_{\z{p}})n_-\delta(\varepsilon_{{\z{p}}'}-\varepsilon_{{\z{p}}}-E)
+f_{{\z{p}}'}(1-f_{\z{p}})n_+\delta(\varepsilon_{{\z{p}}'}-\varepsilon_{{\z{p}}}+E)
\nonumber\\
& &
-f_{{\z{p}}}(1-f_{{\z{p}}'})n_-\delta(\varepsilon_{{\z{p}}}-\varepsilon_{{\z{p}}'}-E)
-f_{{\z{p}}}(1-f_{{\z{p}}'})n_+\delta(\varepsilon_{{\z{p}}}-\varepsilon_{{\z{p}}'}+E)\Bigg\},
\label{IRTLSin}
\end{eqnarray}
where $\z{R}$ is the place of the TLS.

The current correction due to the elastic scattering processes can
be written similarly with two remarks: (i) in an elastic process
the energy of the TLS does not change, thus $E=0$ must be
inserted; (ii) the scattering cross section is different for the
TLS being in the ``$+$'' and the ``$-$'' state. According to the
matrix elements in Eq.~(\ref{matrix_elements_elastic}), the two
scattering strengths are given as:
\begin{eqnarray}\label{Wel}
W^+_{\hat{\z{p}}\hat{\z{p}}'}&=&\frac{2\pi}{\hbar}\left|\langle
\z{p},E_+|H_{\rm e-TLS}|\z{p}',E_+\rangle\right|^2=\frac{2\pi}{\hbar}\left|V^0_{{\z{p}}{\z{p}'}}+(\mu^2-\nu^2)V^z_{{\z{p}}{\z{p}'}}\right|^2  \nonumber\\
W^-_{\hat{\z{p}}\hat{\z{p}}'}&=&\frac{2\pi}{\hbar}\left|\langle
\z{p},E_-|H_{\rm
e-TLS}|\z{p}',E_-\rangle\right|^2=\frac{2\pi}{\hbar}\left|V^0_{{\z{p}}{\z{p}'}}-(\mu^2-\nu^2)V^z_{{\z{p}}{\z{p}'}}\right|^2.
\end{eqnarray}
After introducing these notations the correction to the collision
integral for the elastic scattering is:
\begin{eqnarray}
I_{\z{R},\rm TLS,el}(\z{p},\z{r})&\simeq&\delta(\z{r}-\z{R})
\int{{\rm
d}^3\z{p}'\over{(2\pi\hbar)^3}}\left(W^+_{\hat{\z{p}}\hat{\z{p}}'}n_+
+W^-_{\hat{\z{p}}\hat{\z{p}}'}n_-\right)\left[f_{{\z{p}}'}(1-f_{\z{p}})
-f_{{\z{p}}}(1-f_{{\z{p}}'})\right]\delta(\varepsilon_{{\z{p}}}-\varepsilon_{{\z{p}}'})\label{IRTLSel}.
\end{eqnarray}

The electron distribution function $f_{\z{p}}(\z{r})$ and the
electric potential is expanded in terms of $d/l_{\rm in}$. At a
large distance measured from the orifice the potential is constant
and $\lim_{|{\z{r}}|\to\infty}\Phi(\z{r})=\pm V/2$. Thus, at
$|{\z{r}}|\gg l_{\rm in}$ the electrons are in thermal
equilibrium. The distribution function $f_{\z{p}}(\z{r})$ and the
potential $\Phi(\z{r})$ can be expanded as
\begin{eqnarray}
f_{\z{p}}(\z{r})&=&f_{\z{p}}^{(0)}(\z{r})+
f_{\z{p}}^{(1)}(\z{r})+\dots\nonumber\\
\Phi(\z{r})&=&\Phi^{(0)}(\z{r})+\Phi^{(1)}(\z{r})+\dots,
\end{eqnarray}
and similarly the electric field
\begin{equation}
\z{E}(\z{r})=\z{E}^{(0)}(\z{r})+\z{E}^{(1)}(\z{r})+\dots,
\end{equation}
where the upper indices $(0)$, $(1)$ label the order in the
strength of the electron-TLS coupling. The zero order terms have
been previously calculated and the following treatment is
restricted to the first order terms. The previous results are
\begin{equation}
f_{\z{p}}^{(0)}(\z{r})=f_0\left(\varepsilon_\z{p}-e\Phi(\z{r})+{eV\over2}
\eta(\z{p},\z{r})\right), \label{eq:f0}
\end{equation}
and according to Eq.~(\ref{potent_op})
\begin{equation}
\Phi^{(0)}(\z{r})=-{V\over2}\left(1-{\Omega(\z{r})\over{2\pi}}\right)
{\rm sign}(z),
\end{equation}
for a round orifice perpendicular to the $z$-axis, where $f_0$ is
the Fermi function. The first order terms in the Boltzmann
equation given by Eq.~(\ref{boltzmann_new}) can be written as
\begin{eqnarray}
\z{v}_\z{p}{\partial f_{\z{p}}^{(1)}(\z{r})\over{\partial\z{r}}}-
e\z{E}^{(0)}(\z{r}){\partial
f_{\z{p}}^{(1)}(\z{r})\over{\partial\z{p}}}=
e\z{E}^{(1)}(\z{r}){\partial
f_{\z{p}}^{(0)}(\z{r})\over{\partial\z{p}}}+ I_{\rm
coll}^{(0)}(\z{p},\z{r}),
\end{eqnarray}
where the label $(0)$ of the collision term indicates that the collisions
are calculated with the distribution functions of zeroth order, $f^{(0)}$.

\begin{figure}
\centering
\includegraphics[width=5truecm]{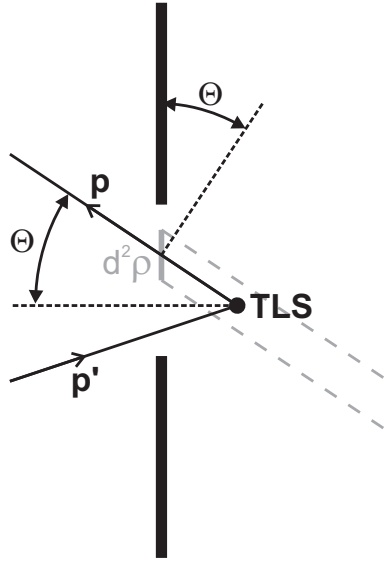}
\caption{\it The orifice with the path of the electron scattered
by the TLS on the right. The current flows through the surface
element $d^2\varrho$ and the angle $\theta$ is also
indicated.}\label{pcTLS1}
\end{figure}

The change in the potential is determined again by the neutrality
condition given by Eq.~(\ref{charge}) for $f^{(0)}$, which is
\begin{equation}
-e\int{{\rm
d}^3\z{p}\over{(2\pi\hbar)^3}}f_{\z{p}}^{(1)}(\z{r})=0.
\end{equation}
The boundary conditions are
\begin{equation}
\lim_{|\z{r}|\to\infty}f_{\z{p}}^{(1)}(\z{r})=0,
\end{equation}
and
\begin{equation}
\lim_{|\z{r}|\to\infty}\Phi^{(1)}(\z{r})=0.
\end{equation}
In  order to determine $f^{(1)}$ and $\Phi^{(1)}$ the trajectory
method is used (see e.g.\ \onlinecite{Jansen1980}) dealing with
phonons. Electron trajectories are considered where the electron
comes from the left or right and arrives to the plane of the
orifice at time $\tau=0$ at point $\z{r}$ with momentum $\z{p}$ to
calculate the current flowing through the orifice. These
trajectories correspond to zeroth order. Moving along the
trajectory $\z{r}(\tau),\z{p}(\tau)$ the distribution function
varies as
\begin{equation}
{\partial f^{(1)}(\z{p}(\tau),\z{r}(\tau))\over{\partial\tau}}=
{\partial\z{p}(\tau)\over{\partial\tau}} {\partial
f^{(1)}(\z{p}(\tau),\z{r}(\tau)) \over{\partial\z{p}(\tau)}}+
{\partial\z{r}(\tau)\over{\partial\tau}} {\partial
f^{(1)}(\z{p}(\tau),\z{r}(\tau)) \over{\partial\z{r}(\tau)}},
\end{equation}
where ${\partial\z{p}(\tau)\over{\partial\tau}}=-e\z{E}$, and
${\partial\z{r}(\tau)\over{\partial\tau}}=\z{v}_\z{p}$. This
equation can be integrated and using the Boltzmann equation one
finds for $f^{(1)}$ at $\tau=0$
\begin{eqnarray}
f^{(1)}(\z{p},\z{r})= \int\limits_{-\infty}^0{\rm d}\tau\Big\{
-e{{\rm d}\Phi^{(1)}(\z{r}(\tau))\over{{\rm d}\z{r}}} {\partial
f^{(0)}(\z{p}(\tau),\z{r}(\tau))
\over{\partial\z{p}(\tau)}}+I_{\rm
coll}^{(0)}(\z{p}(\tau),\z{r}(\tau)) \Big\}. \label{eq:f1}
\end{eqnarray}
Here the collision term describes the adding or taking off
electrons to the trajectory arriving at the contact at $\tau=0$.
In the expression of $f^{(1)}$ only the terms linear in the
voltage are kept, thus ${\partial f^{(0)}(\z{p}(\tau),\z{r}(\tau))
\over{\partial\z{p}(\tau)}}$ in the above equation can be
approximated by the zeroth order term $\z{v}_{\z{p}}{\partial
f_0\over{\partial\varepsilon_{\z{p}}}}$. Taking into account that
${\rm d}\z{r}=\z{v}_{\z{p}}{\rm d}\tau$ the first term of the
integral can be performed, thus
\begin{equation}
f^{(1)}(\z{p},\z{r})= -e\Phi^{(1)}(\z{r}){\partial
f_0\over{\partial\varepsilon_{\z{p}}}}+
\int\limits_{-\infty}^0{\rm d}\tau\; I_{\rm
coll}^{(0)}(\z{p}(\tau),\z{r}(\tau)). \label{eq:f1.2}
\end{equation}
The equation of electrical neutrality taken at $\tau=0$ with
$\z{p}(\tau=0)=\z{p}$ and $\z{r}(\tau=0)=\z{r}$ combined with the
equation above determines $\Phi^{(1)}(\z{r})$
\begin{equation}
e\Phi^{(1)}(\z{r})={\int{{\rm d}^3\z{p}\over{(2\pi\hbar)^3}}
\int\limits_{-\infty}^0{\rm d}\tau\; I_{\rm
coll}^{(0)}(\z{p}(\tau),\z{r}(\tau)) \over {\int{{\rm
d}^3\z{p}\over{(2\pi\hbar)^3}}{\partial
f_0\over{\partial\varepsilon_{\z{p}}}}}}.
\end{equation}
The general expression of the electrical current flowing through the
surface of the contact is
\begin{equation}
I=-2e\int {\rm d}^2\varrho\int{{\rm
d}^3\z{p}\over{(2\pi\hbar)^3}}v_{\z{p}}^zf(\z{p},\z{r}),
\end{equation}
where ${\rm d}^2\varrho$ is the surface element of the contact and
the integral is taken over the orifice, the factor $2$ is due to
the electron spin. The change $\delta I$ in the total current due
to the presence of the TLS can be separated as
$$I(V)=I_0(V)+\delta I(V),$$
and $\delta I(V)$ can be further split, whether the electron scattering is
elastic or inelastic
$$
\delta I(V)=\delta I_{\rm el}+\delta I_{\rm in}.
$$
Making use of the expression (\ref{eq:f1.2}) for $f^{(1)}$ only
the collision term contributes to the current because the other
term is even in the momentum. The integral according to the time
can be transformed to the one along the path using ${\rm
d}\tau={{\rm d}s\over{v_{\z{p}}}}$ where ${\rm d}s$ is the element
of the path. The expression for the change of the current due to
the presence of collision on TLS is
\begin{equation}
\delta I=-2e\int {\rm d}^2\varrho\int{{\rm
d}^3\z{p}\over{(2\pi\hbar)^3}}v_{\z{p}}^z
\int\limits_{-\infty}^0{{\rm d}s\over{v_{\z{p}}}} I_{\rm
coll}^{(0)}(\z{p}(\tau),\z{r}(\tau)).
\end{equation}
The path of the integral is changed due to the collisions but it
may contain reflections by the insulating surface of the contact.
The next step is to introduce the collision term due to a single
TLS at position $\z{R}$, given by Eq.~(\ref{IRTLSin}). As the
paths arriving at the surface element of the orifice ${\rm
d}^2\varrho$ are straight lines, the volume element for the
scattering event is ${\rm d}^3\z{r}={\rm d}^2\varrho {\rm
d}s\cos\theta$ where $\theta$ is the angle between the path and
$z$ direction. Now the integration over different paths and
position of the collision gives
\begin{equation}
\delta I_{\z{R}}=-2e\int\limits_{\Omega_{\z{R}}} {{\rm
d}^3\z{p}\over{(2\pi\hbar)^3}}{v_{\z{p}}^z\over{v_{\z{p}}}}\cos\theta
\cdot I_{{\z{R}},\rm TLS}(\z{p})
\end{equation}
where $I_{{\z{R}},\rm TLS}$ is defined by $I_{{\z{R}},\rm
TLS}(\z{p},\z{r})=\delta(\z{r}-\z{R})I_{{\z{R}},\rm TLS}(\z{p})$
and the momentum integral is restricted to the solid angle
$\Omega_{\z{R}}$ at which the contact can be seen from the TLS at
position $\z{R}$. Furthermore,
$v_{\z{p}}^z/v_{\z{p}}=-\cos\theta\;{\rm sign}(R^z)$ as the
electron passes the contact from the TLS (for an illustration see
Fig.~\ref{pcTLS1}). The final expression is obtained using ${{\rm
d}{\z{p}}^3\over{(2\pi\hbar)^3}}=\varrho_0{{\rm
d}\Omega_{\z{p}}\over{4\pi}}{\rm d}\varepsilon$ where $ {\rm
d}\Omega_{\z{p}}$ is the solid angle element in the momentum space
and $\varrho_0$ is the conduction electron density of states for
one spin direction. Assuming a spherical Fermi surface
\begin{equation}
\delta I_{\z{R}}=2e\;{\rm
sign}(R^z)\varrho_0\int\limits_{\z{p}\in\Omega_{\z{R}}} {{\rm
d}\Omega_{\z{p}}\over{4\pi}}\int {\rm
d}\varepsilon\;\cos^2\theta\cdot I_{{\z{R}},\rm TLS}(\z{p}).
\end{equation}

\subsubsection{Current correction related to inelastic scattering}

Using the expression of the collisison term $I_{{\z{R}},\rm
TLS,in}$ (\ref{IRTLSin}) the change in the current due to the
inelastic scattering can be written as
\begin{eqnarray}\label{dIRin}
&&\delta I_{{\z{R}},{\rm in}}=\frac{2e}{\hbar}{\rm
sign}(R_z)\int\limits_{\z{p}\in\Omega_{\z{R}}} {{\rm
d}\Omega_{\z{p}}\over{4\pi}}\int {\rm d}\varepsilon\;\cos^2\theta
\int{{\rm
d}\Omega_{\z{p}'}\over{4\pi}}w_{\hat{\z{p}}\hat{\z{p}}'}\int {\rm
d}\varepsilon'\nonumber\\
&{ }&\Big\{
\left[f_{\z{p}'}^0(1-f_{\z{p}}^0)n_--f_{\z{p}}^0(1-f_{\z{p}'}^0)n_+\right]
\delta(\varepsilon+E-\varepsilon')
+\left[f_{\z{p}'}^0(1-f_{\z{p}}^0)n_+-f_{\z{p}}^0(1-f_{\z{p}'}^0)n_-\right]
\delta(\varepsilon-E-\varepsilon')\Big\},
\end{eqnarray}
where the distribution functions $f^0$ are given by
Eq.~(\ref{eq:f0}) and taken at $\z{r}=\z{R}$, and the
dimensionless notation
\begin{equation}\label{kisw}
w_{\hat{\z{p}}\hat{\z{p}}'}=\hbar\varrho_0^2W_{\hat{\z{p}}\hat{\z{p}}'}
\end{equation}
is introduced. The potential $\Phi(\z{R})$ appearing in these
expressions can be dropped as the energy integral variables can be
shifted. The essential contribution is given by
${eV\over2}\eta(\z{p},\z{R})=\pm{eV\over2}$, which has a positive
sign if the electron at $\z{r}=\z{R}$ arrives from the left
contact and a negative sign if it is arriving from the right
contact.

The final expression is
\begin{eqnarray}
&&\delta I_{\z{R},{\rm in}}=\frac{2e}{\hbar}{\rm
sign}(R_z)\int\limits_{\z{p}\in\Omega_{\z{R}}} {{\rm
d}\Omega_{\z{p}}\over{4\pi}}w_{\hat{\z{p}}\hat{\z{p}}'}\int {\rm
d}\varepsilon\;\cos^2\theta \int{{\rm
d}\Omega_{\z{p}'}\over{4\pi}}\int {\rm
d}\varepsilon'\nonumber\\
&&\Bigg\{\Big[
f_0(\varepsilon'+{eV\over2}\eta(\z{p}'))(1-f_0(\varepsilon+{eV\over2}\eta(\z{p})))n_--
f_0(\varepsilon+{eV\over2}\eta(\z{p}))(1-f_0(\varepsilon'+{eV\over2}\eta(\z{p}')))n_+
\Big]
\delta(\varepsilon+E-\varepsilon')+\nonumber\\
&&\Big[f_0(\varepsilon'+{eV\over2}\eta(\z{p}'))(1-f_0(\varepsilon+{eV\over2}\eta(\z{p})))n_+-
f_0(\varepsilon+{eV\over2}\eta(\z{p}))(1-f_0(\varepsilon'+{eV\over2}\eta(\z{p}')))n_-
\Big]
\delta(\varepsilon-E-\varepsilon')\Bigg\}.\nonumber\\
\end{eqnarray}

\begin{figure}
\centering
\includegraphics[width=0.7\textwidth]{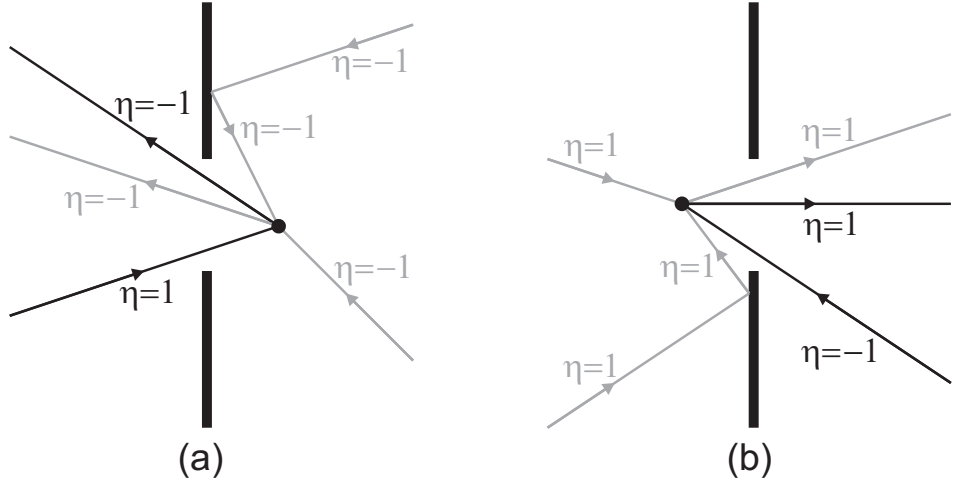}
\caption{The paths for forward (grey line) and backscattering
(black line) are shown with the TLS on the right (a) or left (b).
The $\eta$ signs are also indicated.} \label{fig:paths}
\end{figure}

This expression can be divided into two parts. If the electron
with momentum $\z{p}'$ arrives from the same side of the contact
($|\z{r}|\to\infty$) as the unscattered electron with momentum
$\z{p}$, thus $\eta(\z{p})\eta(\z{p}')=1$, then the electron is
scattered forward, while for $\eta(\z{p})\eta(\z{p}')=-1$ it is
scattered backward (see Fig.~\ref{fig:paths}). The total current
can be divided according to that, thus
\begin{equation}
\delta I_{\z{R},{\rm in}}=\delta I_{\z{R},{\rm in, for}}+\delta
I_{\z{R},{\rm in, back}}.
\end{equation}

The forward scattering cancels out. That can be seen by looking
e.g.\ the first and the last term in the previous expression and
changing $\varepsilon\leftrightarrow\varepsilon'$,
\begin{equation}
\delta I_{\z{R},{\rm in, for}}=0.
\end{equation}

In the calculation of the backscattering the case ${\rm
sign}(R_z)=1$ is taken first. Then the electron contributing to
the current originally comes from the left ($\eta(\z{p}')=1$) and
after the scattering goes from right to the left
($\eta(\z{p})=-1$).
\begin{eqnarray}\label{dIin.right}
\delta I_{\z{R},{\rm in, back}}&=&
\frac{2e}{\hbar}\int\limits_{\z{p}\in\Omega_{\z{R}}} {{\rm
d}\Omega_{\z{p}}\over{4\pi}}\int {\rm d}\varepsilon\;\cos^2\theta
\int\limits_{{-\z{p}\;'\in\Omega_{\z{R}}}}{{\rm
d}\Omega_{\z{p}'}\over{4\pi}}\int {\rm d}\varepsilon'\;
w_{\hat{\z{p}}\hat{\z{p}}'}\cdot
\nonumber\\
&{ }&\Bigg\{
\Big\{f_0(\varepsilon'+{eV\over2})(1-f_0(\varepsilon-{eV\over2}))n_--
f_0(\varepsilon-{eV\over2})(1-f_0(\varepsilon'+{eV\over2}))n_+\Big\}
\delta(\varepsilon+E-\varepsilon')\nonumber\\
&{+}&\Big\{f_0(\varepsilon'+{eV\over2})(1-f_0(\varepsilon-{eV\over2}))n_+-
f_0(\varepsilon-{eV\over2})(1-f_0(\varepsilon'+{eV\over2}))n_-\Big\}
\delta(\varepsilon-E-\varepsilon')\Bigg\}.
\end{eqnarray}

The factor due to the angular integrals
$\int\limits_{\z{p}\in\Omega_{\z{R}}}{{\rm
d}\Omega_{\z{p}}\over{4\pi}}\cos^2\theta\int\limits_{{-\z{p}\;'\in\Omega_{\z{R}}}}{{\rm
d}\Omega_{\z{p}'}\over{4\pi}}w_{\hat{\z{p}}\hat{\z{p}}'}$ plays
the role of the geometrical factor for the bulk phonons. Only the
backscattering contributes like in the case of phonons as it can
be seen from Eq.~(\ref{Kfactorball}) for ballistic regions where
the $K$-factor contains the factor $\Theta(-\hat{p}_z\hat{p}_z')$.

In the following the geometrical factor is introduced which
depends on the position of the TLS, $\z{R}$, and it is independent
of the strength of the interaction
\begin{equation}
K_{\z{R}}={1\over w} \int\limits_{\z{p}\in\Omega_{\z{R}}}{{\rm
d}\Omega_{\z{p}}\over{4\pi}}
\cos^2\theta\int\limits_{{-\z{p}'\in\Omega_{\z{R}}}}{{\rm
d}\Omega_{\z{p}'}\over{4\pi}} w_{\hat{\z{p}}\hat{\z{p}}'},
\end{equation}
where
\begin{equation}\label{w}
w=\int{{\rm d}\Omega_{\z{p}}\over{4\pi}}\int{{\rm
d}\Omega_{\z{p}'}\over{4\pi}}w_{\hat{\z{p}}\hat{\z{p}}'}.
\end{equation}

If the momentum dependence of the interaction is ignored then
$w_{\hat{\z{p}}\hat{\z{p}}'}=w$ and
\begin{equation}\label{KR}
K_{\z{R}}=\int\limits_{\z{p}\in\Omega_{\z{R}}}{{\rm
d}\Omega_{\z{p}}\over{4\pi}}
\cos^2\theta\int\limits_{{-\z{p}'\in\Omega_{\z{R}}}}{{\rm
d}\Omega_{\z{p}'}\over{4\pi}}
\end{equation}
is a pure geometrical factor. Further on this simplification is
assumed. If the TLS is positioned on the geometrical axis of a
circular contact then the above integral can be evaluated as
\begin{equation}
K_\z{R}=\left(\frac{\Omega_\z{R}}{4\pi}\right)^2\left(1-\frac{4}{3}\left(\frac{\Omega_\z{R}}{4\pi}\right)^2\right).
\end{equation}

In the second part of the integral in Eq.~(\ref{dIin.right}) the
variables are changed as $\varepsilon\leftrightarrow\varepsilon'$
and the integral with respect $\varepsilon'$ is performed,
furthermore, in some of the integrals the variable is shifted as
$\varepsilon-{eV\over2}\to\varepsilon$. The result is
\begin{eqnarray}
\delta I_{{\z{R}},{\rm in, back}}=\frac{2e}{\hbar}wK_{\z{R}}\int
{\rm d}\varepsilon\; f_0(\varepsilon)
&\Big\{&\left[f_0(\varepsilon-E+eV)-f_0(\varepsilon-E-eV)\right]n_-\nonumber\\
&+&\left[f_0(\varepsilon+E+eV)-f_0(\varepsilon+E-eV)\right]n_+
\Big\}.\label{eq:112}
\end{eqnarray}

Now the TLS on the left is considered (sign$(R_z)=-1$). Then for
backscattering the signs of $\eta(\z{p})$ and $\eta(\z{p}')$ are
reversed (see Fig.~\ref{fig:paths}b) That is equivalent of
changing the sign of $V$ ($V\to -V$) in Eq.~(\ref{dIin.right}) and
then the expression in the curly bracket also changes its sign as
it can be seen by inserting
$\varepsilon\leftrightarrow\varepsilon'$. That negative sign is
cancelled by sign$(R_z)=-1$, thus the total current is unchanged.
Therefore, the current does not depend on whether the TLS is on
the right or left as it should be.

After inserting the Fermi functions and performing the energy
integral, the following result is obtained for the inelastic
current correction:
\begin{equation}\label{Iinel}
\delta I_{\z{R},\rm in}=n_-C+n_+D,
\end{equation}
where
\begin{eqnarray}\label{CD}
C=\frac{2e}{\hbar}wK_{\z{R}}\left\{ {eV+E\over
{2}}\Bigg[\coth({eV+E\over {2k_BT}})-1\Bigg]- {E-eV\over
{2}}\Bigg[\coth({E-eV\over {2k_BT}})-1\Bigg]\right\},\nonumber\\
D=\frac{2e}{\hbar}wK_{\z{R}}\left\{ {eV-E\over
{2}}\Bigg[\coth({eV-E\over {2k_BT}})-1\Bigg]- {-E-eV\over
{2}}\Bigg[\coth({-E-eV\over {2k_BT}})-1\Bigg]\right\}.
\end{eqnarray}
These expressions are simplified at $T=0$ as
\begin{equation}
\delta I_{\z{R},\rm in}=-\frac{2e}{\hbar}wK_{\z{R}}{\rm
sign}(V)\left\{
\begin{array}{lr}
2e|V|n_+ &e|V| < E \\
 & \\
2En_+ +(e|V|-E) &e|V| \geq E
\end{array}
\right..\label{Iinel2}
\end{equation}

If the TLS was in thermal equilibrium with the bath, at zero
temperature $n_+=0$ and $n_-=1$ would hold. In that case the
voltage dependence of the conductance is easily evaluated, it
contains a jump-like decrease at the excitation energy of the TLS:
\begin{eqnarray}
\delta G_{\z{R}, \rm in}={\partial\delta
I_{\z{R},\rm{in}}\over{\partial
V}}=-\frac{2e^2}{\hbar}wK_{\z{R}}\cdot\Theta(e|V|-E),
\end{eqnarray}
where $\Theta(x)=\Big\{{{1\;{\rm  for }\;x > 0}\atop{{0\;{\rm  for
}\;x \leq 0}}}$. That result means if the TLS is in the ground
state $n_-=1$ then the TLS can be excited only by a voltage
$e|V|>E$, where the backscattering reduces the current. The second
derivative has a Dirac delta peak at the excitation energy
\begin{equation}
{\partial^2\delta I_{\z{R}, \rm in}\over{\partial V^2}}={\partial
\delta G_{\z{R}, \rm in}\over{\partial V}}
=-\frac{2e^3}{\hbar}wK_{\z{R}}{\rm sign}(V)\cdot\delta(e|V|-E).
\end{equation}
This result is similar to the phonon result, that is the
${\partial^2 I\over{\partial V^2}}$ shows the structure of the
excitation spectrum of a scatterer in the contact region. However,
for a TLS situated in the contact region the occupation of the
upper level is not zero, as the system is driven out of
equilibrium. Due to this feature a nontrivial contribution occurs
in the background resistance beside the spectroscopic feature at
the excitation energy. This can be evaluated after calculating the
voltage dependence of the occupation number, $n_+(V)$, which is
performed later in this section.

\subsubsection{Elastic scattering}
\label{elasticscattering}

A similar calculation can be performed for the current correction
due to the elastic scattering on TLSs by using the collision
integral $I_{\z{R}, \rm TLS,el}(\z{p},\z{r})$ (\ref{IRTLSel})
instead of $I_{\z{R}, \rm TLS,in}(\z{p},\z{r})$ (\ref{IRTLSin}).
The formulas for the elastic scattering can be derived easily by
inserting $E=0$ and modifying the scattering strength in the
results for the inelastic case. For elastic scattering the
strength of the interaction can be different for the TLS being in
the ``$+$'' and ``$-$'' state, thus in Eq.~(\ref{CD}) $wK_\z{R}$
must be replaced by $[wK_\z{R}]^-$ and $[wK_\z{R}]^+$ in the
expression for $C$ and $D$, respectively. After these
considerations the result is:
\begin{equation}\label{dIel}
\delta I_{\z{R},\rm
el}=-\frac{2e^2V}{\hbar}\left(n_-\cdot[wK_\z{R}]^-+n_+\cdot[wK_\z{R}]^+\right).
\end{equation}
Note, that in the elastic case the voltage dependent
``$\coth(\cdots)$'' terms cancel out, thus the only voltage
dependence comes from the voltage dependence of the occupation
number. If the TLS is in thermal equilibrium with the bath, then
$n_+=1-n_-=0$ holds, thus the elastic current correction is a
voltage independent constant term. Similarly, if the scattering
cross sections for the two states of the TLS are equal, then the
voltage dependence cancels out due to the condition $n_++n_-=1$.
In the following we use a simplified notation for the elastic
current correction:
\begin{equation}\label{Iel}
\delta I_{\rm el}=-V(\gamma^-n_-+\gamma^+n_+),
\end{equation}
where $\gamma^+$ and $\gamma^-$ are considered as the reduction of
the conductance due to the TLS being in the ``$+$'' and the
``$-$'' state, respectively.\cite{KK} The average change in the
conductance due to the elastic scattering is characterized by
$\frac{\gamma^++\gamma^-}{2}$, which is written as follows
according to formulas (\ref{dIel}, \ref{KR}, \ref{kisw},
\ref{Wel}):
\begin{equation}\label{sgamma}
\frac{\gamma^++\gamma^-}{2}=\frac{2e^2}{h}(2\pi)^2K_\z{R}\varrho_0^2\left(\left[V^0\right]^2+\left[(\mu^2-\nu^2)V^z\right]^2\right),
\end{equation}
where $V^0$ and $V^z$ are the the interaction matrix elements from
Eq.~(\ref{int_slowtls}) assuming isotropic scattering. The
difference between the conductances corresponding to the two
states of the TLS are written as:
\begin{equation}\label{dgamma}
\gamma^+-\gamma^-=\frac{2e^2}{h}(2\pi)^2K_\z{R}\varrho_0^2(4\mu^2-4\nu^2)V^0V^z.
\end{equation}

Again, the voltage dependence of the current correction can only
be determined after calculating the occupation number as a
function of bias voltage, which is done in the following part. It
is important to note that in case $V^0=0$ the scattering strength
is symmetric for the two states, thus $\gamma^+=\gamma^-$.

\subsubsection{Calculation of the occupation number}

In the following  $n_+$ will be calculated in the nonequilibrium
case, where it is determined by the interaction with the
nonequilibrium electrons. The transition probability  from the
``$-$'' to the ``$+$'' state due to an electron scattered from
momentum state $\z{p}$ to $\z{p}'$ has been considered in the
collision term (\ref{IRTLSin}) and it is:
\begin{equation}
W_{\z{R}}^{+-}(\z{p}'\z{p})=W_{\hat{\z{p}}'\hat{\z{p}}}
n_-f_{\z{p}}(\z{R})(1-f_{\z{p}'}(\z{R}))\delta(\varepsilon_{\z{p}}-
\varepsilon_{\z{p}'}-E),
\end{equation}
and similarly:
\begin{equation}
W_{\z{R}}^{-+}(\z{p}'\z{p})=W_{\hat{\z{p}}'\hat{\z{p}}}
n_+f_{\z{p}}(\z{R})(1-f_{\z{p}'}(\z{R}))\delta(\varepsilon_{\z{p}}-
\varepsilon_{\z{p}'}+E).
\end{equation}
The transition probabilities for the TLS are obtained after
integrating for the electron momenta:
\begin{eqnarray}
W_{\z{R}}^{+-\atop{-+}}=n_\mp{1\over\hbar} \int{{\rm
d}\Omega_{\z{p}}\over{4\pi}}\int{{\rm
d}\Omega_{\z{p}'}\over{4\pi}}w_{\hat{\z{p}}'\hat{\z{p}}}\int {\rm
d}\varepsilon
\Big\{f_0(\varepsilon-e\Phi(\z{R})+{eV\over2}\eta(\z{p}))
(1-f_0(\varepsilon \mp E-e\Phi(\z{R})+{eV\over2}\eta(\z{p}'))\Big\},\nonumber\\
\end{eqnarray}
where the expression (\ref{eq:f0}) for $f_{\z{p}}(\z{R})$ is
used.

The kinetic equation for $n_+$ is:
\begin{equation}
{{\rm d}n_+\over{{\rm d}t}}=W_{\z{R}}^{+-}-W_{\z{R}}^{-+}.
\end{equation}
In the Fermi functions the electron momentum is only involved in
$\eta(\z{p})$, which tells which reservoir the electron at
position $\z{R}$ with momentum $\z{p}$ is coming from. Therefore
the integral can be performed by separating the regions of
integration into four cases with respect to the possible values of
$\eta(\z{p})$ and $\eta(\z{p}')$. In all of these four regions the
factor containing the Fermi functions is independent of
$\Omega_\z{p}$ and $\Omega_\z{p}'$. There is an essential
simplification if $w_{\hat{\z{p}}\hat{\z{p}}'}$ is replaced by a
momentum independent averaged one, $w$. Take e.g.\ the case where
the TLS is on the right hand side of the contact ($R^z>0$) and
introduce the solid angle $\Omega_{\z{R}}$ at which the opening
can be seen from point $\z{R}$. In this case the integrals for the
different regions are written as:
\begin{equation}
W_\z{R}^{+-\atop{-+}}=n_\mp\frac{w}{\hbar}\int(\cdots){\rm
d}\varepsilon\cdot\left\{
\begin{array}{ll}
({\Omega_{\z{R}}\over4\pi})^2 &\quad {\rm if} \quad \eta(\z{p})=1,\;\;\eta(\z{p}')=1\\
&\\
(1-{\Omega_{\z{R}}\over4\pi})^2 &\quad {\rm if} \quad
\eta(\z{p})=-1,\;\;\eta(\z{p}')=-1\\
&\\
{\Omega_{\z{R}}\over4\pi}(1-{\Omega_{\z{R}}\over4\pi}) &\quad {\rm if} \quad \eta(\z{p})=1,\;\;\eta(\z{p}')=-1\\
&\\
{\Omega_{\z{R}}\over4\pi}(1-{\Omega_{\z{R}}\over4\pi}) &\quad {\rm
if} \quad \eta(\z{p})=-1,\;\;\eta(\z{p}')=1
\end{array}
\right.,
\end{equation}
This simplification is justified only after taking an average over
large number of TLSs but that simplifies the calculation and makes
the result more transparent. In this simplified case the kinetic
equation for the TLS is obtained after calculating the energy
integrals for the different cases. (As before, $e\Phi(\z{R})$ is
eliminated by shifting the integral variables.) The final result
is independent whether the TLS is on the right or left, and it can
be written in the form:
\begin{equation}\label{dndt}
{{dn_+}\over{dt}}=n_-A-n_+B=A-n_+(A+B),
\end{equation}
where the coefficients $A$ and $B$ are:
\begin{eqnarray}\label{AB}
A={w\over\hbar}\Bigg\{
\left(1-\frac{\Omega_\z{R}}{4\pi}\right)\frac{\Omega_\z{R}}{4\pi}\left[\frac{eV+E}{2}\left(\coth
(\frac{eV+E}{2k_BT})-1\right)+ \frac{-eV+E}{2}\left(\coth
(\frac{-eV+E}{2k_BT})-1\right)\right]+\nonumber\\
+\left( \left(1-\frac{\Omega_\z{R}}{4\pi}\right)^2+ \left(
\frac{\Omega_\z{R}}{4\pi}\right)^2\right) \frac{E}{2} \left[\coth(\frac{E}{2k_BT})-1\right] \Bigg\} ,\nonumber\\
B={w\over\hbar}\Bigg\{\left({1-{\Omega_\z{R}\over
{4\pi}}}\right){\Omega_\z{R}\over
{4\pi}}\left[{{eV-E\over{2}}\left({\coth
({eV-E\over{2k_BT}})-1}\right)+ {-eV-E\over{2}}\left({\coth
({-eV-E\over{2k_BT}})-1}\right)}\right]+\nonumber\\
+\left({\left({1-{\Omega_\z{R}\over {4\pi}}}\right)^2+
\left({{\Omega_\z{R}\over{4\pi}}}\right)^2}\right) {-E\over
2}\left[{\coth ({-E\over{2k_BT}})-1}\right]\Bigg\},
\end{eqnarray}
respectively. The relaxations are of Korringa types, which depend
on $V$ only if the two electrons involved have different
chareacters in terms of $\eta(\z{p})$. The stationary value of
$n_+$ is obtained as
\begin{equation}
n_+={A\over{A+B}},\label{nplus}
\end{equation}
which is an even function of $V$. At $T=0$ this expression
simplifies essentially:
\begin{equation}\label{nT0}
n_+=\Bigg\{\begin{array}{lr}
0 & e|V| < E \\
\frac{1}{2}-\frac{E}{2E+4\kappa (e|V|-E)} & e|V| \geq E
\end{array},
\end{equation}
where
$\kappa={\Omega_{\z{R}}\over4\pi}(1-{\Omega_{\z{R}}\over4\pi})$.
$n_+$ is a continuous function of $V$. Far from the opening
$\kappa\to0$ thus $n_+\to0$ at voltages for which
$(e|V|-E)/E\ll\kappa^{-1}$ as the electron gas is in thermal
equilibrium. For $\kappa={\rm const.}$ and
$(e|V|-E)/E\gg\kappa^{-1}$ $n_+\to{1\over2}$ as for large enough
voltage the levels are equally occupied $n_-=n_+={1\over2}$. If
the TLS is in the middle of the contact, then $\kappa=1/4$.

If the TLS is far away from the contact then it is in thermal
equilibrium, which can be obtained by taking $\Omega_\z{R}\to0$
and the deviation from equilibrium is large when the TLS is in the
middle of the contact region $\Omega_\z{R}=2\pi$. Here other
relaxation mechanism different from the scattering of electrons is
not taken into account as the generation of bulk phonons is very
weak as the relevant phase space is very small at low energies. At
large concentration of TLSs the TLSs are interacting and the
collective effects may modify the stationary values of the
occupation numbers.

\subsubsection{Conductance with inelastic and elastic scattering}

Now the inelastic and elastic contributions to the current and the
conductance are calculated using the stationary occupation numbers
obtained. The expressions will be given when the TLS is just in
the middle of the round opening, thus $\Omega_\z{R}=2\pi$
($\kappa={1\over 4}$), but the calculation for arbitrary
$\Omega_\z{R}$ can be easily performed as well.

At zero temperature the inelastic current correction is obtained
by inserting the stationary value of $n_+$ given by
Eq.~(\ref{nT0}) into (\ref{Iinel2}):
\begin{equation}
\delta I_{\z{R}=0,\rm in}=-\frac{2e}{\hbar}wK_{\z{R}=0}{\rm
sign}(V)\left\{
\begin{array}{lr}
0 & e|V| < E \\
e|V|-{2E^2\over{E+e|V|}} & e|V| \geq E
\end{array}
\right.,
\end{equation}
which is a continuous function of $V$. After differentiation with
respect to the voltage the correction to the conductance is
obtained as:
\begin{eqnarray}
\delta G_{\z{R}=0,\rm in}&=&{\partial\delta I_{\z{R}=0,\rm
in}\over{\partial V}}=-\frac{2e^2}{\hbar}wK_{\z{R}=0}
\left\{\begin{array}{lr}
0 & e|V| < E\\
1+2{E^2\over{(E+e|V|)^2}} & e|V| \geq E
\end{array}\right.,
\end{eqnarray}
and the second derivative of the current is:
\begin{equation}
{\partial\delta G_{\z{R}=0,\rm in}\over{\partial V}}=
-\frac{2e^3}{\hbar}wK_{\z{R}=0}{\rm sign}(V)\left\{
{3\over2}\delta(E-e|V|)-\Theta(e|V|-E){4E^2\over{(E+e|V|)^3}}\right\}.
\end{equation}
At positive voltages the second derivative of the current shows a
negative Dirac delta peak at $eV=E$, which reflects the energy
spectrum of a single TLS. Above the excitation energy a positive
background is seen due to the nonequilibrium distribution of the
TLS occupation number.

As a next step the elastic contribution is determined at $T=0$ for
a single TLS positioned in the middle of the contact. The change
in the conductance due to the elastic scattering can be calculated
by differentiating Eq.~(\ref{Iel}). The elastic current correction
contains a linear term, which causes a constant, voltage
independent reduction of the conductance. Experimentally it is
hard to separate this constant contribution, thus we calculate
only the voltage dependent part by subtracting the zero bias
conductance:
\begin{equation}
\delta G_{\z{R}=0,\rm el}=\frac{\partial\delta I_{\z{R}=0,\rm
el}}{\partial V}-\left.\frac{\partial\delta I_{\z{R}=0,\rm
el}}{\partial V}\right|_{V=0}
=-(\gamma^+-\gamma^-)\left\{\begin{array}{lr}
0 & e|V| < E \\
\frac{1}{2}-\frac{E^2}{(E+e|V|)^2} & e|V| \geq E
\end{array}
\right.,
\end{equation}
The second derivative of the $I(V)$ curve is
\begin{equation}
{\partial\delta G_{\z{R}=0,\rm el}\over{\partial
V}}=-e(\gamma^+-\gamma^-){\rm
sign}(V)\left\{\frac{1}{4}\delta(E-e|V|)+
\Theta(e|V|-E)\frac{2E^2}{(E+e|V|)^3}\right\}.
\end{equation}
Again, a Dirac delta peak reflects the spectrum of the TLS, and a
continuous background arises at $e|V|>E$ due to the nonequilibrium
distribution. Contrary to the inelastic case, the Dirac delta peak
and the background have the same sign; furthermore, in the elastic
case the sign of the peak can either be positive or negative
depending on the sign of $\gamma^+-\gamma^-$. The amplitude
$\gamma^+-\gamma^-$ is related to the universal conductance
fluctuation $2e^2/h$ (see e.g.~\onlinecite{Lee1986}).

For arbitrary position of the TLS the occupation number $n_+$ is
zero at $V=0$ and $1/2$ at $V\to\infty$. Therefore, the total
amplitudes for the change in the conductance in the elastic and
inelastic case can be generally written as:
\begin{eqnarray}
  \Delta G_{\z{R},\rm in} = -\frac{2e^2}{\hbar}wK_\z{R} &=& \frac{2e^2}{h}(2\pi)^2K_\z{R}\varrho_0^2 (2\mu\nu)^2(V^z)^2\\
  \Delta G_{\z{R},\rm el} = -\frac{1}{2}(\gamma^+-\gamma^-) &=&
  \frac{2e^2}{h}(2\pi)^2K_\z{R}\varrho_0^2 (2\mu^2-2\nu^2)V^0V^z,
\end{eqnarray}
where isotropic scattering was assumed, and the formulas
(\ref{scatteringstrength}, \ref{kisw}, \ref{KR}, \ref{dgamma})
were used. The results are given in the unit of the universal
conductance quantum, $2e^2/h$. According to Eq.~(\ref{munu}) for a
highly asymmetric TLS, where the energy splitting $\Delta$ is much
larger than the transition term $\Delta_0$, the equations
$\mu\simeq 1$ and $\nu\simeq 0$ hold. In this case the inelastic
term is suppressed. In the opposite case, where
$\Delta\ll\Delta_0$ the elastic term is suppressed as
$\mu\simeq\nu\simeq 1/2$. It must be also noted that the inelastic
term depends only on the matrix element $V^z$, whereas the elastic
term is influenced both by $V^z$ and $V^0$. The inelastic term can
be roughly estimated as $\Delta G_{\z{R},\rm in}\lesssim
0.02\,\frac{2e^2}{h}$ using $K_{\z{R}=0}=1/6$, $\mu=\nu=1/2$,
$\varrho_0 V^z\sim 0.1$ (see \onlinecite{VZa,VZc}). The amplitude
sharply drops by moving the TLS further from the orifice than its
diameter because of the geometrical factor. The estimation of the
elastic term is more difficult as $V^0$ is strongly model
dependent.

\begin{figure}
\centering
\includegraphics[width=0.5\textwidth]{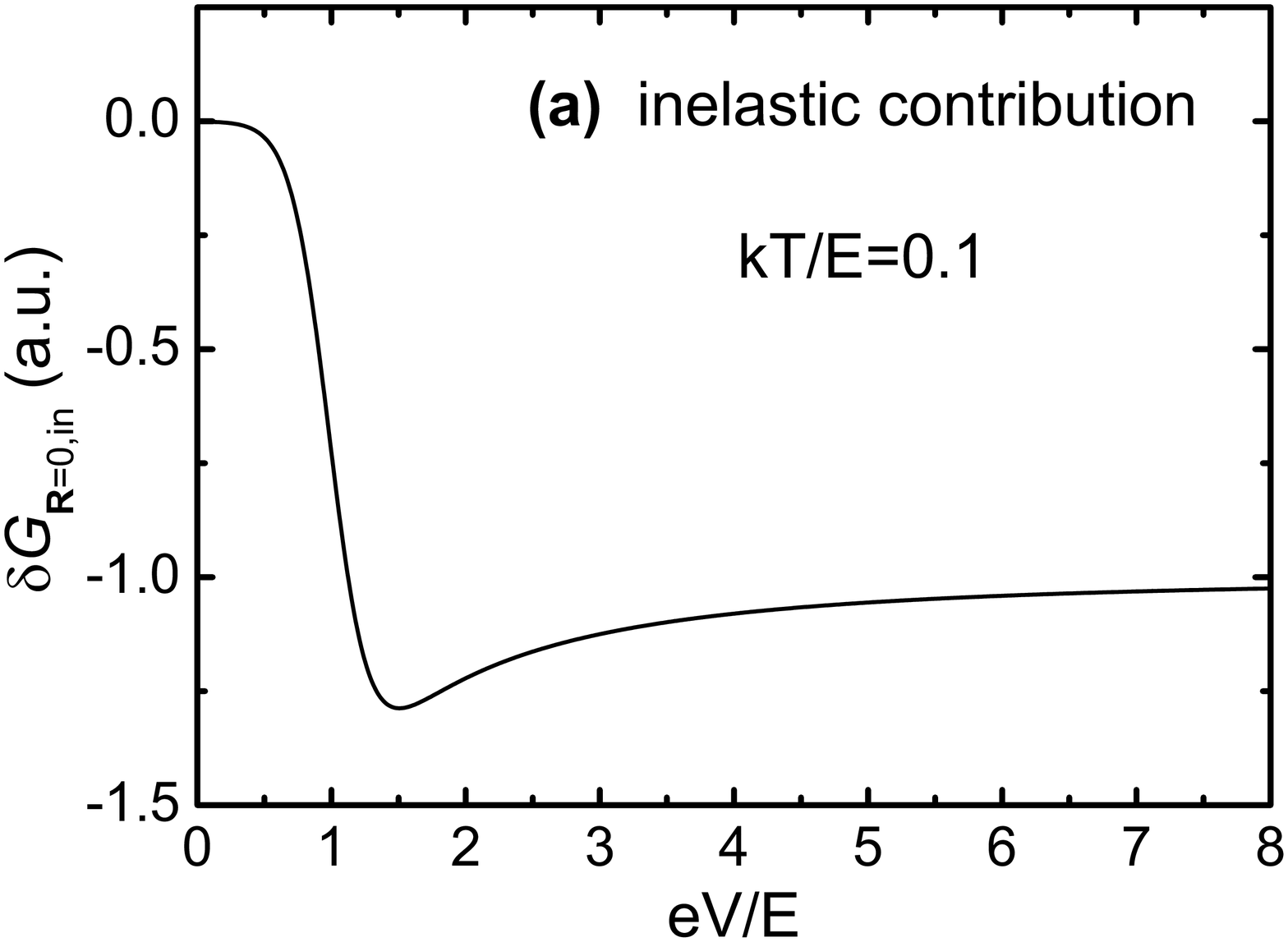}\includegraphics[width=0.5\textwidth]{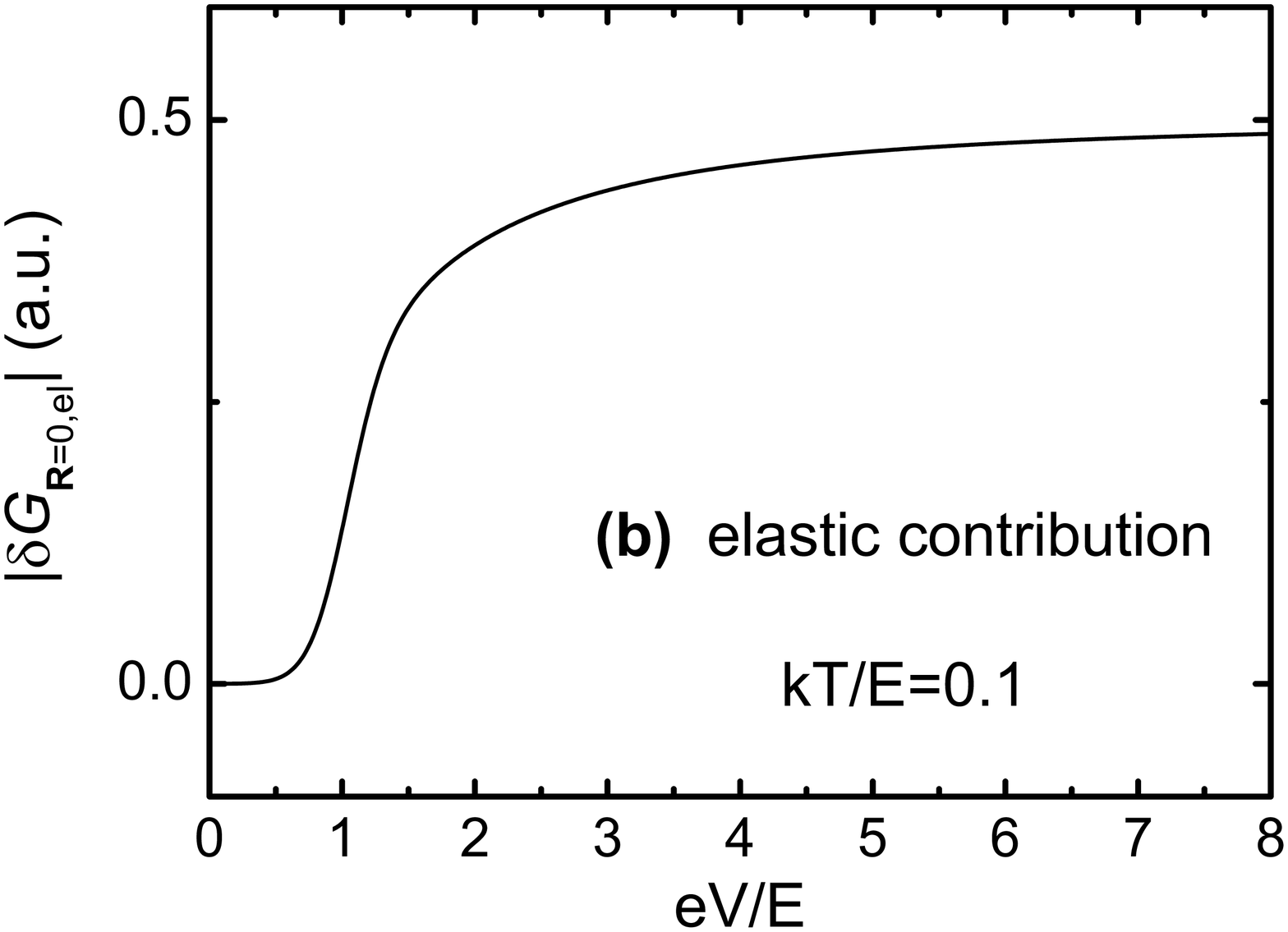}
\caption{\it The contribution of the elastic and inelastic
scattering to the resistivity in the ballistic limit when the TLS
is in the center of the contact ($\delta G_{\z{R}=0,\rm in}$ and
$\delta G_{\z{R}=0,\rm el}$, respectively). The temperature is
$kT=0.1E$. The sign of the elastic contribution can be either
positive or negative depending on the sign of
($\gamma^+-\gamma^-$). On the other hand, the inelastic scattering
always causes a decrease in the conductance.}\label{ball_cent}
\end{figure}

\begin{figure}
\centering
\includegraphics[width=0.5\textwidth]{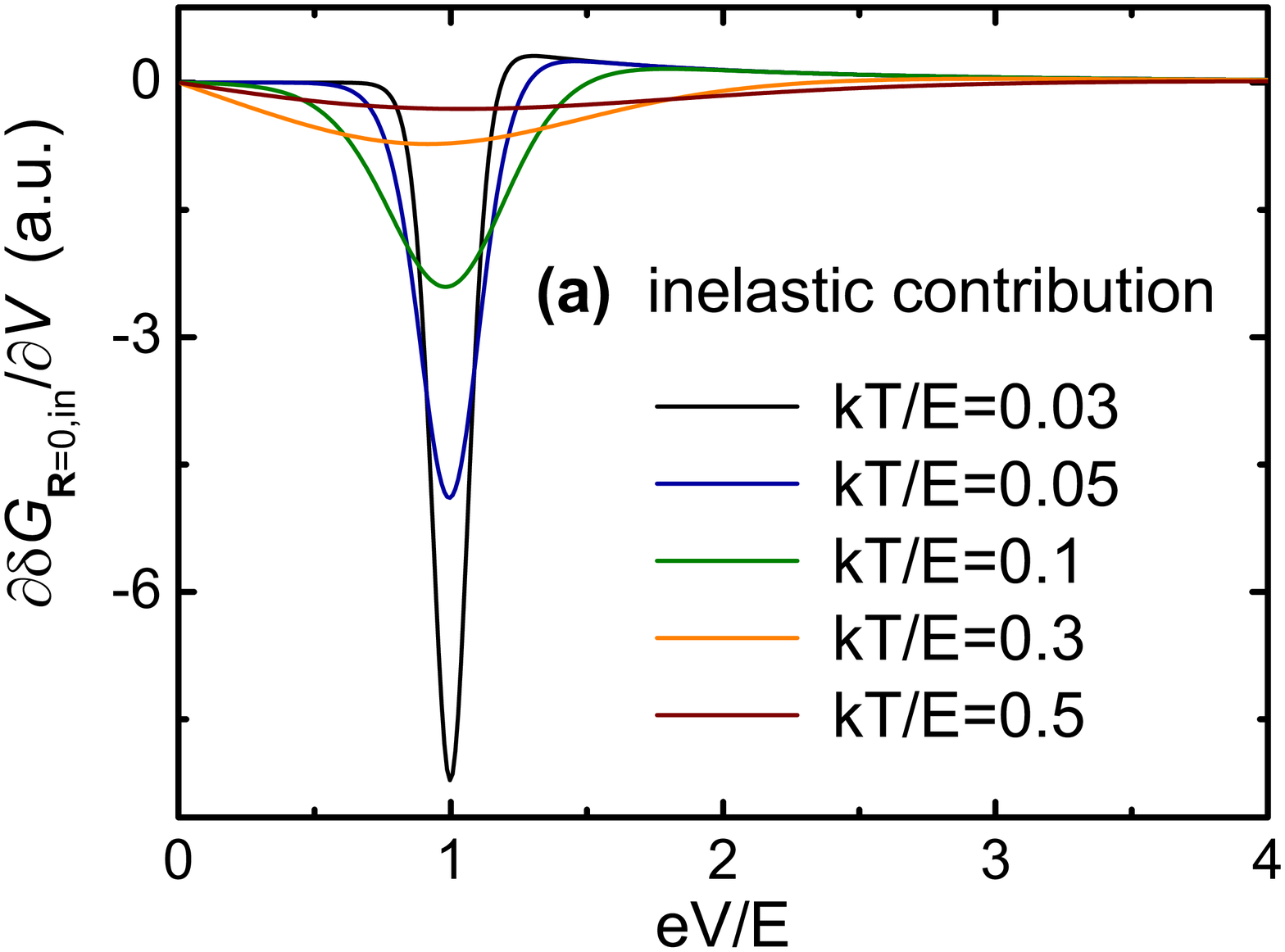}\includegraphics[width=0.5\textwidth]{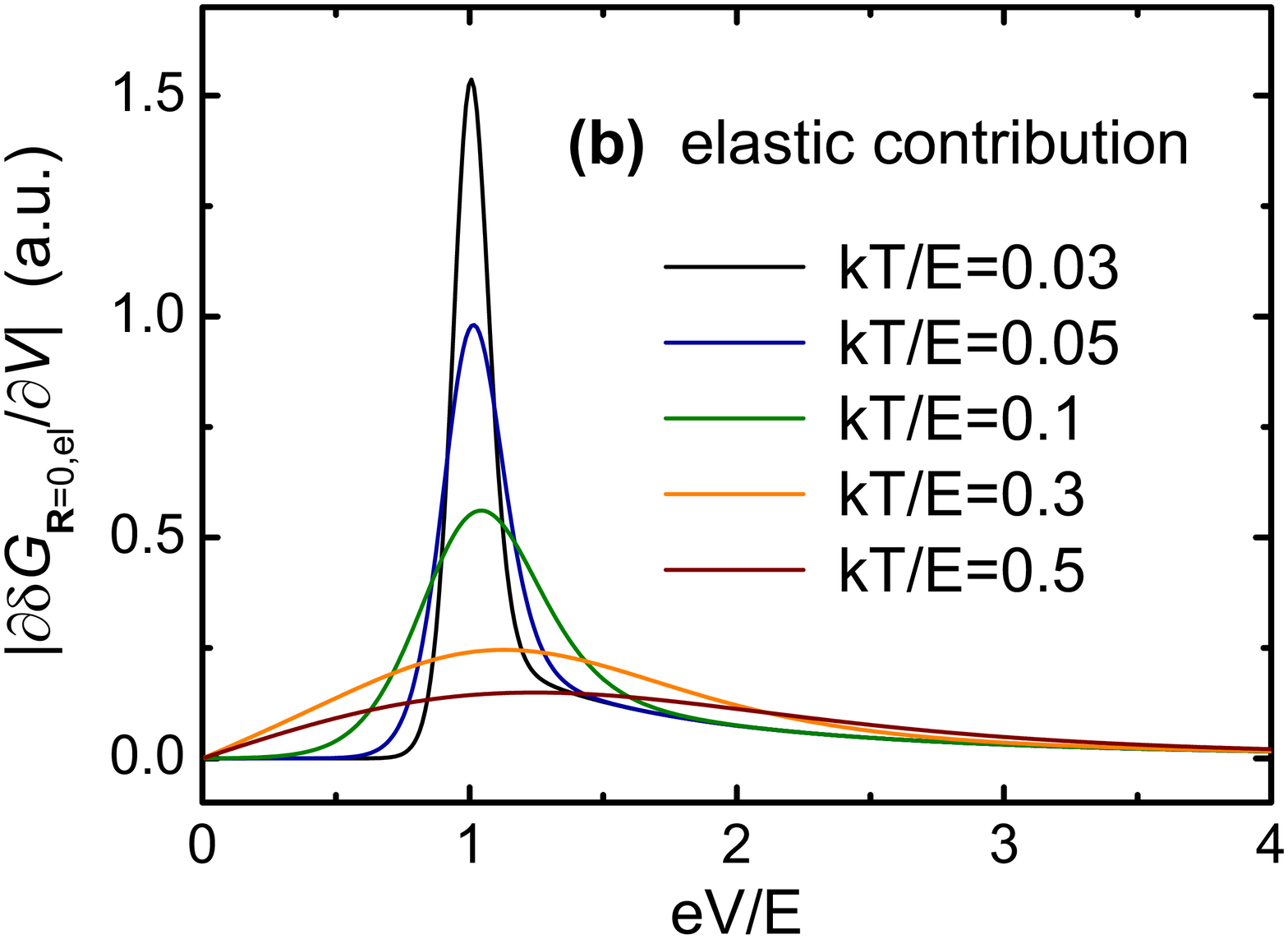}
\caption{\it The second derivative of the $I(V)$ curve for
inelastic and elastic scattering at various temperatures for a
ballistic point contact ($\partial\delta G_{\z{R}=0,\rm
in}/\partial V$, and $\partial\delta G_{\z{R}=0,\rm el}/\partial
V$, respectively). The inelastic contribution shows a negative
peak at the excitation energy of the TLS and a positive background
tail at higher voltages. The elastic contributions similarly shows
a peak at $eV=E$ and a background tail at higher voltages, but
here the sign of the peak and the tail is the same. In the elastic
case the sign of the signal can either be positive or negative
depending on the sign of ($\gamma^+-\gamma^-$).}
\label{secder_ball}
\end{figure}

\begin{figure}
\centering
\includegraphics[width=0.5\textwidth]{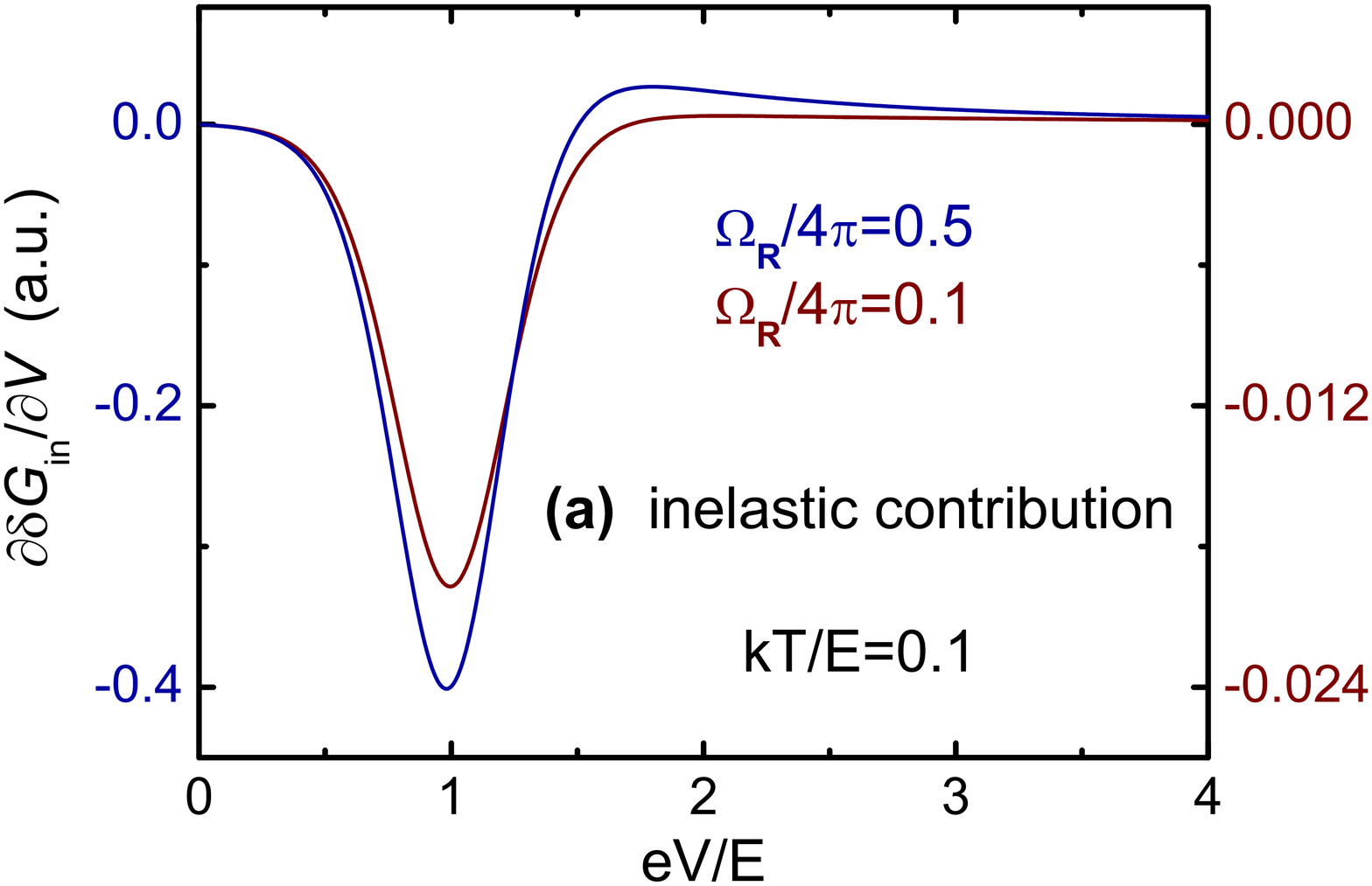}\includegraphics[width=0.5\textwidth]{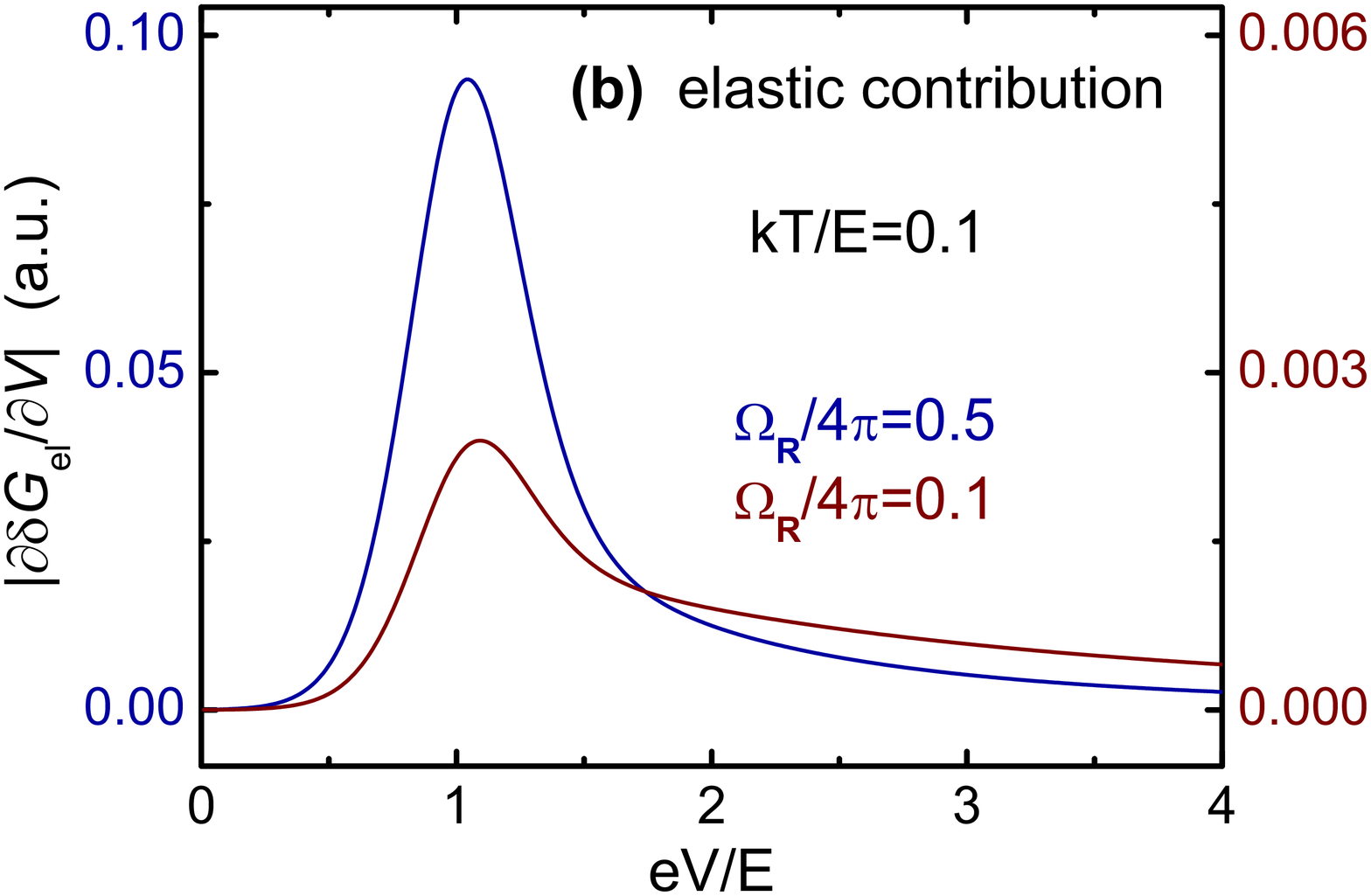}
\caption{\it The second derivative of the $I(V)$ curve for
inelastic and elastic scattering for two different positions of
the TLS: in one case the TLS is in the contact center, in the
other case it is farther away on the contact axis, so that the
opening is seen in a solid angle $\Omega_\z{R}=4\pi/10$. The TLS
being farther away from the contact is closer to thermal
equilibrium, which is reflected by the reduced amplitude of the
background tail with respect to the spectroscopic peak in the
inelastic contribution.} \label{secder_ballq}
\end{figure}

In the following the effect of a TLS is considered at finite
temperature. The formulas for the conductance and the second
derivative of the $I(V)$ curve can be explicitly calculated using
equations (\ref{Iinel}, \ref{CD}, \ref{dIel}, \ref{AB},
\ref{nplus}); however, these equations are very complicated, thus
the results are demonstrated by figures. Figure \ref{ball_cent}
shows the voltage dependence of the conductance for the inelastic
and elastic case respectively. In Fig.~\ref{secder_ball} the
second derivative of the $I(V)$ curve, the so-called point contact
spectrum is presented at various temperatures both for the elastic
and inelastic case. In Fig.~\ref{secder_ballq} both the inelastic
and elastic contributions are compared for a TLS positioned in the
contact center ($\Omega_\z{R}=2\pi$) and for a TLS being farther
away ($\Omega_\z{R}=2\pi/5$).

\subsection{Slow TLS in a diffusive contact} \label{tls:detdiff}

As it has been discussed in Sec.~\ref{maxwell} the
phenomenological treatment must be replaced by a theory based on
kinetic equation, where effect of the inelastic scatterings are
taken into account in the distribution function. In the following
the elastic and inelastic scatterings are treated in equal
footing. As a first step the distribution function is determined
in the presence of elastic scattering and next the contributions
of the TLS are treated as a perturbation.

Consider first the elastic scattering in the limit $l_{\rm el}\to
0$, where the resistance is very large. Because of the very strong
elastic scattering, the electrons are immediately redistributed
concerning the direction of their momenta. Therefore, in the limit
$l_{\rm el}\to 0$ the distribution function depends only on the
energy of the electrons, $f_\z{p}(\z{r})\to f_\varepsilon
(\z{r})$. As the electron arriving at the contact is either coming
from the left or the right reservoirs and the electron energies
are changed due to the external potential $\Phi(\z{r})$ the
distribution function must be the superposition of the
distribution of the electrons coming from the left or right with
amplitudes $\alpha_0(\z{r})$ and $1-\alpha_0(\z{r})$,
respectively: \be
f_{\varepsilon}(\z{r})=\alpha_0(\z{r})f_0(\varepsilon_{\z{p}}-e\Phi(\z{r})
+{eV\over 2})+\Big(1-\alpha_0(\z{r})\Big)f_0(\varepsilon_{\z{p}}-
e\Phi(\z{r})-{eV\over 2}). \label{diffdistr} \ee Using the charge
neutrality condition ($\int f_\epsilon (\z{r}){\rm d}\varepsilon
=\int f_0 (\epsilon_\z{p}){\rm d}\varepsilon)$ $\alpha_0(\z{r})$
can be determined by a similar treatment used earlier in case of
Eq.~(\ref{potent_op}): \be \alpha_0(\z{r})={1\over
2}+{\Phi(\z{r})\over V}. \label{alpha} \ee This formula is valid
in case of arbitrary geometry because only the shape of the
potential function contains information about the details of the
geometry. For an opening type circular contact the potential was
determined in Sec.~\ref{maxwell} by solving the Laplace equation
using a hyperbolic coordinate system.

\begin{figure}
\epsfxsize=0.6\textwidth \centering \epsfbox{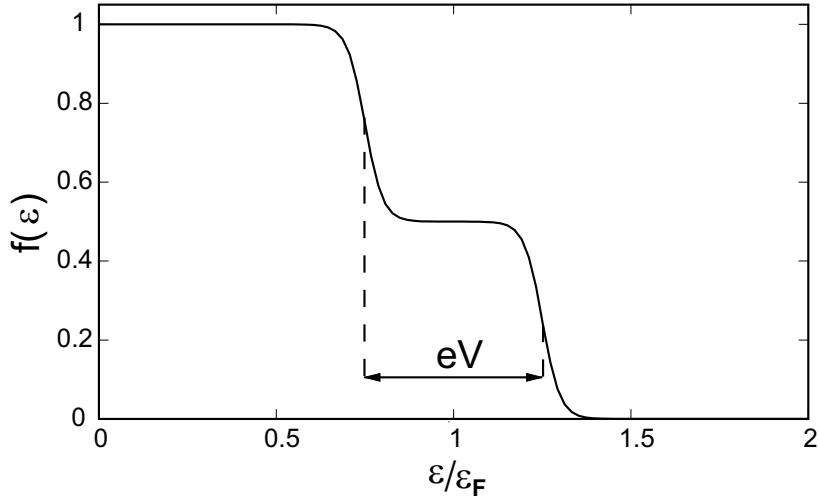}
\vskip0.7truecm \caption{\it Visualization of the distribution
function in the diffusive limit. Without inelastic processes the
smearing of the two steps are due to the finite temperature, $T$.}
\label{distrmax}
\end{figure}

The distribution function contains two sharp steps due to the two
different Fermi energies in the two reservoirs (see
Fig.~\ref{distrmax}). Such steps are measured by the Saclay group
studying short metallic wires of type Fig.~\ref{PCmodels.fig}b by
attaching an extra tunneling diode on the side of the wire. (See
\onlinecite{Pothier1997,Gougam}).

In the ballistic limit the distribution function is very similar
only the factors $\alpha_0(\z{r})$ and $1-\alpha_0(\z{r})$ are
replaced by the geometrical factors $\frac{\Omega(\z{r})}{4\pi}$
and $1-\frac{\Omega(\z{r})}{4\pi}$ determining the solid angles in
which the ballistic electron arrives from the left or right
reservoirs to the point $\z{r}$.

The electron distribution function (\ref{diffdistr}) obtained for
$l_{\rm el}\to 0$ does not result in any current in agreement that
the resistivity is infinite in this limit. For finite $l_{\rm
el}\ll d$ the factor $\alpha_0(\z{r})$ must be replaced by a
momentum dependent one $\alpha_\z{p}(\z{r})$, and the current is
due to
$\delta\alpha_\z{p}(\z{r})=\alpha_\z{p}(\z{r})-\alpha_0(\z{r})$,
which has a strong asymmetric momentum dependence and
$\delta\alpha_\z{p}(\z{r})\to 0$ as $l_{\rm el}\to 0$.

The $\delta\alpha_\z{p}(\z{r})$ will be determined by the very
elegant theory of Kulik, Shekter and Shkorbatov,\cite{KSS} who
have extended the treatment of the ballistic limit to the
diffusive one treating the scattering on phonons and defects which
are large in space. The strong elastic scattering on defects is
combined with a weak inelastic scattering. The limit considered is
specified as $(l_{\rm el}l_{\rm in})^{1/2}\gg d$ where $d$ is the
size of the contact and $l_{\rm el}$ and $l_{\rm in}$ are the
elastic and inelastic mean free paths of the electrons,
respectively. Introducing the corresponding relaxation times,
$\tau_{\rm el}$ and $\tau_{\rm in}$ the inequality can be
rewritten as $(\tau_{\rm el}\tau_{\rm in})^{1/2}\gg {d\over v_F}$.
That means that an electron can diffuse out from the contact
region of size $d$ with small probability of inelastic scattering,
thus double inelastic scatterings can be neglected. In the
following $\tau_{\rm in}$ is related to the relaxation time due to
TLSs.

Also in the diffusive limit the current correction due to the TLSs
is derived by solving the Boltzman equation (\ref{boltzmann_new})
with an appropriate collision term. In this case the collision
integral $I(\z{p},\z{r})$ has two contributions as the elastic
impurity part $I_{\rm el}(\z{p},\z{r})$ and the contribution of
individual TLSs at position $\z{R}$, $I_{\rm
TLS,\z{R}}(\z{p},\z{r})$ thus
\begin{equation}
I(\z{p},\z{r})=I_{\rm el}(\z{p},\z{r})+\sum\limits_n I_{\rm
TLS,\z{R}_n}(\z{p},\z{r}),
\end{equation}
where the sum is due to different TLSs, but in the following it is
assumed that the concentration of the TLSs is so low that first a
single TLS at position $\z{R}$ is considered and in the final
result the contributions of the different TLSs are additive.

The kinetic equation can be arranged as
\begin{equation}
\z{v}_\z{p}{\partial f_\z{p}(\z{r})\over{\partial\z{r}}}-
e\z{E}{\partial f_\z{p}(\z{r})\over{\partial\z{p}}}- I_{\rm
el}(\z{p},\z{r})= I_{\rm TLS,\z{R}}(\z{p},\z{r}).
\end{equation}
First the left hand side is treated in the diffusive limit and the
right hand side is considered as a weak perturbation.

The impurity part of the collision integral is
\begin{equation}
I_{\rm el}(\z{p},\z{r})=
\frac{1}{(2\pi\hbar)^3}\int\limits_{\varepsilon_\z{p}=\varepsilon_{\z{p}'}}
{{\rm d}S_{\z{p}'}\over{v_{\perp}'}}W_{\z{p}\z{p}'}^{\rm imp}
\left(f_{\z{p}'}(\z{r})-f_{\z{p}}(\z{r})\right), \label{eq:I_i}
\end{equation}
where $W_{\z{p}\z{p}'}^{\rm imp}$ is the elastic transition
probability, $v_{\perp}'$ is the velocity at momentum $\z{p}'$
perpendicular to the equienergetic surface for which the integral
is performed. The electric field can be expressed by the electric
potential $\Phi(\z{r})$ as $\z{E}=-\nabla_{\z{r}}\Phi(\z{r})$,
where $\Phi$ is determined by the neutrality condition as in the
ballistic case.

The distribution function $f_\z{p}(\z{r})$ satisfies the following
boundary condition: At the boundary, $\Sigma$ between the
insulator and the metal we assume mirror reflection
\begin{equation}
f_\z{p}(\z{r}\in\Sigma)=f_{\tilde{\z{p}}}(\z{r}\in\Sigma),
\end{equation}
where an incoming electron with momentum $\z{p}$ is reflected with
the momentum $\tilde{\z{p}}$. Furthermore, very far from the
contact $|\z{r}|\to\infty$ the equilibrium distribution is
recovered with chemical potential $\mu$, thus
\begin{equation}
f_\z{p}(|\z{r}|\to\infty)=f_0(\varepsilon_\z{p})=
{1\over{e^{{\varepsilon_\z{p}-\mu}\over{k_BT}}+1}},
\end{equation}
and $\mu$ satisfies the charge neutrality.

The distribution function in the absence of the collision term due
to TLS is denoted by $f_\z{p}^{(0)}(\z{r})$ and the first order
correction due to the TLS is $f_\z{p}^{(1)}(\z{r})$, thus
\begin{equation}
f_\z{p}(\z{r})=f_\z{p}^{(0)}(\z{r})+f_\z{p}^{(1)}(\z{r})+\cdots,
\end{equation}
where the higher order corrections are neglected.

The kinetic equation for $f_\z{p}^{(0)}(\z{r})$ contains the
static impurity contributions, thus
\begin{equation}
\z{v}_\z{p}{\partial f_\z{p}^{(0)}(\z{r})\over{\partial\z{r}}}-
e\z{E}^{(0)}{\partial f_\z{p}^{(0)}(\z{r})\over{\partial\z{p}}}-
I_{\rm el}(f_\z{p}^{(0)}(\z{r}))=0,
\end{equation}
and $f_\z{p}^{(1)}(\z{r})$ satisfies the equation
\begin{equation}
\z{v}_\z{p}{\partial f_\z{p}^{(1)}(\z{r})\over{\partial\z{r}}}-
e\z{E}^{(0)}{\partial f_\z{p}^{(1)}(\z{r})\over{\partial\z{p}}}-
I_{\rm el}(f_\z{p}^{(1)}(\z{r}))= I_{\rm
TLS,\z{R}}(f_\z{p}^{(0)}(\z{r}))+ e\z{E}^{(1)}{\partial
f_\z{p}^{(0)}(\z{r})\over{\partial\z{p}}}, \label{eq:kineq_sep}
\end{equation}
where the electric field is also expanded as
$\z{E}=\z{E}^{(0)}+\z{E}^{(1)}+\cdots$. The term $I_{\rm
TLS,\z{R}}$ is linearized in the collision, thus $f_\z{p}(\z{r})$
is replaced by $f_\z{p}^{(0)}(\z{r})$.

Similarly, the current is $I=I^{(0)}+I^{(1)}+\cdots$, where
\begin{equation}
I^{(i)}=-2e\int {\rm d}\z{S}\int{{\rm
d}^3\z{p}\over{(2\pi\hbar)^3}}
f_\z{p}^{(i)}(\z{r})\z{v}_\z{p}\qquad\qquad(i=0,1,\cdots),
\end{equation}
and the integral with respect ${\rm d}\z{S}$ is taken on a
dividing surface representing the point contact.

According to the introductory remarks of this section, at finite
elastic mean free path the distribution function can be expressed
with a momentum dependent parameter, $\alpha_\z{p}(\z{r})$:
\begin{equation}
f_\z{p}^{(0)}(\z{r})=\alpha_\z{p}(\z{r})f_0^++(1-\alpha_\z{p}(\z{r}))f_0^-,
\label{falpha}\end{equation} where the electrons arriving from far
left (right) at the contact have distribution function $f_0^+$
($f_0^-$), thus
\begin{equation}
f_0^\pm=f_0(\varepsilon_\z{p}-e\Phi(\z{r})\pm{eV\over2}).
\end{equation}
As the collision term (\ref{eq:I_i}) is linear in the distribution
function the equation for $f_\z{p}^{(0)}(\z{r})$ is satisfied with
$\alpha_\z{p}(\z{r})$ as well:
\begin{equation}
\z{v}_\z{p}{\partial\alpha_\z{p}\over{\partial\z{r}}}-
e\z{E}^{(0)}{\partial\alpha_\z{p}\over{\partial\z{p}}}- I_{\rm
el}(\alpha_\z{p})=0. \label{eq:alpha_kineq}
\end{equation}
In the $l_{\rm el}\to 0$ limit the solution of this equation is
$\alpha_0(\z{r})$, given by (\ref{alpha}). At finite elastic mean
free path $\alpha_\z{p}(\z{r})$ can be expanded as:
\begin{equation}
\alpha_\z{p}(\z{r})=\alpha_0(\z{r})+\delta\alpha_\z{p}(\z{r}),
\label{deltaalpha}\end{equation} where $\delta\alpha_\z{p}(\z{r})$
is the first order term in the small parameter, $\frac{l_{\rm
el}}{d}$. The momentum dependent correction, $\delta\alpha_\z{p}$
can be determined by a simple argumentation. At non-zero
relaxation time the probability $\alpha_\z{p}$ can be considered
as the momentum independent probability $\alpha_0$ taken at the
position of the last elastic collision, that is:
\begin{equation}
\alpha_\z{p}(\z{r})=\alpha_0(\z{r}-\tau_{\rm el}\z{v}_\z{p}).
\end{equation}
The expansion of this formula in $\tau_{\rm el}$ gives:
\begin{equation}
\delta\alpha_\z{p}=-\tau_{\rm
el}\z{v}_\z{p}{\partial\alpha_0\over{\partial\z{r}}}.
\label{deltaalpha2}\end{equation} The value of $\alpha_0$ is
determined by the potential, which drops in the contact region,
thus $\frac{\partial\alpha_0}{\partial \z{r}}\sim \frac{1}{d}$,
and $\delta\alpha_\z{p}\sim \frac{l_{\rm el}}{d}$.

The value of $\delta\alpha_\z{p}$ can also be obtained by
inserting (\ref{deltaalpha}) into the Boltzmann equation for
$\alpha_\z{p}$ (\ref{eq:alpha_kineq}). After neglecting the higher
order terms in the small parameters $\frac{l_{\rm el}}{d}$ and
$\frac{eV}{\varepsilon_F}$ the following simple formula is
achieved:
\begin{equation}
\z{v}_\z{p}{\partial\alpha_\z{p}\over{\partial\z{r}}}- I_{\rm
el}(\alpha_\z{p})=0. \label{boltzalpha}\end{equation} For
isotropic scattering $W^{\rm imp}_{\z{p},\z{p}'}=W^{\rm
imp}_{\z{p},-\z{p}'}$ holds, the relaxation time approximation is
appropriate, thus the collision integral is expressed as:
\begin{equation}
I_{\rm
el}(\alpha_\z{p})=-\frac{\delta\alpha_\z{p}}{(2\pi\hbar)^3}\int{{\rm
d}S_{\z{p}'}\over v_{\perp}'}W^{\rm
imp}_{\z{p}\z{p}'}=-{\delta\alpha_\z{p}\over{\tau_{\rm el}}},
\end{equation}
which provides the relaxation time, $\tau_{\rm el}$ due to the
elastic scattering by the impurities. Inserting this result into
(\ref{boltzalpha}) the formula (\ref{deltaalpha2}) for
$\delta\alpha_\z{p}$ is regained.

The next step is to determine $f_\z{p}^{(1)}$ using the equation
(\ref{boltzalpha}) for $\alpha_\z{p}$. The boundary condition for
$\alpha_\z{p}(|\z{r}|\to\infty)=\Theta(-z)$ as very left the
electron comes from the left contact where the electrons are in
equilibrium thus
\begin{equation}
f_\z{p}^{(1)}(|\z{r}|\to\infty)=0.
\end{equation}
Furthermore, the collision term contributes to the distribution
function only if the system is out of equilibrium, thus
$f_\z{p}^{(1)}$ is proportional to the electric field. Therefore,
the second term $-e\z{E}^{(0)}{\partial
f_\z{p}^{(1)}\over{\partial\z{p}}}$ in equation
(\ref{eq:kineq_sep}) can be dropped as it is proportional to
$|\z{E}^{(0)}|^2$.

The correction to the kinetic equation is determined by
Eq.~(\ref{eq:kineq_sep}) where after linearizing each term in the
field the Ansatz with the function $\chi_\z{p}(\z{r})$
\begin{equation}
f_\z{p}^{(1)}(\z{r})=-e\Phi^{(1)}(\z{r}) {\partial
f_\z{p}^{(0)}\over{\partial\varepsilon}}+\chi_\z{p}(\z{r})
\label{chi}\end{equation} can be used. The terms proportional to
$\z{E}^{(1)}=-\partial\Phi^{(1)}/\partial\z{r}$ cancel out in
Eq.~(\ref{eq:kineq_sep}). Thus
\begin{equation}
\z{v}_\z{p}{\partial\chi_\z{p}\over{\partial\z{r}}}- I_{\rm
el}(\chi_\z{p})=I_{\rm TLS,\z{R}}(f_\z{p}^{(0)}(\z{r})),
\label{eq:chi_kineq}
\end{equation}
with the boundary condition $\chi_\z{p}(|\z{r}|\to\infty)=0$ and
$\chi_\z{p}(\z{r}\in\Sigma)=\chi_{\tilde{\z{p}}}(\z{r}\in\Sigma)$.
As it will be shown below the change in the distribution due to
the collision term containing the interaction with the TLS can be
expressed as linear response with the aid of a Green's function
$g_{\z{p}\z{p}'}(\z{r},\z{r}')$:
\begin{equation}
\chi_\z{p}(\z{r})=\int {\rm d}^3\z{r}'\int {\rm d}^3 \z{p}'
g_{\z{p}\z{p}'}(\z{r},\z{r}')I_{\rm
TLS,\z{R}}(f_{\z{p}'}^{(0)}(\z{r}')), \label{chipr1}
\end{equation}
where
\begin{equation}
\z{v}_\z{p}{\partial
g_{\z{p}\z{p}'}(\z{r},\z{r}')\over\partial\z{r}} +I_{\rm
el}(g_{\z{p}\z{p}'}(\z{r},\z{r}'))=-\delta(\z{p}-\z{p}')
\delta(\z{r}-\z{r}'), \label{chipr2}
\end{equation}
and
\begin{equation}
I_{\rm el}(g_{\z{p}\z{p}'}(\z{r},\z{r}'))=
\frac{1}{(2\pi\hbar)^3}\int\limits_{\varepsilon_\z{p}=\varepsilon_{\z{p}''}}
{{\rm d}S_{\z{p}''}\over{v_{\bot}''}}W_{\z{p}\z{p}''}^{\rm imp}
\left\{g_{\z{p}''\z{p}'}(\z{r},\z{r}')-g_{\z{p}\z{p}'}(\z{r},\z{r}')\right\},
\label{chipr3}
\end{equation}
with the boundary condition
$g_{\z{p}\z{p}'}(|\z{r}|\to\infty,\z{r}')=0$ and
$g_{\z{p}\z{p}'}(\z{r}\in\Sigma,\z{r}')=g_{\tilde{\z{p}}\z{p}'}(\z{r}\in\Sigma,\z{r}')$.
As $g_{\z{p}\z{p}'}(\z{r},\z{r}')$ is closely related to the
correlation function $\langle
n_\z{p}(\z{r})n_{\z{p}'}(\z{r}')\rangle$ the time reversal
symmetry can be applied as
\begin{equation}
g_{\z{p}\z{p}'}(\z{r},\z{r}')=g_{-\z{p}'-\z{p}}(\z{r}',\z{r}).
\label{gpp}
\end{equation}

The function $G_\z{p}(\z{r})$ is defined as
\begin{equation}
G_\z{p}(\z{r})=\int {\rm d}S' {{\rm
d}^3\z{p}'\over{(2\pi\hbar)^3}}
v_{\z{p}'}^zg_{\z{p}\z{p}'}(\z{r},\z{r}'), \label{eq:G}
\end{equation}
where the integral with respect to ${\rm d}S'$ is a surface
integral of variable $r'$ taken on the orifice. This function
satisfies the equation
\begin{equation}
\z{v}_\z{p}{\partial G_\z{p}\over{\partial\z{r}}}+ I_{\rm
el}\{G_\z{p}(\z{r})\}=-v_\z{p}^z\delta(z), \label{eq:G_kineq}
\end{equation}
which follows from the equation (\ref{chipr2}) defining
$g_{\z{p}\z{p}'}(\z{r},\z{r}')$. The boundary conditions are the
usual ones $G_\z{p}(|\z{r}|\to\infty)=0$ and
$G_\z{p}(\z{r}\in\Sigma)=G_{\tilde{\z{p}}}(\z{r}\in\Sigma)$. This
equation for $G_\z{p}(\z{r})$ is closely related to
Eq.~(\ref{boltzalpha}) for $\alpha_\z{p}(\z{r})$ introduced
earlier, but without the inhomogeneous term. (The sign in front of
the collision integral can be changed by replacing $\z{p}\to
-\z{p}$ and taking the relation $\z{v}_{-\z{p}}=-\z{v}_\z{p}$ into
account.) By comparing the two equations the expression
\begin{equation}
G_\z{p}(\z{r})=1-\alpha_{-\z{p}}(\z{r})-\Theta(z)
\label{eq:Galpha}
\end{equation}
is obtained, which automatically satisfies the boundary condition
for $G_\z{p}(\z{r})$.

Using the expression (\ref{eq:Galpha}) for $G_\z{p}(\z{r})$ the
current can be calculated and only the $\z{p}$-dependent part is
contributing, as the total number of electrons at position $\z{r}$
is not changed by the collision. Therefore:
\begin{equation}
\int {\rm d}^3\z{p}\; I_{\rm TLS,\z{R}}(f_\z{p}^{(0)}(\z{r}))=0.
\label{qwerty}\end{equation}

The kinetic equation (\ref{eq:chi_kineq}) for $\chi_\z{p}$ can be
checked using equations (\ref{chipr1}), (\ref{chipr2}),
(\ref{chipr3}) and inserting twice the identity (\ref{gpp}). The
next task is to calculate the correction to the current due to the
inelastic scattering on the TLS. Using the expression (\ref{chi})
for $f_\z{p}^{(1)}$ (dropping the first term which is even in
momentum $\z{p}$ in leading order):
\begin{equation}
I^{(1)}=-2e {1\over{(2\pi\hbar)^3}}\int\limits_S {\rm d}S\int {\rm
d}^3\z{p}\; v_\z{p}^z\chi_\z{p}(\z{r}), \label{eq:I1chi}
\end{equation}
where the variable $\z{r}$ is taken over the contact area, $S$.
Inserting the expression of $\chi$ the result can be rewritten as
\begin{equation}
I^{(1)}=-2e {1\over{(2\pi\hbar)^3}}\int\limits_S {\rm d}S {\rm
d}^3\z{p}\; v_\z{p}^z \int{\rm d}^3\z{p}'{\rm d}^3\z{r}'
g_{\z{p}\z{p}'}(\z{r},\z{r}') I_{\rm
TLS,\z{R}}(f_{\z{p}'}^{(0)}(\z{r}')).
\end{equation}
That result can be further rewritten in terms of $G_\z{p}(\z{r})$
as
\begin{equation}
I^{(1)}=-2e {1\over{(2\pi\hbar)^3}} \int {\rm d}^3\z{p}'\int {\rm
d}^3\z{r}'\;G_{\z{p}'}(\z{r}') I_{\rm
TLS,\z{R}}(f_{\z{p}'}^{(0)}(\z{r}')).
\end{equation}
Using Eq.~(\ref{eq:Galpha}) for $G_\z{p}(\z{r})$ the momentum
independent terms do not contribute due to Eq.~(\ref{qwerty}). The
remaining term contains $\delta\alpha_\z{p}$ which was determined
earlier and $\delta\alpha_{-\z{p}}=-\delta\alpha_\z{p}$ holds.
Furthermore, $I_{\rm TLS,\z{R}}(\z{p},\z{r})= \delta(\z{r}-\z{R})
I_{\rm TLS,\z{R}}(\z{p})$ and in this way the final expression for
the current is
\begin{equation}
I^{(1)}=-2e {1\over{(2\pi\hbar)^3}} \int {\rm
d}^3\z{p}\;\delta\alpha_\z{p}I_{\rm TLS,\z{R}}(\z{p}),
\end{equation}
where $\delta\alpha_\z{p}$ is taken at the position of the TLS,
$\z{R}$. Because of the odd parity of $\delta\alpha_\z{p}$ only
the odd part of the collision integral contributes. Using the
equation (\ref{falpha}) for $f_\z{p}^{(0)}(\z{r})$ the collision
integral for the inelastic scattering can be written similarly to
the ballistic case (Eq.~\ref{IRTLSin}):
\begin{eqnarray}
I_{\z{R},\rm TLS,in}(\z{p})&=&{\varrho_0\over{(2\pi\hbar)^3}}\int
{\rm d}\varepsilon' {\rm d}\Omega_{\z{p}'}
W_{\hat{\z{p}}\hat{\z{p}}'}\nonumber\\
&{
}&\left\{\left[\alpha_{\hat{\z{p}}'}f_0(\varepsilon'+{eV\over2})
+(1-\alpha_{\hat{\z{p}}'})f_0(\varepsilon'-{eV\over2})\right]
\left[\alpha_{\hat{\z{p}}}(1-f_0(\varepsilon+{eV\over2}))
+(1-\alpha_{\hat{\z{p}}})(1-f_0(\varepsilon-{eV\over2}))\right]
\right.\cdot\nonumber\\
&{ }&\;\;\;\bigg[n_-\delta(\varepsilon'-\varepsilon-E)+
n_+\delta(\varepsilon'-\varepsilon+E)\bigg]\nonumber\\
&{ }&\;-\bigg[\quad\hat{\z{p}}'\leftrightarrow\hat{\z{p}}\quad
\varepsilon'\leftrightarrow\varepsilon\quad\bigg]
\bigg[\quad\hat{\z{p}}'\leftrightarrow\hat{\z{p}}\quad
\varepsilon'\leftrightarrow\varepsilon\quad\bigg] \bigg[\quad
\varepsilon'\leftrightarrow\varepsilon\quad\bigg] \bigg\}
\end{eqnarray}
is obtained, where in the second set of brackets the variables are
changed as $\varepsilon\leftrightarrow\varepsilon'$,
$\z{p}\leftrightarrow \z{p}'$ and all the energy variables
$\varepsilon$ and $\varepsilon'$ are shifted by $-e\Phi(\z{r})$
thus $\Phi(\z{r})$ drops out. Now using Eq.~(\ref{deltaalpha})
$\alpha_{\hat{\z{p}}}=\alpha_0+\delta\alpha_{\hat{\z{p}}}$ and
keeping only the odd part of the collision integral, after
changing the variables the current is
\begin{eqnarray}
I^{(1)}&=&-\frac{2e}{\hbar}\int {\rm d}\varepsilon {{\rm
d}\Omega_{\z{p}}\over{4\pi}} \delta\alpha_{\hat{\z{p}}} \int {\rm
d}\varepsilon' {{\rm
d}\Omega_{\z{p}'}\over{4\pi}}w_{\hat{\z{p}}\hat{\z{p}}'}\nonumber\\
&{ }&\left\{\left[ \delta\alpha_{\hat{\z{p}}'}
\left[f_0(\varepsilon'+{eV\over2})-f_0(\varepsilon'-{eV\over2})\right]
\left[\alpha_0(1-f_0(\varepsilon+{eV\over2}))+
(1-\alpha_0)(1-f_0(\varepsilon-{eV\over2}))\right]
\right.\right.\nonumber\\
&{ }&\;\;\;-\delta\alpha_{\hat{\z{p}}}
\left[\alpha_0f_0(\varepsilon'+{eV\over2}))+
(1-\alpha_0)f_0(\varepsilon'-{eV\over2})\right]
\left[f_0(\varepsilon+{eV\over2})-f_0(\varepsilon-{eV\over2})\right]
\bigg]\times\nonumber\\
&{ }&\;\;\;\left[n_-\delta(\varepsilon'-\varepsilon-E)+
n_+\delta(\varepsilon'-\varepsilon+E)\right]\nonumber\\
&{ }&\;\;-\bigg[\quad\hat{\z{p}}'\leftrightarrow\hat{\z{p}}\quad
\varepsilon'\leftrightarrow\varepsilon\quad\bigg]\bigg\},
\end{eqnarray}
where $\alpha_0$ and $\delta\alpha_\z{p}$ are taken at the site of
the TLS ($\z{r}=\z{R}$).

In this expression two squared averaged matrix elements occur
\begin{eqnarray}
w^{(\z{R})}_1&=&\int {{\rm d}\Omega_{\z{p}}\over{4\pi}}\int {{\rm
d}\Omega_{\z{p}'}\over{4\pi}}
(\delta\alpha_{\hat{\z{p}}})^2w_{\hat{\z{p}}\hat{\z{p}}'}=
\tau_{\rm el}^2\int {{\rm d}\Omega_{\z{p}}\over{4\pi}}\int {{\rm
d}\Omega_{\z{p}'}\over{4\pi}}w_{\hat{\z{p}}\hat{\z{p}}'}
\left({\z{p}\over{m}}{\partial\alpha_0\over{\partial\z{r}}}\right)^2_{\z{r}=\z{R}}
= \tau_{\rm
el}^2{v_F^2\over3}\left|{\partial\alpha_0\over{\partial\z{r}}}
\right|^2\int {{\rm d}\Omega_{\z{p}}\over{4\pi}}\int {{\rm
d}\Omega_{\z{p}'}\over{4\pi}}
w_{\hat{\z{p}}\hat{\z{p}}'}\nonumber\\
w_2^{(\z{R})}&=&\int {{\rm d}\Omega_{\z{p}}\over{4\pi}}\int {{\rm
d}\Omega_{\z{p}'}\over{4\pi}}
\delta\alpha_{\hat{\z{p}}}\delta\alpha_{\hat{\z{p}'}}= \tau_{\rm
el}^2\int {{\rm d}\Omega_{\z{p}}\over{4\pi}}\int {{\rm
d}\Omega_{\z{p}'}\over{4\pi}}w_{\hat{\z{p}}\hat{\z{p}}'}
\left({\z{p}\over
m}{\partial\alpha_0\over{\partial\z{r}}}\right)_{\z{r}=\z{R}}
\left({\z{p}'\over
m}{\partial\alpha_0\over{\partial\z{r}}}\right)_{\z{r}=\z{R}},
\end{eqnarray}
where the expression (\ref{deltaalpha2}) for $\delta\alpha_\z{p}$
was used. Furthermore $w_1\geq |w_2|$ holds as
$w_{\hat{\z{p}}\hat{\z{p}}'}=w_{\hat{\z{p}}'\hat{\z{p}}}$ and
$${1\over2}\left(\left({\z{p}\over
m}{\partial\alpha\over{\partial\z{r}}}\right)^2+\left({\z{p}'\over
m}{\partial\alpha\over{\partial\z{r}}}\right)^2\right)\leq\left|{\z{p}\over
m}{\partial\alpha\over{\partial\z{r}}}\right|\left|{\z{p}'\over
m}{\partial\alpha\over{\partial\z{r}}}\right|.$$ Integrating over
the variable $\varepsilon'$ the following expression is obtained
\begin{eqnarray}\label{Idiff}
&&I^{(1)}=-\frac{2e}{\hbar}wK_\z{R}^{\rm diff}\int {\rm
d}\varepsilon
\left(f_0(\varepsilon-{eV\over2})-f_0(\varepsilon+{eV\over2})\right)\times\\
&&\left\{ 1+(n_--n_+)\left[
\alpha_0\left(f_0(\varepsilon+E+{eV\over2})-
f_0(\varepsilon-E+{eV\over2})\right)+
(1-\alpha_0)\left(f_0(\varepsilon+E-{eV\over2})-
f_0(\varepsilon-E-{eV\over2})\right)\right]\right\},\nonumber
\end{eqnarray}
where the geometrical factor $K_\z{R}^{\rm diff}$ is introduced
for the diffusive limit as
\begin{equation}\label{Kdiff}
K_\z{R}^{\rm diff}=\frac{w_1^{(R)}-w_2^{(R)}}{w}>0.
\end{equation}
The expression (\ref{Idiff}) can be brought to a form which is
comparable to the result (\ref{eq:112}) in the ballistic limit.
The key of the algebra applied is in the integral
$$\int {\rm d}\varepsilon\; f_0(\varepsilon)(1-f_0(\varepsilon+a))=
\int {\rm d}\varepsilon\;
f_0(\varepsilon+b)(1-f_0(\varepsilon+a+b))$$ the variable can be
shifted by an arbitrary energy $b$ as the contribution comes from
the neighborhood of the Fermi energy and the bandwidth cutoff does
not play any role. The result of a lengthy algebra is \be
I^{(1)}=-\frac{2e}{\hbar}wK_\z{R}^{\rm diff}\int {\rm
d}\varepsilon\;
f_0(\varepsilon)\left\{\left(f_0(\varepsilon-E-eV)-
f_0(\varepsilon-E+eV)\right)n_-+\left(f_0(\varepsilon+E-eV)-
f_0(\varepsilon+E+eV)\right)n_+\right\}.\ee This expression
coincides with the one obtained in the ballistic limit
(\ref{eq:112}) after replacing the geometrical factor
$K_\z{R}\rightarrow K_\z{R}^{\rm diff}$. Therefore, the final
results in the ballistic limit (\ref{Iinel}, \ref{CD},
\ref{Iinel2}) are also valid in the diffusive limit, just the
geometrical factor, $K_\z{R}$ is different.

The elastic contribution can be obtained by inserting $E=0$ and
considering different scattering strengths for the two states of
the TLS in the diffusive limit as well. Therefore, the ballistic
result for the elastic case (\ref{dIel}) is also valid in the
diffusive regime with the replaced geometrical factor.

The next step is to calculate $n_+$. In the present case whether
the electrons come from the left or right reservoir is determined
by $\alpha_0$ or $(1-\alpha_0)$, thus the previous results in the
ballistic regime (\ref{nplus}, \ref{AB}, \ref{nT0}) can be adopted
after changing
\begin{equation}
{\Omega_\z{R}\over{4\pi}}\to\alpha_0.
\end{equation}
The momentum dependent correction $\delta\alpha_\z{p}$ to
$\alpha_0$ contributes only if the momentum dependence of
$w_{\z{p}\z{p}'}$ is kept but that is small in case $l_{\rm el}\ll
d$.

Concluding, in the diffusive limit the shape of the current
correction due to the scattering on a TLS is the same as in the
ballistic case, the only essential difference is in the
geometrical factors, which are independent of the strength of the
interaction between the electrons and the TLSs. In the ballistic
case $K_\z{R}$ is in the range of unity if the TLS is situated in
the contact region. However, $K_\z{R}^{\rm diff}$ contains other
quantities proportional to $(\tau_{\rm
el}\frac{p}{m})^2(\frac{\partial \alpha_0}{\partial r})^2\sim
l_{\rm el}^2(\frac{\partial\alpha_0}{\partial r})^2$ which
indicates how much $\alpha$ is changed in space on the scale of
the impurity scattering mean free path. We have seen that
$\frac{\partial \alpha}{\partial r}$ scales with the inverse of
the contact diameter, $d$, thus
\begin{equation}
K_\z{R}^{\rm diff}\sim \left(\frac{l_{\rm el}}{d}\right)^2
\end{equation}
in case where the TLS is nearby the orifice. Far from the orifice
($|\z{r}|\to\infty$) $\frac{\partial \alpha}{\partial r}\to 0$.
The amplitudes of the change in the conductance for the important
TLS (being close to the contact) are scaled down by
\begin{equation}
\frac{\delta G^{\rm ball}}{\delta G^{\rm
diff}}\sim\left(\frac{l_{\rm el}}{d}\right)^2\ll 1,
\end{equation}
which can be a very small factor in the dirty limit. The largest
contribution arises from the TLSs for which $\frac{\partial
\alpha}{\partial r}$ is the largest, thus for the TLS just nearby
the edge of the orifice. (See Eq.~\ref{Ez})

The change of the geometrical factor due to the impurity
scattering length is similar to the case of electron-phonon
interaction, where integration is taken according to location of
the interaction.\cite{KY} The $K$-factors for electron-phonon
scattering are given in (\ref{Kfactorball}, \ref{Kfactordiff}) for
the ballistic and diffusive case, respectively. In the phonon case
the amplitude of the signal in the diffusive limit is reduced only
by $(\frac{l_{\rm el}}{d})$.

\subsection{The case of time-dependent applied voltage}
\label{time_v}

Up to this point the scattering on TLSs in point contacts was
investigated as a function of bias voltage, which provides
information about the excitation spectrum of the TLS, and also
about the nonequilibrium distribution of the TLS states. A further
possibility to study the dynamical properties of the TLSs is the
study of the response to a harmonic excitation. Let us consider
that a voltage of
\begin{equation}
V(t)=V_0+V_1\cos(\omega t)
\end{equation}
is applied on the point contact, where the DC bias voltage, $V_0$
is varied on the scale of the excitation energy of the TLS to
cover the spectroscopic peak at $e|V|=E$, and also the background
tail at $e|V|>E$. The amplitude of the harmonic excitation is
considered to be much smaller than the excitation energy, $eV_1\ll
E$. Furthermore, it is assumed that the energy $\hbar\omega$ is
much smaller than the excitation energy of the TLS, thus the
quantum nature of the radiation should not be taken into account.
In this case the current flowing through the contact can be
expanded in second order in terms of the time dependent voltage
contribution:
\begin{equation}\label{Iexpansion}
I(V_0+V_1\cos(\omega
t))=\underbrace{I(V_0)+\frac{1}{4}\left.\frac{\partial^2
I}{\partial
V^2}\right|_{V=V_0}V_1^2}_{I^0}+\underbrace{\left.\frac{\partial
I}{\partial V}\right|_{V=V_0}V_1\cos(\omega
t)}_{I^\omega}+\underbrace{\frac{1}{4}\left.\frac{\partial^2
I}{\partial V^2}\right|_{V=V_0}V_1^2\cos(2\omega
t)}_{I^{2\omega}}+...,
\end{equation}
where the equation $\cos^2(\varphi)=(\cos (2\varphi+1))/2$ is
used. The voltage dependence of the harmonic response, $I^\omega$
provides the differential resistance curve, $\partial I/\partial
V$. The second harmonic response, $I^{2\omega}$ directly provides
the point contact spectrum, i.e.\ the second derivative of the
$I(V)$ curve. The DC component, which can be measured as a time
average of the response signal: $I^0=\langle I(V_0+V_1\cos(\omega
t))\rangle$ also shows the second derivative of the $I(V)$ curve
after subtracting the current without harmonic excitation,
$I(V_0)$. Further on this term is referred to as the DC shift
signal, denoted by $\delta I^0$.

The above considerations are, however, only valid if the system is
fast enough to follow the time dependent voltage. In the case of
two level systems, the dynamical behavior is determined by the
kinetic equation (\ref{dndt}):
\begin{equation}\label{mw1}
\frac{dn_+}{dt}=A-n_+\cdot(A+B),
\end{equation}
where the coefficients $A$ and $B$ are given by Eq.~(\ref{AB}).
Both of the coefficients depend on the actual voltage $V(t)$ at
given $t$. For simplicity the notation $A+B=\bar{B}$ is
introduced, which is the inverse of the relaxation rate of the TLS
($\bar B=\tau^{-1}$). For slow two level systems the value of the
relaxation time can be estimated using
Eq.~(\ref{dndt},\ref{kisw},\ref{scatteringstrength}) as follows:
\be \bar B\simeq\frac{2\pi}{\hbar} (2\mu\nu)^2(\varrho_0
V^z)^2\label{relaxtime}\ee Inserting $\mu=\nu=1/2$, $E=1$\,meV,
and $\varrho_0 V^z\sim 0.1$\cite{VZa,VZc} into this equation a
relaxation time of $\tau\sim 10^{-10}$\,s is obtained.

If the alternating voltage is faster than the relaxation of the
TLS ($\omega\tau\gg 1$) then a more complicated procedure is
necessary to determine the time dependent response of the system.
This phenomenon, however can be used to trace the typical
relaxation time of the TLS by studying the response signal as a
function of the frequency $\omega$. The first calculation with
time dependent applied voltage was performed by Kozub and
Kulik.\cite{KK} In the next part a general solution is given for
the time dependent occupation number $n_+(t)$ at arbitrary
frequency and amplitude of the harmonic excitation signal. After
that various limits are treated, where the formulas are
essentially simplified.

\subsubsection{General solution for $n_+(t)$}

The kinetic equation (\ref{mw1}) is a linear inhomogeneous
differential equation which has a general solution
\begin{equation}\label{mw2}
n_+(t)=e^{-\int\limits_0^t {\rm
d}t'\;\bar{B}(V(t'))}\cdot\left(C_1+\int\limits_0^t {\rm
d}t'\;A(V(t'))e^{\int\limits_0^{t'}{\rm
d}t''\;\bar{B}(V(t''))}\right),
\end{equation}
where the coefficient $C_1$ is determined by the initial condition
at time $t=0$. According to the definitions given by
Eq.~(\ref{AB}) $A\geq 0$ and $B\geq 0$ hold, thus the exponent in
the last term of Eq.~(\ref{mw2}) monotonically increases with
time.  This means that $C_1$ is essential only for the transient
behavior and for long time it can be dropped. For periodic voltage
$A$ and $\bar{B}$ are also a periodic function of time:
$A(V(t))=A(t)=A(t+T_{\omega})$,
$\bar{B}(V(t))=\bar{B}(t)=\bar{B}(t+T_{\omega})$. The time
integrals can be divided according to the periods as:
\begin{equation}\label{mw3}
\int\limits_0^t {\rm d}t...=\int\limits_0^{T_{\omega}}{\rm
d}t+\int\limits_{T_{\omega}}^{2T_{\omega}}{\rm d}t+...+
\int\limits_{(n-1)T_{\omega}}^{nT_{\omega}}{\rm
d}t+\int\limits_{nT_{\omega}}^{\bar{t}}{\rm d}t,
\end{equation}
where $t=nT_{\omega}+\bar{t}$, $n$ is an integer and
$0<\bar{t}<T_{\omega}$. In this way:
\begin{equation}\label{mw4}
\int\limits_0^t{\rm d}t\;\bar{B}=n\Delta Q+Q(\bar t),
\end{equation}
where
\begin{equation}\label{mw5}
\Delta Q=\int\limits_0^{T_{\omega}}{\rm d}t\;\bar{B}
\end{equation} and
\begin{equation}\label{mw6}
Q(\bar t)=\int\limits_0^{\bar{t}}{\rm d}t'\;\bar{B}(t').
\end{equation}
Using these identities, one gets:
\begin{equation}\label{mw7}
\int\limits_0^t{\rm
d}t'\;A(t')e^{\int\limits_0^{t'}\bar{B}(t''){\rm
d}t''}=\sum\limits_{n'=0}^{n-1}e^{n'\Delta
Q}\int\limits_0^{T_{\omega}}{\rm d}t'\;A(t')e^{Q(t')}+ e^{n\Delta
Q}\int\limits_0^{\bar{t}}{\rm d}t'\;A(t')e^{Q(t')}.
\end{equation}
The final result is:
\begin{equation}\label{mw8}
n_+(t)=\frac{1}{e^{Q(\bar{t})}}\left(\frac{1}{e^{n\Delta
Q}}\frac{e^{n\Delta Q}-1}{e^{\Delta
Q}-1}\int\limits_0^{T_{\omega}}{\rm d}t'\;
A(t')e^{Q(t')}+\int\limits_0^{\bar t}{\rm
d}t'\;A(t')e^{Q(t')}\right),
\end{equation}
which gives the following stationary solution in the long time
$(n\gg1)$ limit:
\begin{equation}\label{mw9}
n_+(t)=\frac{1}{e^{Q(\bar{t})}}\left(\frac{1}{e^{\Delta
Q}-1}\int\limits_0^{T_{\omega}}{\rm d}t'\;
A(t')e^{Q(t')}+\int\limits_0^{\bar t}{\rm
d}t'\;A(t')e^{Q(t')}\right).
\end{equation}
Turning to dimensionless variables like $\varphi=2\pi
t/T_{\omega}$, and $\bar{\varphi}=2\pi\bar t/T_{\omega}$ and also
new notations, $a(\varphi)=A(V(t))/2\pi$, $b(\varphi)=\bar
B(V(t))/2\pi$, $q(\varphi)=\int\limits_0^{\varphi}b(\varphi'){\rm
d}\varphi'=Q(t)/T_{\omega}$ and $\Delta q=q(2\pi)=\Delta
Q/T_{\omega}$ one gets:
\begin{equation}\label{mw10}
n_+(t)=\frac{1}{e^{T_{\omega}q(\bar{\varphi})}}\left(\frac{1}{e^{T_{\omega}\Delta
q}-1}T_{\omega}\int\limits_0^{2\pi}{\rm d}\varphi'\;
a(\varphi')e^{T_{\omega}q(\varphi')}+T_{\omega}\int\limits_0^{\bar
\varphi}{\rm
d}\varphi'\;a(\varphi')e^{T_{\omega}q(\varphi')}\right).
\end{equation}
It is important to note that $q(\varphi)$ is proportional to the
relaxation rate $1/\tau$ and $a(\varphi)$ to the excitation rate
of the upper level, which are independent of the frequency
$\omega$.

In the following two simple limits are treated:
$\omega\tau\rightarrow 0$ ($T_{\omega}/\tau\rightarrow\infty$);
and $\omega\tau\rightarrow \infty$ ($T_{\omega}/\tau\rightarrow
0$).

\paragraph{Low frequency limit ($\omega\tau\ll 1$)}

Assuming that $T_{\omega}q(\bar{\varphi})\gg 1$
($T_{\omega}/\tau\gg 1$) the first part of the right hand side of
Eq.~(\ref{mw10}) disappears as $q(\varphi')<\Delta q=q(2\pi)$.
(Note that $q(\varphi)$ is a monotonous function.) The
contribution to the second part also disappears as far as
$q(\varphi')-q(\bar{\varphi})<0$ is finite. In the vicinity of the
upper limit of the integral ($\bar{\varphi}=\varphi$), however,
$q({\varphi})$ can be expanded as
$q(\varphi')-q(\bar{\varphi})=\left.\frac{{\rm d}q}{{\rm
d}\varphi'}\right|_{\varphi'=\bar{\varphi}}(\varphi'-\bar{\varphi})$,
where $\frac{{\rm d}q}{{\rm d}\varphi'}=b(\varphi')$, thus
\begin{equation}\label{mw11a}
T_{\omega}\int\limits_{}^{\bar{\varphi}}{\rm
d}\varphi'\;a(\varphi')e^{T_{\omega}(q(\varphi')-q(\bar\varphi))}
\rightarrow\frac{a(\bar\varphi)}{b(\bar\varphi)},
\end{equation}
where the term arising from the lower limit is dropped as it is
exponentionally decreasing with $T_{\omega}\rightarrow\infty$. The
final result for the limit $T_{\omega}\rightarrow\infty$ is:
\begin{equation}\label{mw11b}
    n_+(t)=\frac{a(\bar\varphi)}{b(\bar\varphi)}=\frac{a(\varphi)}{b(\varphi)}=\frac{A(V(t))}{\bar B(V(t))}
\end{equation}
That result also holds for $\bar\varphi\ll 1$, where the condition
$T_{\omega}q(\bar\varphi)\gg 1$ is not satisfied any more. Without
making a rigorous proof, it can be argued, that $\bar\varphi\ll 1$
is at the beginning of the time period $\bar\varphi=0$. In the
long time limit ($t\gg T_{\omega}$) the time can be shifted ,
therefore, that time is not special, thus the result (\ref{mw11b})
must be generally valid.

In this way the static result is recovered with $A$ and $\bar B$
taken with the actual voltage at time $t$. Therefore, at arbitrary
time the system is in the stationary state, and the the expansion
of the current in (\ref{Iexpansion}) is valid.

\paragraph{High frequency limit ($\omega\tau\gg 1$)}

Taking $T_{\omega}\rightarrow 0$ only the first part of the right
hand side of Eq.~(\ref{mw10}) contributes. The denominator can be
expanded as $e^{T_{\omega}\Delta q}-1\sim T_{\omega}\Delta q$ and
the result is
\begin{equation}\label{mw11c}
n_+(t)=\frac{\int\limits_0^{2\pi}{\rm d}\varphi\;
a(\varphi)}{\int\limits_0^{2\pi}{\rm d}\varphi\;
b(\varphi)}=\frac{\langle A \rangle}{\langle \bar B\rangle},
\end{equation}
where $\langle ... \rangle$ stands for an average for a time
period and the definition of $\Delta q$ is used. As the TLS reacts
slowly on that time scale, thus it reacts on the average value of
the parameters $A$ and $\bar B$. Note that $\langle
A\rangle/\langle \bar B\rangle\ne\langle A/\bar B\rangle$.

\subsubsection{Dynamical conductance}

\begin{figure}[t]
\centering
\includegraphics[height=5.2truecm]{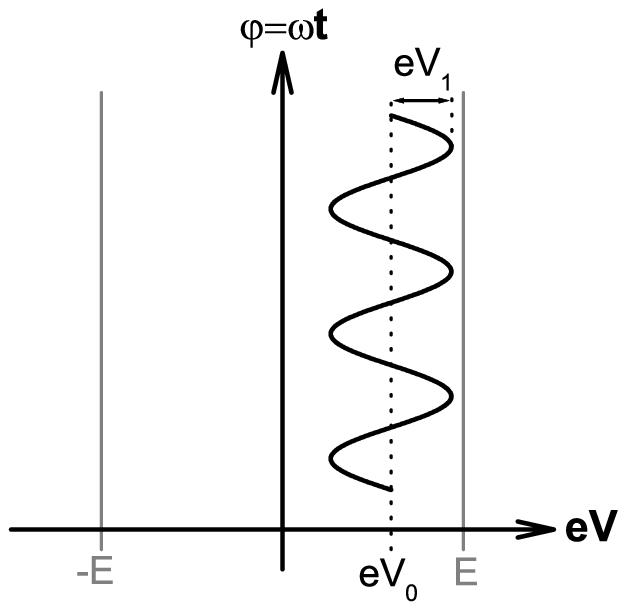}
\includegraphics[height=5.2truecm]{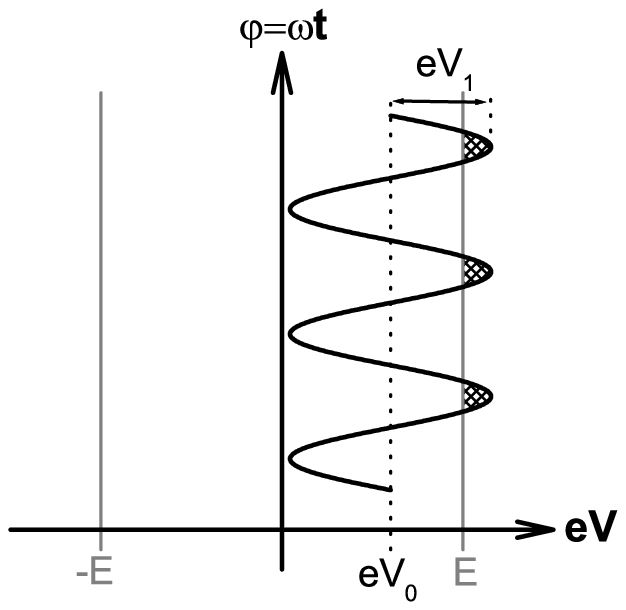}
\includegraphics[height=5.2truecm]{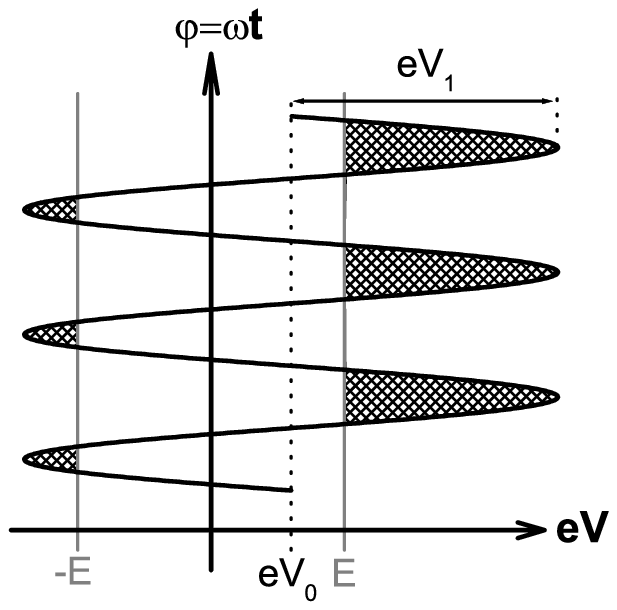}
\flushleft
\hspace{2.6truecm}(a)\hspace{5.4truecm}(b)\hspace{4.9truecm}(c)
\caption{\it Three different cases for the harmonic excitation
with respect to the excitation energy of the TLS.}
\label{rectification1.fig}
\end{figure}

In order to calculate the time dependent current first the
occupation number $n_+(t)$ must be calculated, which is not an
easy task in the general case as the formula given by
Eq.~(\ref{mw10}) contains double integrals. The next step is to
insert the obtained $n_+(t)$ into the equation for the current
(\ref{Iinel}). If one is interested in the average current (DC)
and the harmonics of the current, it is not enough just to
determine the the averaged $n_+$ and its harmonics.

That can be demonstrated easily at $T=0$ by Eq.~(\ref{Iinel2}). At
$e|V|\geq E$ the expression contains a term, which is independent
of the occupation number. This causes a nonlinearity in the $I(V)$
curve in every case, regardless of the voltage dependence of the
occupation number. For instance, if the TLS was in thermal
equilibrium with the bath then $n_+=0$ would hold at any bias
voltage. In this case the $I(V)$ curve is linear both at $e|V|<E$
and at $e|V|>E$, but it has a breaking point at $e|V|=E$. If the
alternating voltage is smaller than $E$ in the whole time period
($e|V_0+V_1\cos(\omega t)|<E$), then the response to the
excitation is completely linear, and no higher harmonics are
generated. This situation is demonstrated in
Fig.~\ref{rectification1.fig}a. Similarly, the response is linear
if $e|V_0+V_1\cos(\omega t)|>E$ at any time. However, if the
condition $e|V_0+V_1\cos(\omega t)|<E$ is only valid in a part of
the time period, the system gives a rectified response to the
harmonic excitation, thus higher harmonics are also generated.
This case is shown in Fig.~\ref{rectification1.fig}b. The TLS can
be excited only in the part of the time that is demonstrated by
the shadowed areas, where the conductance is reduced due to the
possibility of backscattering. In Fig.~\ref{rectification1.fig}c a
third case is presented, where the amplitude of the alternating
voltage is so large, that rectification occurs both at $eV=E$ and
at $eV=-E$. In this case the contributions from the both
polarities work against each other, and the higher harmonic
generation is reduced.

In a general case both the nonlinearity of the occupation number
$n_+(V)$ and the above discussed intrinsic nonlinearity of the
$I(V)$ curve cause higher harmonic generation. However, if the
period time of the excitation is much shorter than the
characteristic relaxation time of the TLS then the occupation
number cannot follow the alternating voltage, thus it sets at a
constant steady state value. In this situation the higher harmonic
generation is still present due to the intrinsic nonlinearity of
the $I(V)$ curve.

In the following section these features are discussed in two
special cases. In Sec.~\ref{rectification.sec} the $T=0$ the limit
is treated in the high frequency limit
($\omega\tau\rightarrow\infty$), where the rectified current and
the harmonics can be easily calculated at arbitrary value of the
amplitude of the excitation, $V_1$. These results are also valid
at finite temperature if $eV_1\gg kT$. In the second case
(Sec.~\ref{expansion.sec}) an expansion with respect to $V_1$ is
applied. This approach is valid if the amplitude of the excitation
is small  ($eV_1<kT$), however in this limit the frequency
dependence can be investigated in a wide range of frequencies.

\subsubsection{Rectification at zero temperature in the
$\omega\tau\rightarrow\infty$ limit} \label{rectification.sec}

The rectification due to a TLS in the contact can be nicely
demonstrated at $T=0$ in the $\omega\tau\rightarrow\infty$ limit.
We have seen that at $\omega\tau\rightarrow\infty$ the occupation
number $n_+$ becomes time independent which makes it possible to
perform the calculation analytically. According to Eq.~\ref{mw11c}
$\lim\limits_{\omega\rightarrow\infty}n_+=\frac{\langle
A\rangle}{\langle \bar B \rangle}$, where $\langle A\rangle$ and
$\langle \bar B \rangle$ are the time average of these
coefficients. Inserting zero temperature into the formula
(\ref{AB}) the voltage dependence of $A$ and $\bar{B}$ can be
written as:
\begin{eqnarray}
A(t)&=&\frac{w}{\hbar}\cdot\left\{
\begin{array}{lr}
0 &\quad {\rm if}\;\;e|V_0+V_1\cos(\omega t)| < E \\
&\\
\kappa(e|V_0+V_1\cos(\omega t)|-E) &\quad {\rm
if}\;\;e|V_0+V_1\cos(\omega t)| \geq E
\end{array}\right.\nonumber\\
\nonumber\\
\bar{B}(t)&=&\frac{w}{\hbar}\cdot\left(E+2A(t)\right),
\end{eqnarray}
Therefore, in case of amplitudes for which $e|V_0+V_1\cos(\omega
t)|<E$ holds for the total period of time the occupation number is
zero. If that is not satisfied, then the actual voltage at a
certain part of the time can excite the TLS out of the
groundstate. Two cases must be distinguished: (i) where
$V_0+V_1>E$ but $V_0-V_1\geq -E$; (ii) where $V_0+V_1>E$ and
$V_0-V_1< -E$.

\begin{figure}[t]
\centering
\includegraphics[height=6truecm]{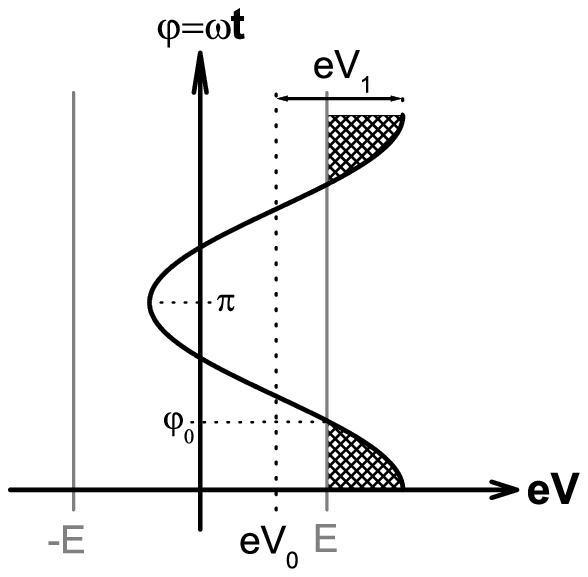}
\includegraphics[height=6truecm]{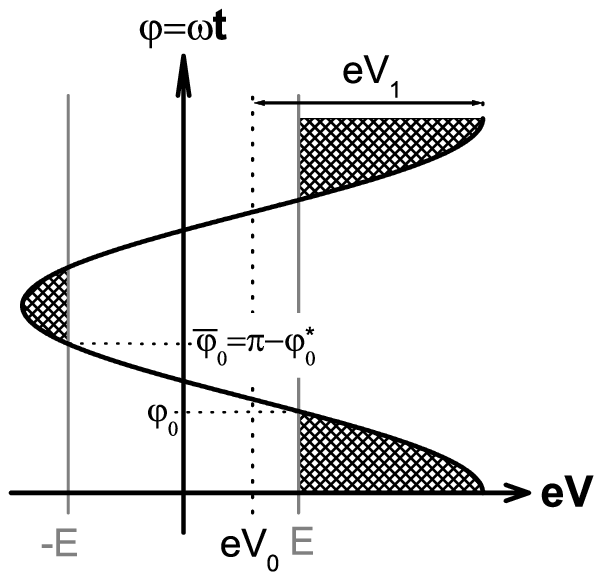}
\flushleft \hspace{4.3truecm}case (i)\hspace{5.5truecm}case (ii)
\caption{\it Illustrations for the calculation of rectification.}
\label{rectification2.fig}
\end{figure}

Figure \ref{rectification2.fig} show the actual voltage as a
function of time using $\varphi=\omega t$ in these two cases. The
shadowed areas indicate the regions where $e|V_0+V_1\cos(\omega
t)|>E$. The phases, where the shadowed areas are ending are
denoted by $\varphi_0$ (in case (i)) and $\bar
\varphi=\pi-\varphi_0^{\star}$ (in case (ii)). All of the
following results for case (i) can be applied also for case (ii)
by inserting $\varphi_0^{\star}\equiv 0$. These phases can be
expressed by $V_0$, $V_1$ and $E$ as follows:

\begin{equation}\label{mw12}
\cos(\varphi_0)=\frac{E-eV_0}{eV_1};\ \ \ \ \
\cos(\bar\varphi)=-\cos(\varphi_0^{\star})=\frac{-E-eV_0}{eV_1}.
\end{equation}

The time average of $A$ is calculated as:
\begin{eqnarray}\label{mw13}
\nonumber \langle A
\rangle=\frac{\kappa}{\pi}\left\{\int\limits_0^{\varphi_0}\left[e(V_0+V_1\cos\varphi)-E\right]{\rm
d}\varphi+
\int\limits_{\bar\varphi_0}^{\pi}\left[-e(V_0+V_1\cos\varphi)-E\right]{\rm
d}\varphi\right\}=\\
=\frac{\kappa}{\pi}e\left[
V_0(\varphi_0-\varphi_0^{\star})+V_1(\sin\varphi_0+\sin\varphi_0^{\star})\right]-\frac{\kappa}{\pi}E(\varphi_0+\varphi_0^{\star}),
\end{eqnarray}
and the occupation number is written as
\begin{equation}\label{mw14}
n_+=\frac{\langle A\rangle}{E+2\langle A\rangle}
\end{equation}
After having the result for $n_+$ the inelastic current correction
can be calculated using Eq.~\ref{Iinel2}. At the points, where
$e|V_0+V_1\cos(\omega t)|=E$ the time derivative of the current is
discontinuous, which results in higher harmonic generation. The
second harmonic generation with $\cos(2\omega t)$ and
$\sin(2\omega t)$ and also the shift of the DC signal can be
easily calculated:
\begin{eqnarray}\label{mw16}
  I^0 &=& \frac{1}{2\pi}\int\limits_0^{2\pi}I(V(\varphi)){\rm
d}\varphi =2e\varrho_0^2wK_R\frac{eV_1}{\pi}(1-2n_+)(\sin\varphi_0-\varphi_0\cos\varphi_0-\sin\varphi^{\star}_0+\varphi^{\star}_0\cos\varphi^{\star}_0) \\
  I^{2c} &=& \frac{1}{2\pi}\int\limits_0^{2\pi}I(V(\varphi))\cos(2\varphi){\rm
d}\varphi =2e\varrho_0^2wK_R\frac{eV_1}{3\pi}(1-2n_+)(\sin^3\varphi_0-\sin^3\varphi^{\star}_0)\\
  I^{2s} &=& \frac{1}{2\pi}\int\limits_0^{2\pi}I(V(\varphi))\sin(2\varphi){\rm
d}\varphi =0
\end{eqnarray}
The results for the $\cos(2\omega t)$ are plotted for different
values of the excitation amplitude, $V_1$. In this case the width
of the peak at $e|V_0|=E$ is determined by the amplitude $V_1$,
and these results are also valid at finite temperature if $kT\ll
eV_1$.

\begin{figure}[t]
\centering
\includegraphics[width=10truecm]{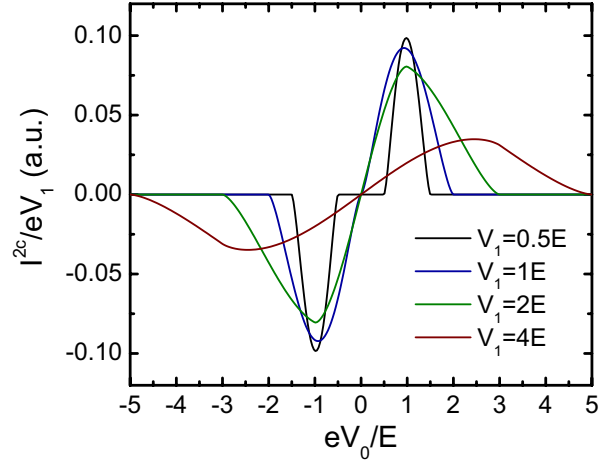}
\caption{\it Second harmonic generation at zero temperature in the
$\omega\tau\to\infty$ limit. The curves are plotted for different
amplitudes of the excitation signal.} \label{h2c.fig}
\end{figure}

\subsubsection{Expansion with respect to $V_1$}
\label{expansion.sec}

At finite frequencies the occupation number is already time
dependent ($n_+(t)$), thus the above considerations cannot be
applied. A calculation for intermediate frequencies can be
performed in the $eV_1\ll kT$ limit. In this case the integrals in
the general solution for $n_+$ (Eq.~\ref{mw2}) are calculated
after expanding the coefficients $A(V_0+V_1\cos\omega t)$ and
$\bar B(V_0+V_1\cos\omega t)$ with respect to $V_1$. The expansion
for $A$ is:
\begin{equation}\label{mw17}
A(V(t))=A_0+A_0^{'}V_1\cos\omega
t+\frac{1}{2}A_0^{''}V_1^2\cos^2\omega t+...
\end{equation}
and similarly for $\bar B$. The derivatives are taken at $V=V_0$.
In the general solution for $n_+$ first the quantity $e^{\int {\rm
d}t\;\bar B(t)}$ must be calculated. The non-oscillating part of
the exponent is linearly increasing with time, thus the expansion
can be carried out only for the oscillating part assuming that
$B_0^{'}V_1/\omega\ll 1$ and $B_0^{''}V_1^2/8\omega\ll 1$:
\begin{equation}\label{mw18}
e^{\int\limits_0^t {\rm d}t'\;\bar B(t')}\simeq e^{\beta
t}\cdot\left(1+B_0^{'}V_1\frac{\sin\omega
t}{\omega}+\frac{1}{2}(B_0^{'})^2V_1^2\left(\frac{\sin\omega
t}{\omega}\right)^2+\frac{1}{4}\bar B_0^{''}V_1^2\frac{\sin2\omega
t}{2\omega}\right),
\end{equation}
where
\begin{equation}\label{mw19}
\bar B_0+\frac{1}{4}\bar B_0^{''}V_1^2=\beta=\frac{1}{\tau}
\end{equation}
The shift of the relaxation rate by a term proportional to $V_1^2$
is due to the nonlinear behavior. The next step is to calculate
the integrals $\int {\rm d}t\;A(t)e^{Q(t)}$. All the terms
obtained are proportional to the exponential, $e^{\beta t}$, which
is cancelled out by the factor $e^{-\beta t}$ in the expression
(\ref{mw2}). In the remaining part the higher order terms
$O(V_1^3)$ are dropped. The next tedious task is to collect all
the terms in the coefficients of the expansion. The occupation
number can be written in the following form:
\begin{equation}\label{mw20}
n_+=n_+^{0}+V_1(n_+^s\sin\omega t+n_+^c\cos\omega t)+V_1^2(\delta
n_+^{0}+n_+^{2s}\sin2\omega t+n_+^{2c}\cos2\omega t)+O(V_1^3),
\end{equation}
where $n_+^0$ is the stationary occupation number at $V_1=0$
($n_+^0=A_0/\bar B_0$), $n_+^s$ and $n_+^c$ define the amplitudes
of the harmonic response to the excitation, $\delta n_+^0$ is the
shift of the stationary value due to the nonlinear behavior, while
$n_+^{2s}$ and $n_+^{2c}$ define the amplitudes of the second
harmonic generation. The values of these coefficients are given by
the following equations:
\begin{eqnarray}
  n_+^s &=& \frac{A_0^{'}\omega^2+A_0\bar B_0^{'}\beta}{\omega(\beta^2+\omega^2)}-\frac{A_0\bar B_0^{'}}{\omega\beta}\\
  n_+^c &=& \frac{A_0^{'}\beta-A_0\bar B_0^{'}}{\beta^2+\omega^2}\\
  \delta n_+^{0} &=& \frac{A_0^{''}}{4\beta}-\frac{A_0\bar B_0^{''}}{4\beta \bar B_0}+\frac{A_0(\bar B_0^{'})^2}{2\omega^2 \beta }-
  \frac{A_0^{'}\bar B_0^{'}\omega^2+A_0(\bar B_0^{'})^2\beta}{2\omega^2(\beta^2+\omega^2)} \\
  n_+^{2s} &=& \frac{4A_0^{''}\omega^2-4A_0(\bar B_0^{'})^2+A_0\bar B_0^{''}\beta
  +4A_0^{'}\bar B_0^{'}\beta}{8\omega(\beta^2+4\omega^2)}-\frac{A_0^{'}\bar B_0^{'}\beta-A_0(\bar B_0^{'})^2}{2\omega(\beta^2+\omega)}-\frac{A_0\bar B_0^{''}}{8\omega\beta} \\
  n_+^{2c} &=& \frac{A_0^{''}\beta\omega^2-A_0(\bar B_0^{'})^2\beta-A_0\bar B_0^{''}\omega^2-4A_0^{'}\bar B_0^{'}\omega^2}
  {4\omega^2(\beta^2+4\omega^2)}+\frac{A_0^{'}\bar B_0^{'}\omega^2+A_0(\bar B_0^{'})^2\beta}{2\omega^2(\beta^2+\omega^2)}-\frac{A_0(\bar B_0^{'})^2}{4\omega^2\beta}
\end{eqnarray}
The conditions for the expansion have excluded the narrow
frequency region around $\omega=0$. Accordingly, the above results
contain singular terms in $\omega$, which diverges as
$\omega\rightarrow 0$. On the other hand, it has been seen that
the $\omega\rightarrow 0$ limit is well defined for $n_+$, and
indeed, it can be shown that the singular terms cancel out in the
low frequency limit.

\begin{figure}[t]
\centering
\includegraphics[width=0.6\textwidth]{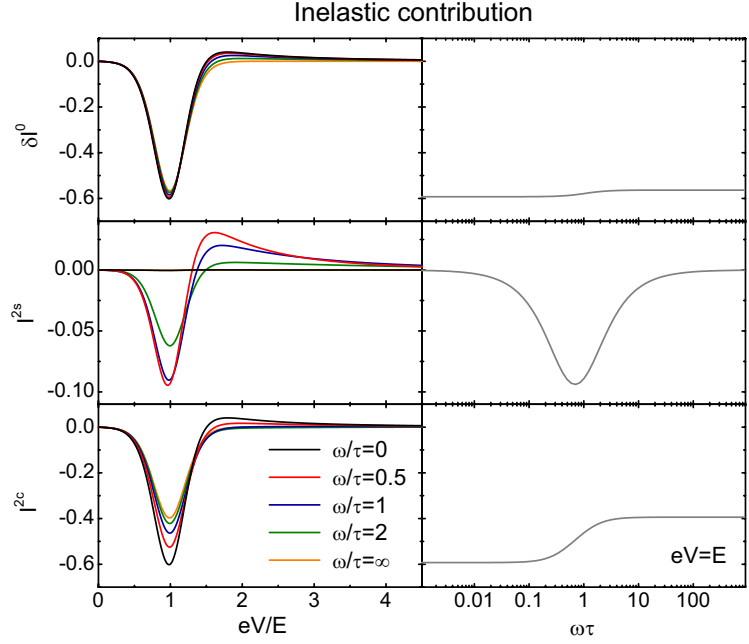}
\caption{\it The left panels show the voltage dependence of DC
current shift ($\delta I^0_{in}$) and the second harmonic terms
($I^{2s}_{in}$, $I^{2c}_{in}$) as a function of bias voltage at
different excitation frequencies in the inelastic case. The right
panels presents the frequency dependence of the corresponding
terms at a fixed voltage of $eV=E$. The temperature is $kT=0.1E$.}
\label{mwin.fig}
\end{figure}

The expression for the current correction can be expressed
similarly to (\ref{mw20}):
\begin{equation}\label{mw23}
I=I^0+V_1(I^s\sin\omega t+I^c\cos\omega t)+V_1^2(\delta
I^0+I^{2s}\sin 2\omega t+I^{2c}\cos 2\omega t)+O(V_1^3),
\end{equation}
in the inelastic case the current correction given by
Eq.~(\ref{Iinel}) can be written as
\begin{equation}\label{mw22}
I_{\rm in}=n_+D+n_-C=n_+\bar D+C,
\end{equation}
where $\bar D=D-C$. These must be expanded similarly to equation
(\ref{mw17}) given for $A$. After this step the coefficients in
(\ref{mw23}) can be given as follows:
\begin{eqnarray}\label{mw24}
I_{\rm in}^s &=& n_+^s\bar D_0 \\
I_{\rm in}^c &=& n_+^c\bar D_0+n_+^0\bar D_0^{'}+C_0^{'}\\
\delta I_{\rm in}^0 &=& \delta n_+^0\bar D_0+\frac{1}{4}n_+^0\bar
D_0^{''}+\frac{1}{4}C_0^{''}+\frac{1}{2}n_+^c\bar D_0^{'}\\
I_{\rm in}^{2s} &=& n_+^{2s}\bar D_0+\frac{1}{2}n_+^s\bar D_0^{'}\\
I_{\rm in}^{2c} &=& n_+^{2c}\bar
D_0+\frac{1}{4}C_0^{''}+\frac{1}{4}n_+^0\bar
D_0^{''}+\frac{1}{2}n_+^c\bar D_0^{'}
\end{eqnarray}
These results can be analytically calculated, and for
demonstration the experimentally important quantities ($\delta
I^0$, $I^{2s}$, and $I^{2c}$) are plotted in Fig.~\ref{mwin.fig}.
The left panels show the voltage dependence at different $\omega$
values, while the right panels demonstrate the $\omega$ dependence
at the fixed bias voltage value of $eV=E$. At low frequencies, the
TLS occupation number can follow the alternating voltage, thus
both $I^{2c}$ and $\delta I^0$ are proportional to the second
derivative of the $I(V)$ curve (see Eq.~\ref{Iexpansion}). In this
limit there is no phase shift, and thus $I^{2s}=0$. In accordance
with the static results, the curves show a negative peak at
$eV=E$, and a positive tail at higher voltages. In the high
frequency limit the TLS cannot follow the excitation, thus the
occupation number sets at a constant value. Due to the intrinsic
nonlinearity of the $I(V)$ curve, even in the high frequency limit
a pronounced negative peak is observed at $eV=E$, but the positive
background disappears. (At zero temperature the amplitude of the
negative dirac delta peak at $eV=E$ in $I^{2c}$ is exactly $2/3$
times smaller in the $\omega\rightarrow\infty$ limit than in the
$\omega\rightarrow 0$ limit. This factor of $2/3$ is well
demonstrated by the $\omega$ dependence of $I^{2c}(eV=E)$ in
Fig.~\ref{mwin.fig} as well.) Around the characteristic frequency
of the TLS ($\omega\tau\simeq 1$) a significant phase shift is
observed, as demonstrated by the curves for $I^{2s}$.

\begin{figure}
\centering
\includegraphics[width=0.6\textwidth]{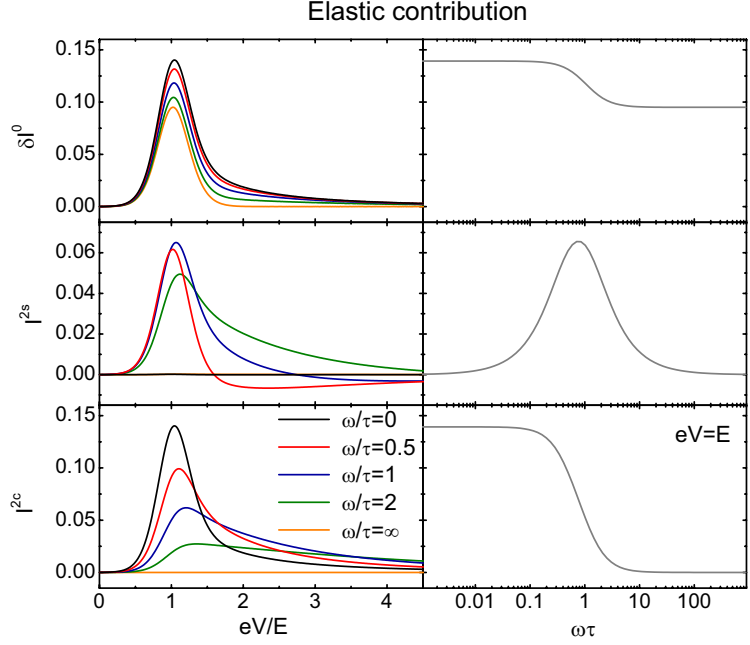}
\caption{\it The left panels show the voltage dependence of DC
current shift ($\delta I^0_{el}$) and the second harmonic terms
($I^{2s}_{el}$, $I^{2c}_{el}$) as a function of bias voltage at
different excitation frequencies in the elastic case. The right
panels presents the frequency dependence of the corresponding
terms at a fixed voltage of $eV=E$. The temperature is $kT=0.1E$.}
\label{mwel.fig}
\end{figure}

In the elastic case the different current components can be
expressed by using Eq.~(\ref{Iel}):
\begin{eqnarray}\label{mw24b}
I_{\rm el}^s &=& -(\gamma^+-\gamma^-)V_0n_+^s\\
I_{\rm el}^c &=& -(\gamma^+-\gamma^-)\left(n_+^0+V_0n_+^c\right)-\gamma^-\\
\delta I_{\rm el}^0 &=& -(\gamma^+-\gamma^-)\left(\frac{1}{2}n_+^c+V_0\delta n_+^0\right)\\
I_{\rm el}^{2s} &=& -(\gamma^+-\gamma^-)\left(\frac{1}{2}n_+^s+V_0n_+^{2s}\right)\\
I_{\rm el}^{2c} &=&
-(\gamma^+-\gamma^-)\left(\frac{1}{2}n_+^c+V_0n_+^{2c}\right)
\end{eqnarray}
In Fig.~\ref{mwel.fig} the second order terms are plotted for the
elastic case. The nonlinearity of the $I(V)$ curve of the elastic
contribution is basically determined by the nonlinearity of
$n_+(V)$. Above the characteristic frequency of the TLS the
occupation number cannot follow the excitation, thus both
$I^{2s}_{\rm el}$ and $I^{2s}_{\rm el}$ are suppressed. The shift
of the DC component ($\delta I^{0}_{\rm el}$) shows the
spectroscopic peak at $eV=E$ even at $\omega\tau\to\infty$, just
the amplitude is reduced by $\sim30\%$. Around the characteristic
frequency of the TLS a significant $I^{2s}_{\rm el}$ contribution
is observed.

\begin{figure}
\centering
\includegraphics[width=0.35\textwidth]{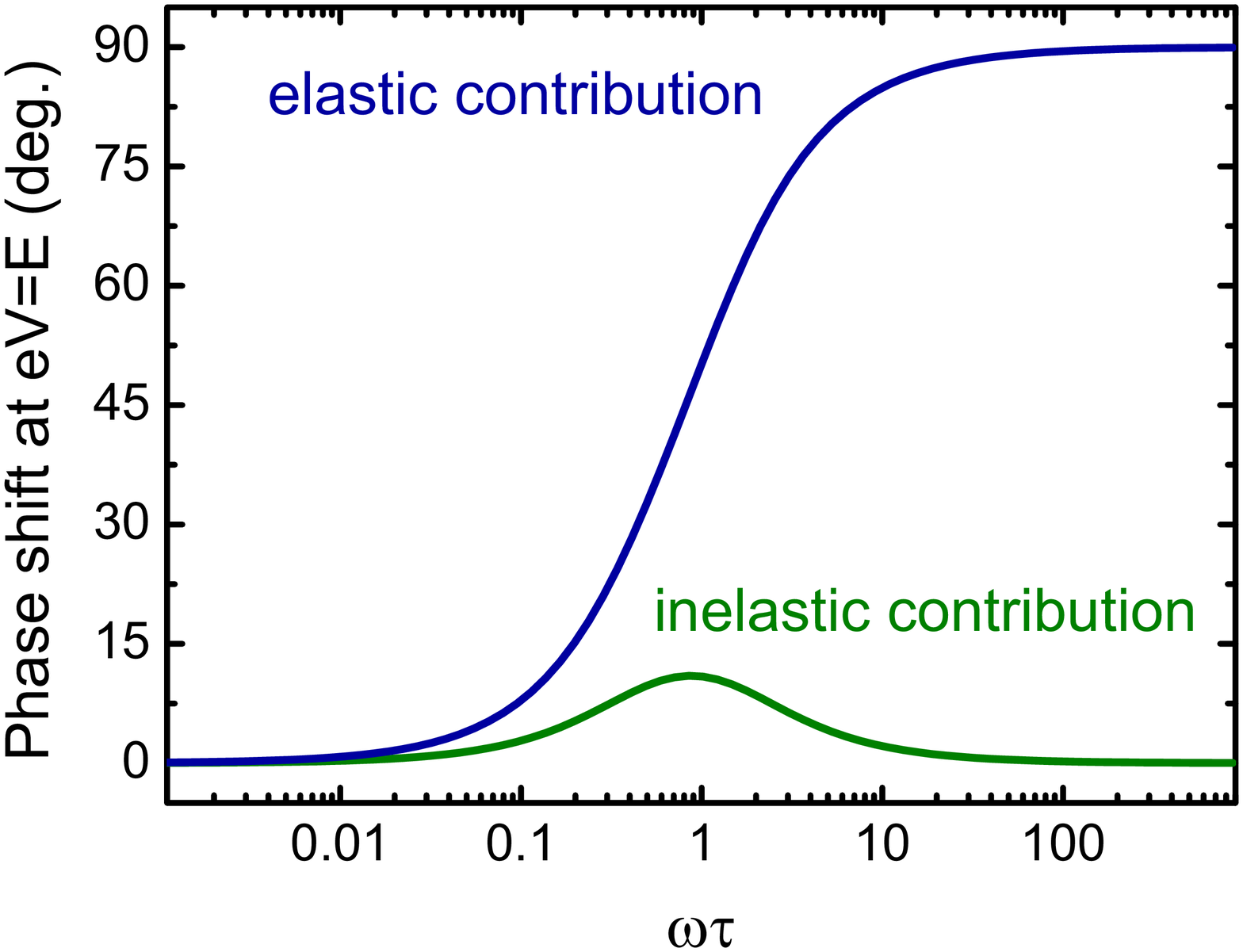}
\caption{\it Phaseshift in the second harmonic current correction
at $eV=E$ for the elastic and the inelastic contribution,
respectively. The temperature is $kT=0.1E$.}
\label{phaseshift.fig}
\end{figure}

The dynamical properties of the system are also demonstrated by
the phaseshift in the second harmonic contributions: $I^{2c}_{\rm
el}\cos(2\omega t)+I^{2s}_{\rm el}\cos(2\omega
t)=I^{(2)}\cos(2\omega t+\varphi)$. In Fig.~\ref{phaseshift.fig}
the phaseshift $\varphi$ is plotted at $eV=E$ both for the
inelastic and elastic case. In the elastic contribution the
phaseshift grows from $0$ to $90^\circ$ between the $\omega\tau\to
0$ and the $\omega\tau\to\infty$ limit. Contrary, in the inelastic
case the phaseshift is zero in both limits, and even at
$\omega\tau\sim 1$ a only minor phaseshift of $\sim 10\%$ is
observed, as the nonlinearity is basically related to the
intrinsic nonlinearity of $I(V)$ curve and the nonlinearity of
$n_+(V)$ plays a less important role.

\subsection{Conclusions}

In this section the theory of the scattering on slow TLSs in point
contacts was reviewed. In Sec.~\ref{ballistic} it was seen that
the electron distribution functions and the electric potential
around a ballistic contact can be determined by solving the
homogeneous Boltzmann equation for a point contact geometry. The
solution yields a voltage independent conductance, which is known
as the ``Sharvin conductance'' (Eq.~\ref{sharvinconductance}). The
effect of the scattering on a slow TLS near the contact can be
considered by perturbation in the collision term of the
homogeneous Boltzmann equation. The electron-TLS interaction may
result in backscattering processes, where an electron that has
already crossed the contact is scattered back through it. These
processes give a strongly voltage dependent correction to the
current, for instance at zero temperature backscattering can only
occur if the voltage bias exceeds the excitation energy of the
TLS.

The nonlinear current correction due to the inelastic scattering
of the electrons on a single TLS can be explicitly given, if the
occupation number of the upper level $n_+$, the electron-TLS
coupling strength $w$, and the geometrical factor $K_{\z{R}}$ are
known. It has crucial importance that the interaction of a TLS
with the neighborhood is dominated by the electron-TLS scattering,
and the phonon contribution is significantly smaller because of
the relatively small phonon phase space at energies much smaller
than the Debye energy. Therefore, the excited TLS mainly relaxes
its energy through the electrons, which have a nonequilibrium
distribution near the contact. As a consequence, the TLS
occupation number is also not the equilibrium one, and it is
strongly voltage dependent. The value of $n_+(V)$ can be
determined by a simple dynamical equation using Fermi's golden
rule. Due to voltage dependence of $n_+$ a nonlinear current
correction arises from the elastic scattering of the electrons on
the TLS as well, provided that the scattering cross sections are
different for the two states of the TLS. Both for the elastic and
inelastic scattering the point contact spectrum (i.e.\ the second
derivative of the $I-V$ curve) shows a sharp peak at the
excitation energy of the TLS, and a background tail at higher
voltages due to the nonequilibrium occupations.

The current correction due to the scattering on a single TLS can
also be determined for a diffusive point contact. In this case the
equations are more difficult, as the collision integral due to the
elastic scattering on defects should also be included in the
Boltzmann equation. Surprisingly, the lengthy calculations give
exactly the same results as in ballistic contacts, just the
geometrical prefactors are different.

The response to a time dependent bias voltage was also reviewed.
Due to the nonlinearity of the $I-V$ curve the application of a
small AC modulation beside the DC bias causes two important
features in the response signal: (i) a second harmonic signal is
generated, (ii) the average value of the current is shifted, which
is referred to as the DC shift signal. If the frequency of the AC
modulation is much smaller than the relaxation frequency of the
TLS and its amplitude is small compared to the energy splitting,
then both of these signals are proportional to the second
derivative of the $I-V$ curve. On the other hand, if the
modulating signal is fast compared to the relaxation of the TLS
then the occupation number $n_+$ cannot follow the variation of
the bias voltage, which has important consequences. Previously it
was believed that in the high frequency limit both the elastic and
inelastic contributions from the scattering on the TLS are
completely suppressed in the response signal. Our more detailed
calculations have shown that the DC shift signal contains a
frequency independent contribution due to the rectification at
voltages in the range of the TLS energies thus it is not decisive
for the relaxation dynamics of the TLS. We have demonstrated that
both the elastic and inelastic contributions are almost
independent of the irradiation frequency in the DC shift signal. A
more appropriate quantity for these  studies is the second
harmonic generation signal. In this case the elastic contribution
is completely suppressed and the inelastic contribution shrinks by
a factor of $2/3$ if the irradiation frequency exceeds the
relaxation frequency of the TLS. We have also investigated the
phase shift in the second harmonic signal which might be examined
in subsequent experimental studies.

There is another suggestion by Kozub and Rudin\cite{Kozub1997} to
explain zero bias anomalies based on the adiabatic renormalization
of the splitting of the TLS by electrons. That gives, however, a
weak temperature dependence because the main contribution comes
from the electrons far from the Fermi energy, thus that is not
related to infrared singularities. The interaction between the TLS
due to the Friedel oscillation was also considered and magnetic
field dependence is suggested. However, it is hard to compare the
theory with experiment in its present form.

It is also interesting to note that TLSs and magnetic impurities
interacting with metallic electrons show several similar features.
Both could lead to logarithmic corrections known as the Kondo
effect. The main similarity is that the spin $S=1/2$ has also two
states. Without external magnetic field or internal magnetic field
characteristic to the spin glass systems the two states $S^z=\pm
1/2$ are degenerate, just like a TLS without splitting
($\Delta=0$). In that case the occupations of the two spin states
$n_{1/2}=n_{-1/2}=1/2$ and that cannot be affected by the applied
voltage on the point contact because of spin degeneracy. In Born
approximation that gives a single peak at zero bias in the point
contact spectra $\frac{\partial^2 I}{\partial V^2}$, and no
background occurs. The logarithmic Kondo corrections, however,
broaden the peak which looks very similar to background in case of
phonons and TLS, but that is not due to a nonequilibrium
distribution.\cite{Omelyanchouk1985} The situation is different,
where the spin degeneracy is lifted by the magnetic field. In that
case $n_{1/2}\ne n_{-1/2}$, thus the occupation numbers depend on
the applied voltage and background occurs even without Kondo
corrections.

In the following section the former experimental results are
discussed and confronted with the model of slow two level systems.

\section{Discussion of the experimental observations}

This review is aimed at collecting and summarizing the main
results concerning the theory of the scattering on slow two-level
systems in point contacts, and to compare those with the
experiments. As a point contact probes the restricted area around
the contact region, it can be used to investigate a few or even a
single TLS located in the contact area. Therefore, some
measurements can be directly used to study the nature of TLSs in
the contact region, and to check the relevance of the model
presented.

\begin{figure}
\centering \epsfxsize=6cm \epsfbox{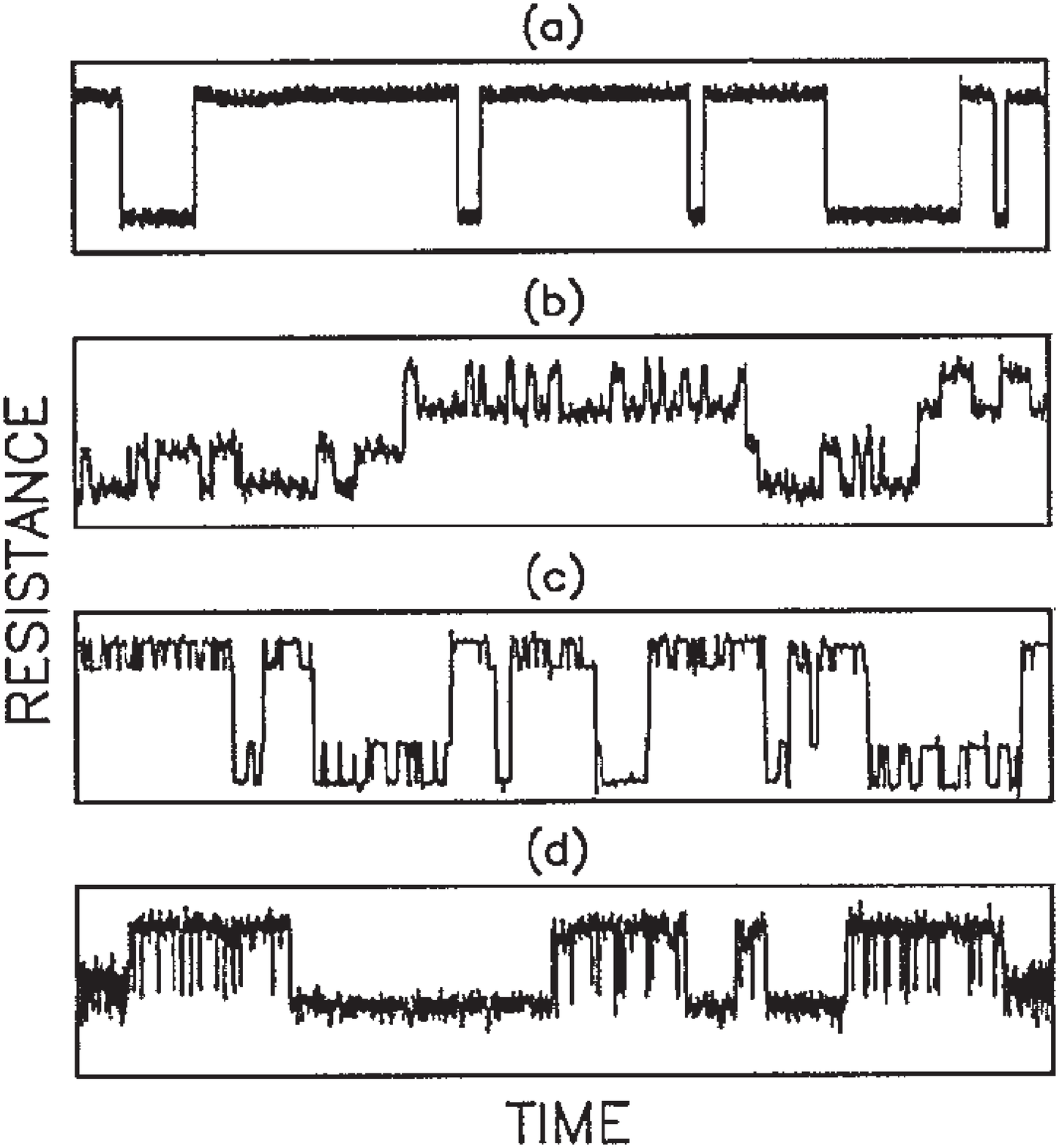}
\caption{Resistance vs.\ time in copper nanobridges for $T<150$K
showing several types of behavior. The fluctuations range from
0.005\% to 0.2\% of the total resistance. The time scales depend
on the temperature at which the fluctuation is observed. (a) A
single TLF. (b) Two independent TLFs. (c) Amplitude modulation.
(d) Frequency modulation of one TLS by another. [Taken from
\onlinecite{Ralls}, with kind permission of the American Institute
of Physics]} \label{Buhrmantls}
\end{figure}

In disordered metallic point contacts TLSs with a wide variety of
relaxation times can be present. A spectacular sign of the
presence of TLSs in a point contact is the telegraph fluctuation
of resistance, caused by defects fluctuating between metastable
configurations on an experimentally resolvable
timescale.\cite{Ralls} Examples for such so-called two level
fluctuators are shown in Fig.~\ref{Buhrmantls}. Panel (a) shows
the simplest telegraph fluctuation: the resistance is switching
between two values, as discussed by considering the elastic
processes (see Sec.~\ref{elasticscattering}), while the switching
itself is an inelastic process. In panel (b) two independent
telegraph fluctuations are superimposed upon each other, while on
panel (c) and (d) one two-level fluctuator modulates the amplitude
or the average switching time of the other. The screening of these
two-level systems by electrons causes a slow-down of the TLS
motion which can be described by the renormalization of the
tunneling amplitude.\cite{Meisenheimer1987,Giordano1991} This
phenomenon can be used to estimate the electron-TLS coupling
constant, as it is discussed in \onlinecite{Golding} and
\onlinecite{Zimmerman}.

This review concentrates on slow TLSs where the relaxation time is
much larger than the electron-TLS interaction time, but still the
TLS is fast enough not to be able to resolve it as a telegraph
fluctuation of the resistance. It was seen, that such ``slow''
TLSs can be studied by point-contact spectroscopy, a singularity
in the PC spectrum ($\partial^2I/\partial V^2$) is expected if the
voltage bias coincides with the energy splitting of the TLS close
to the contact. Figures \ref{ball_cent} and \ref{secder_ball}
demonstrated that the contributions of both the elastic and
inelastic scattering exhibit this anomaly at $eV=E$. In the
inelastic case this anomaly always corresponds to a maximum of the
differential conductance, $\partial I/\partial V$, which is
reflected by a negative peak in the point contact spectrum
$\partial^2 I/\partial V^2$ at $eV=E$. In the elastic case a
similar peak is observed, however its sign can either be negative
or positive, depending on the sign of $\gamma^+-\gamma^-$ (see
Eq.~\ref{Iel}).

A similar peak is expected in the PC spectrum for fast Kondo-like
TLSs, for which the electron assisted transition processes are not
negligible (see Sec.~\ref{fastTLS}). For the fast TLSs the anomaly
appears as a minimum in the differential conductance around zero
bias, and the width of the zero bias minimum is determined by the
Kondo temperature $T_K$. This feature is shown as a positive peak
in the $\partial^2 I/\partial V^2$ curve at
$eV=kT_K$.\cite{CZ,Ralph1992,Ralph1994} That peak shows very
strong similarity to the Kondo peak in the presence of magnetic
impurities,\cite{Yanson1995} but no magnetic field dependence is
observed, and the peak is changing by annealing.

The sign of the anomaly has crucial importance in distinguishing
the slow TLSs from fast TLSs. Along this discussion the anomaly
exhibiting a conductance maximum at zero bias is called
``positive'' zero bias anomaly (ZBA), while the conductance
minimum is called ``negative'' ZBA. We note that the labelling of
the sign of the anomaly can be confusing, as in some cases the
voltage dependence of either the differential conductance or the
differential resistance is plotted, in other studies the point
contact spectrum is presented. Furthermore, concerning the PC
spectrum, in theory the $\partial^2 I/\partial V^2$ curves have
direct physical meaning (showing the excitation spectrum), whereas
in experimental studies usually the second derivative of the
inverse curve, $\partial^2 I/\partial V^2$ is plotted. Regardless
of the way how the PC spectrum is plotted, the sign of the anomaly
can be determined by comparing it with the sign of the
contribution of the phonons. For the inelastic scattering on TLSs
the increase of the bias voltages causes an enhancement of the
back-scattering, thus this anomaly has the same sign as the phonon
part of the spectrum. In contrast, for Kondo-like anomalies a
resonant scattering is observed, which is restricted to the energy
range of the Kondo temperature. In this case the back-scattering
is enhanced at zero bias, and it decreases as the bias is elevated
above $eV\simeq kT_K$, thus the anomaly has negative sign compared
to the phonon contribution. In terms of this labelling, the fast
TLSs should always show a negative anomaly, the inelastic
contribution from slow TLSs shows up as a positive anomaly, while
the elastic contribution has an indefinite sign.

In the experiments different zero bias anomalies were observed. A
basic distinction should be made between the experiments performed
on pure metals, and those, performed on disordered systems, like
metallic glasses. In the first case the temperature dependence of
the bulk resistance does not show any anomalous behavior, but due
to the magnifying effect of point contacts, the scattering on a
few defects near the contact might be seen in the point contact
spectrum. In this case the appearance of the anomaly depends on
the way the contact is created, and usually the anomaly can be
destroyed by annealing. In metallic glasses, however, a high
degree of disorder is initially present, and the anomalous feature
seen in the PC spectrum is also reflected by the anomalous
low-temperature behavior of the bulk resistivity\cite{Harris1981}
(see also \onlinecite{CZ}).

Several metallic contacts show anomalies with different signs
depending on the way the contact is prepared. In the following we
concentrate on a few typical and most interesting cases.

\subsection{Measurements on Cu point contacts}

\begin{figure}
\centering
\includegraphics[height=5truecm]{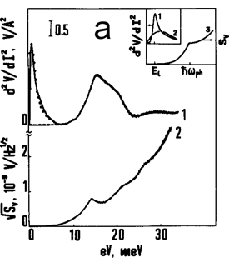}
\includegraphics[height=5truecm]{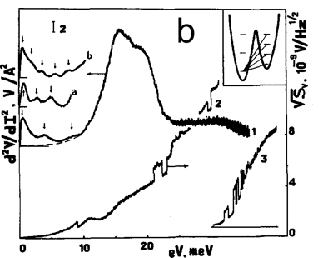}
\caption{(a) Point contact spectrum (curve 1) and spectral density
of noise (curve 2) for a copper point contact. The circles show
the theoretical prediction for the elastic scattering on a single
slow TLS. The inset shows theoretical curves for the point contact
spectra and the spectral density of noise. (b) Point contact
spectra for copper contacts showing several low energy peaks
(marked by arrows). The inset shows the possible transitions in a
double-well potential leading to the peak structure observed.
[Taken from \onlinecite{Akimenko1993} with kind permission of
Elsevier Science.]} \label{Akimenko.fig}
\end{figure}

The zero bias anomalies were widely studied in copper nanocontacts
created with different methods. The first experiments by touching
two Cu electrodes\cite{Akimenko1993} reported a pronounced
positive anomaly at $\simeq 1$\,meV [Fig.~\ref{Akimenko.fig}a].
The ZBA has a highly asymmetric shape, which can be well fitted
with the theoretical prediction for the elastic scattering on a
single slow TLS. The inelastic scattering could be excluded, as in
this case the anomaly should show a basically symmetric peak with
a background of opposite sign [see Fig.~\ref{secder_ballq}].
Furthermore, as the ZBA has consequently shown a positive sign, a
Kondo-like explanation could also be excluded. In some occasions
further low-energy peaks were also observed
[Fig.~\ref{Akimenko.fig}b], which were argued to be due to the
higher excitations of the TLS.

\begin{figure}
\centering \epsfxsize=6cm \epsfbox{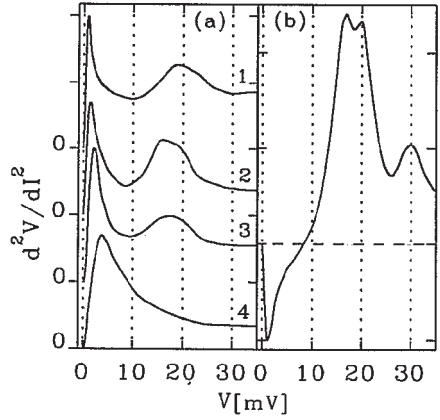} \caption{(a)
Point-contact spectra of copper thin-film MCB junctions with
resistances 1-1.2$\Omega$, 2-22$\Omega$, 3-64$\Omega$ and
4-400$\Omega$ ($T=1.3$K). The curves have been shifted for
clarity. (b) Point-contact spectrum for a Cu single crystal MCB
junction ($R=30\Omega$, $T=2$K). The peaks at 17-18mV and 30mV are
due to the electron-phonon interaction; the feature at bias
voltages between 0 and 5mV [positive for (a), negative for(b)] is
attributed to interactions between electrons and TLS. Similar
anomalies were also observed in Au, Cu, Ag, and Pt contacts.
[Taken from \onlinecite{anomaly} with kind permission of the
American Institute of Physics.]} \label{posneganomaly}
\end{figure}

Similar ZBAs were seen in the experiments of Keijsers et
al.\cite{anomaly} on copper thin film mechanically controllable
break junctions [Fig.~\ref{posneganomaly}a]. These samples were
created by vapor deposition, and judging from the residual
resistance ratio and the intensity of the phonon peaks in the PC
spectrum they were of low crystalline quality implying that the
contacts were in the diffusive regime. However, the authors claim
that the ZBA is not due to the intrinsic disorder of the sample,
but it correlates with the amount of bending and stretching of the
film. In these thin film samples the ZBA has consequently shown a
positive sign, and it could be well fitted with the elastic
scattering on a slow TLS, similarly to the above discussed
measurements of Akimentko et al.

The same work of Keijsers et al.\cite{anomaly} has also presented
measurements on single crystal Cu break junctions
[Fig.~\ref{posneganomaly}b]. In this case the well resolved
electron-phonon interaction spectrum implies a large elastic mean
free path. In these measurements the ZBA has consequently shown a
negative sign. Here, the elastic scattering on a slow TLS is still
a reasonable explanation, however, the Kondo-like behavior also
has to be considered.

\begin{figure}
\centering
\includegraphics[height=5truecm]{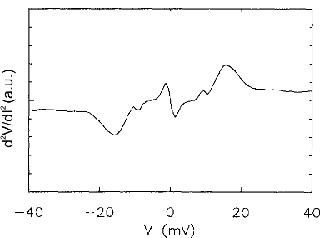}
\includegraphics[height=5truecm]{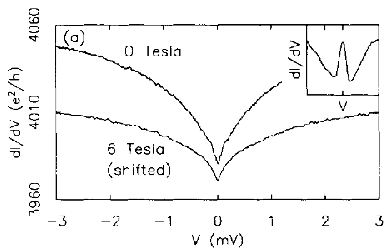}
\caption{(a) Point contact spectrum for an unannealed Cu
nanoconstriction. The negative zero bias anomaly at $\sim 2$\,mV
is attributed to the scattering on fast two level systems, while
the structures at higher voltages originate from the scattering on
phonons. [Taken from \onlinecite{Ralph1994} with kind permission
of the American Institute of Physics.] (b) Differential
conductance of a Cu nanoconstriction at $T=100$\,mK. The curve
measured in $6$\,T magnetic field is shifted down by $20\,e^2/h$
for clarity, and only a slight dependence on the field is
observed, which supports the nonmagnetic origin of the anomaly.
For comparison the inset shows the conductance of Cu constriction
with $200$\,ppm Mn at $100$\,mK, showing a pronounced Zeeman
splitting in magnetic field. [Taken from \onlinecite{Ralph1992}
with kind permission of the American Institute of Physics.]
Similar zero bias anomaly was also observed in Ti
nanojunctions.\cite{Upadhyay1997}} \label{ralphZBA.fig}
\end{figure}

Ralph et al.\cite{Ralph} have investigated nanofabricated Cu
contacts. In this case the size of the contact naturally could not
be varied during the measurements, but as a great advantage, the
contact was ``fixed'', and thus a detailed study of the
temperature and magnetic field dependence could be performed. In
these experiments a well defined negative anomaly was observed if
the sample was quenched [Fig.~\ref{ralphZBA.fig}], and the anomaly
disappeared after annealing. The detailed theoretical analysis of
the curves (including scaling arguments)\cite{Ralph2} has shown
very good agreement with the theory of fast Kondo-like TLSs, even
if their existence is still an open
question.\cite{Aleiner2001a,Aleiner2001b,Borda2003} The relatively
high Kondo temperature ($T_K\sim 10$\,K) can be obtained in the
framework of recent theory\cite{Borda2003} only if the center is
light (e.g.\ H or He) as the energy scale is the kinetic
energy.\cite{Zawadowski2004}

In all of the above experiments the ZBAs were attributed to a
single or a few TLSs. In the measurements, where consequently a
positive ZBA was observed both the inelastic scattering on a slow
TLS and the Kondo-like behavior can be ruled out due to shape and
the sign of the anomaly, respectively. Therefore, the elastic
scattering on a slow TLS remains as an explanation, which indeed
gives good fit for the curves. However, it should be noted that
for elastic scattering the sign of the anomaly is expected to be
contact dependent. A consequent positive sign might be explained
by the following argument: the atom in the higher level of the TLS
can be considered to be more far away from the ordered arrangement
of the atoms, thus its scattering cross section is larger than
that of the low energy state, which results in an enhanced
backscattering as the bias is increased.

It is interesting to note that the positive ZBAs in
\onlinecite{anomaly} exhibit a pronounced contact size dependence:
as the diameter of the junction is decreased the ZBA shifts to
higher energy values [Fig.~\ref{posneganomaly}a]. The ZBAs showing
Kondo effect in magnetically doped contacts have exhibited a
similar size effect.\cite{Yanson1995} The latter phenomenon was
explained by Zar\'and and Udvardi as a result of density of states
oscillations nearby the surface.\cite{Zarand1996a} Similar size
effect is expected for the Kondo-like scattering on fast TLSs, but
we stress that this explanation does not apply to the ZBAs in
Fig.~\ref{posneganomaly}a due to the positive sign of the anomaly.

\subsection{Zero bias anomalies in metallic glasses}

\begin{figure}
\centering \epsfxsize=6cm \epsfbox{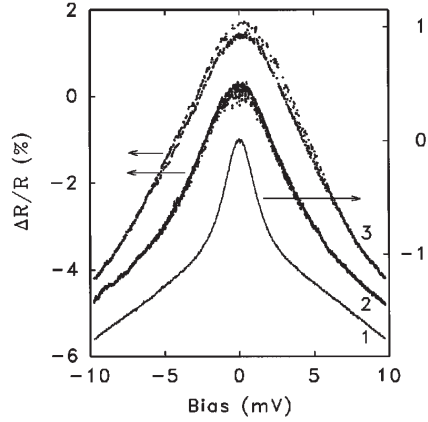}
\caption{Differential resistance $R_d$ as a function of bias
voltage for $Fe_{80}B_{20}$ (1 and 2) and
$Fe_{32}Ni_{36}Cr_{14}P_{12}B_{6}$ (3) MCB junctions ($T=1.2$K).
(1) A 6.6$\Omega$ contact displaying almost no noise. (2) A
366$\Omega$ contact that shows clear noise around zero bias. The
noise amplitude decreases as the bias voltage increases. (3) A
145$\Omega$ contact, showing a clear two-level switching behavior
between two different $R_d$ peaks (the time dependence of the
resistance at a fixed bias shows a telegraph noise similar to
those in Fig.~\ref{Buhrmantls}). Curve 3 has been shifted for
clarity. [Taken from \onlinecite{Keijsers1996} with kind
permission of the American Institute of Physics.] The same paper
has also reported about the telegraph fluctuation of the zero bias
anomaly in silver break junctions. A similar telegraph fluctuation
was was observed in the superconducting characteristics of
Ni$_x$Nb$_{1-x}$ metallic glass break junctions.\cite{Halbritter}}
\label{telegraphZBA}
\end{figure}

The zero bias anomalies are also characteristic for metallic glass
samples, which were always found to show negative
anomaly\cite{Keijsers1996,Halbritter} [for examples see
Fig.~\ref{telegraphZBA}] in accordance with the anomalous
low-temperature increase of the bulk resistance.\cite{Harris1981}
In metallic glasses TLSs with a wide distribution of parameters
can be present, thus the interpretation of the results is even
more difficult. Usually the ZBA is attributed to the scattering on
TLSs, however the magnetic origin also cannot be excluded. The
anomalies cannot be suppressed by magnatic field in contrast to
the magnetic Kondo effect, however the anomalous part of the
temperature dependence of the resistance shows a small but well
defined magnetoresitance, which is not present at higher
temperatures.\cite{Halbritter}

\begin{figure}
\centering \epsfxsize=6cm \epsfbox{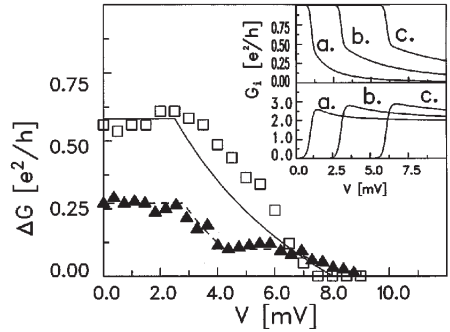}
\caption{Conductance difference for mechanically controlled
metallic glass break junctions. The open squares are obtained from
the measured resistance change values shown in curve 3 of
Fig.~\ref{telegraphZBA}, with uncertainties of order 0.1$e^2/h$.
The solid line is obtained from the theory of Vlad\'ar and
Zawadowski\cite{VZa} using a Kondo scale of $T_K\approx35$K, and
two different splittings for the fast TLS of $E_1=3.5$meV and
$E_2=8$meV, with the assumption that the modulation by the slow
center induces these two different splitting values. The other
curve (filled triangles) represents another measurement from
\onlinecite{Keijsers1996}, but in this case the theoretical fit
implies that two fast TLSs are being modulated by a slow one. The
inset shows conductance curves generated by the theory of Kozub
and Kulik,\cite{KK} which are also shown in Fig.~\ref{ball_cent}
in the present review. [Taken from \onlinecite{Zarand1998}, with
kind permission of the American Institute of Physics.]}
\label{diffZBA}
\end{figure}

Another interesting feature is the telegraph fluctuation of the
ZBA demonstrated by the upper curves in Fig.~\ref{telegraphZBA}.
This telegraph noise is only present in the narrow voltage region
of the ZBA, that is the amplitude of the fluctuation vanishes as
the voltage is increased (Fig.~\ref{diffZBA}). This extraordinary
behavior may be understood by assuming that a slow two level
fluctuator is not directly changing the resistance (like in
Fig.~\ref{Buhrmantls}), but it is modulating the parameters of a
fast TLS in the neighborhood. Therefore, the amplitude of the
fluctuation only depends on the parameters of a {\it single} or a
few TLSs with the slowly moving defect being in one or the other
metastable position, and the contributions of electron-phonon
interaction, or other TLSs being more far away from this
particular defect are less significant. Comparing the voltage
dependence of the fluctuation with theory, it turns out that the
model of fast Kondo-like TLSs is in good agreement with the
measurement, while the theory of slow TLSs gives a poor fit
because of the long tail of the PC spectrum.\cite{Zarand1998}


\subsection{High frequency behavior of the zero bias anomalies}

The common feature in the yet reported ZBAs is that the anomaly
appears in the voltage region $\sim 0.1-5$\,mV. Usually the
anomaly shows a negative sign, however even in these cases the
elastic contribution from slow TLSs should be considered. A direct
distinction between the fast and slow TLSs could be made by
studying the relaxation dynamics of the system. In terms of the
two models the position of the singularity reflects the Kondo
temperature ($T_K$) or the energy splitting ($E$), respectively.
Inserting the typical voltage range of the anomaly into the
models, the standard Korringa-like relaxation rate of slow TLSs
due to the creation of an electron and hole pair is $\sim
10^{-10}-10^{-7}$\,s (see Eq.~\ref{relaxtime}), while in the two
channel Kondo model the relaxation time is in the range of $\sim
10^{-11}$s.\cite{CZ} Therefore the boundary between the two cases
can be investigated with microwave irradiation measurements in the
GHz frequency range.

Experimentally the high frequency measurements were usually
performed by irradiating the contact with a chopped microwave
signal, and measuring the change of the DC current as the
microwave is on and off. According to Eq.~\ref{Iexpansion} the
voltage dependence of the DC shift signal is proportional to the
second derivative of the $I-V$ curve if the frequency of the
irradiation signal is slow compared to the relaxation of the TLS,
and the perturbation due to the alternating voltage is weak.
Previously, it was believed that above the characteristic
frequency of the TLS the DC shift signal should vanish, however
the calculations were only performed for the second generation
signal in the case of elastic scattering.\cite{KK} Measurements on
various systems have shown that the peak in the DC shift signal
cannot be suppressed in the frequency range of the measurements
($\le 60$\,GHz).\cite{vanKempen1,Balkashin2001} Therefore, these
studies concluded that the anomaly comes from fast TLSs with a
characteristic frequency $\omega_{TLS}\gtrsim 10^{11}$\,s$^{-1}$.
However, according to our results in Sec.~\ref{expansion.sec}
these measurements are not decisive as the DC shift signal {\it
should not vanish for slow TLSs either}, it should be only
slightly reduced. (see upper panels in Fig.~\ref{mwin.fig} and
Fig.~\ref{mwel.fig}). The arbitrariness of the substraction of the
background signal makes it even more difficult to draw a clear
conclusion concerning the frequency dependence of the peak.

\begin{figure}
\centering \epsfxsize=6cm \epsfbox{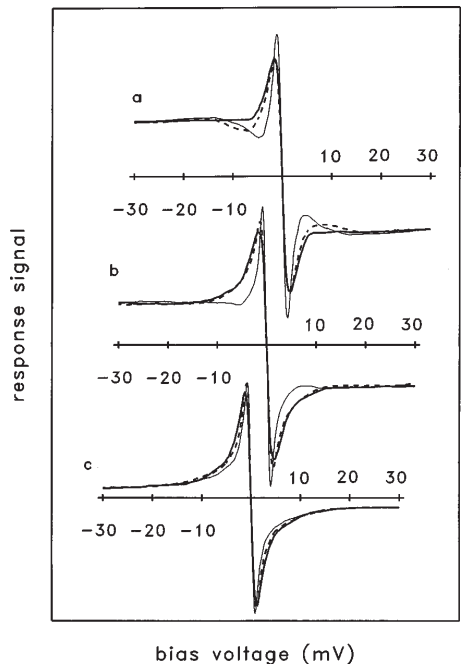} \caption{RF
response signals at 60 GHz (thick solid line) and 0.6 Ghz (dashed
line) and $d^2V/dI^2(V)$ dependence (thin solid line) for (a) a
26-$\Omega$ Fe$_{80}$B$_{20}$ point contact, (b) a 8-$\Omega$
Fe$_{78}$Mo$_2$B$_{20}$ point contact, and (c) a 15-$\Omega$
Fe$_{32}$Ni$_{36}$Cr$_{14}$P$_{12}$B$_{6}$ point contact. [Taken
from,\cite{vanKempen1} with kind permission of the American
Institute of Physics.]} \label{Balkashin1}
\end{figure}

More conclusive results can be obtained by measuring the second
harmonic response to the high frequency irradiation, which is,
however, experimentally more difficult. According to our
calculations, in the second harmonic signal the peak due to the
elastic scattering should disappear at high frequency, while the
peak due to inelastic scattering only decreases by a factor of
~2/3 (bottom panels in Fig.~\ref{mwel.fig} and \ref{mwin.fig},
respectively). Recent second harmonic generation measurements on
Ni$_{59}$Nb$_{41}$ metallic glasses have
shown,\cite{Balkashin2001b} that the ZBA cannot be suppressed with
frequencies up to $\sim 5$\,GHz. This measurement excludes the
elastic contribution of slow TLSs due to the lack of frequency
dependence, and it also excludes the inelastic contribution due to
slow TLSs as the sign of the peak is negative. However, the
authors propose that the frequency dispersion of the smooth
background signal might come from the scattering on slow TLSs.

\begin{figure}
\centering
\includegraphics[width=6truecm]{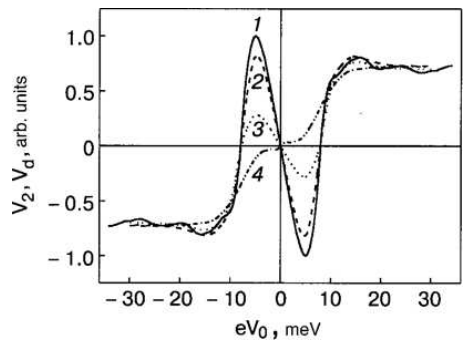}
\caption{Evolution of the PC spectra of a contact of the annealed
alloy Ni$_{59}$Nb$_{41}$ with Ag in measurements at different
frequencies: a sonic frequency (1) and microwave frequencies
$\omega=$ 0.48\,GHz (2), 2.036\,GHz (3), and 4.72\,GHz (4). The
anomaly is strongly suppressed with the microwave irradiation,
whereas the ZBA in unannealed samples is unaffected by the same
irradiation. [Taken from \onlinecite{Balkashin2003} with kind
permission of the American Institute of Physics.]}
\label{annealing.fig}
\end{figure}

An interesting result has been presented by Balkashin et al.\ on
annealed Ni$_{59}$Nb$_{41}$ metallic glass
samples.\cite{Balkashin2003} The DC shift response signal was
measured at various microwave irradiation frequencies both before
and after annealing. The unannealed samples have shown
insignificant frequency dispersion up the maximum frequency of the
measurements ($\sim 5$\,GHz), in agreement with the previous
studies. In the annealed samples, however, the ZBA was completely
suppressed at high microwave frequencies
[Fig.~\ref{annealing.fig}]. As we have discussed, the ZBA in the
DC shift response signal should not dissappear for slow TLSs
either, thus the experimental observations on the annealed samples
cannot be interpreted in terms of the presented theory for slow
TLSs. This observation questions whether the ZBA originates from
the scattering on TLSs in the studied Ni$_x$Nb$_{1-x}$ metallic
glasses.

Up to now, the high frequency measurements were only performed on
metallic glass samples. Unfortunately in metallic glasses various
interactions coexist, which makes the interpretation of the
results difficult. Especially serious problem is the normalization
of the measured response signals. In pure samples the curves can
be normalized to the phonon peaks, as the electron-phonon
relaxation time is very small ($\tau_{\rm el-ph}\sim
10^{-13}$\,s). In metallic glasses, however the phonon peaks
cannot be identified, and a smooth background is observed, which
can contain various contributions. In disordered systems the
disorder induced electron-electron interaction (see e.g.\
\onlinecite{Altshuler1991}) and the scattering on nonequilibrium
phonons\cite{Kulik1985,Yanson1985} are also important, furthermore
the scattering on a variety of TLSs with different excitation
energies can also contribute to the background. Additionally, the
materials under study contain magnetic ions, thus the interactions
of magnetic origin also cannot be excluded. Due to the complexity
of the system the background signal can also have frequency
dispersion in the frequency range of the measurements, which
hampers the comparison of the response signals measured at
different frequencies.

A more clear picture could be obtained by investigating the high
frequency behavior of the ZBAs observed in clean contacts of pure
metals. In this case the ZBA is attributed to the scattering on a
single or a few TLSs, and due to the ballistic nature of the
contact clear phonon peaks are observed at higher energies. In
this case the microwave response signals are easily compared by
normalizing the curves to the phonon peaks, and thus the
measurement of the second harmonic response signal provides
explicit information about the relaxation dynamics of the system.
Especially interesting would be the study of positive ZBAs in
copper nanojunctions (or in other metals like Ag, Au and Pt),
which are explained by the elastic scattering on slow TLSs.

\subsection{Conclusions}

A large variety of point contact spectroscopy measurements show
ZBAs, which may be attributed to the scattering of electrons on
TLSs. Certain measurements on Cu nanocontacts show positive ZBAs,
which can be well explained by the elastic scattering on slow TLS.
In this case, however, the reason for the consequently positive
sign of the anomaly is not well understood. Another group of
measurements show negative ZBAs, and among these the experiments
on Cu nanocinstrictions\cite{Ralph} are well interpreted in the
framework of the two channel Kondo model. In these studies the
ZBAs are attributed to a few or even a single TLS, which allows a
direct comparison of the experimental observations with the
theoretical predictions. ZBAs are frequently observed in metallic
glass samples, however, in this case the complexity of the system
makes the interpretation of the results more difficult.
Interesting attempts were made by investigating the frequency
dispersion of the ZBAs in metallic glasses, however some of the
results were misinterpreted due to the lack of the theoretical
calculation for the frequency dependence of the DC shift signal.
Novel second harmonic generation measurements on Ni$_x$Nb$_{1-x}$
metallic glasses has proved that in this system the ZBA cannot be
explained by the scattering on slow TLSs, however slow TLSs can
give contribution to the background. The study of the high
frequency behavior of a few TLSs near clean metallic contacts
might provide a more clear information about the relaxation
dynamics of TLSs.

In the cases cited above the origins of the TLSs are not well
known. TLSs can occur, however, in crystalline material with
deviation from the stochiometric composition or disorder as well.
The TLS is formed in crystalline neighborhood where the size or
the shape of the cavity in which the atom is sitting can be
modified by the surroundings. In such cases clear indication of
orbital Kondo effect was found in PbTe if some Pb atoms are
replaced by smaller Ge atoms.\cite{Katayama1987} However, this
material is not so easy to produce in a controlled way. Recently a
new family of material was discovered, which are metallic and they
are composed of U, Th, As, Se and S.\cite{Henkie2001} The
systematic study of these well-controlled materials may also
provide the possibility to understand the TLS-electron interaction
in more detail.

\section{Final remarks}

The point contact spectroscopy is a very powerful method to
investigate the dynamics of slow TLSs and magnetic Kondo
impurities and resonances of unknown, nonmagnetic origin,
presumably due to some structured defects at zero bias.  That
spectroscopy is complementary of tunneling spectroscopy with oxide
barrier\cite{Wolf1989} or break junction in the tunneling
limit,\cite{Keijsers1996b,Keijsers2000} but there are similarities
and differences as well. The local excitations in the barrier as
molecular vibrations or magnetic spin flip processes open new
tunneling channels resulting in extra current at voltages above
the excitation energy. That gives just the opposite sign compared
to the point contact case, as those anomalies are the negative
ones. The magnetic impurities in the electrodes nearby the barrier
can also depress the local tunneling density of states around the
Fermi level in the energy range of the Kondo
temperature.\cite{Mezei1971,Bermon1978} That gives a negative zero
bias anomaly just like in the point contact case.

In the future it would be very important to make systematic
studies concerning the TLS, starting with the slow TLS with
contributions of arbitrary signs due to dominating elastic
processes. The detailed studies of the frequency dependence based
on the theoretical part of the present review would be especially
useful and that would clarify some of the contradictions of the
present time.

It is very important to study further the resonances of unknown
origin which have strong resemblance to nonmagnetic Kondo
behavior\cite{CZ} and to make attempts to determine their origin
in at least a few cases.

The origin of those resonances has been the subject of theoretical
debates in the recent years. Originally it was proposed that Kondo
effect could arise from tunneling TLSs with non-commutative
couplings e.g. due to the screening by electrons and electron
assisted tunneling (see e.g.~\onlinecite{CZ}). It was shown,
however, that the adiabatic screening of the moving atom by
electrons depresses the tunneling rate in such a strong
manner,\cite{Kagan,Kaganbook,Aleiner2001a,Aleiner2001b} that the
actual Kondo temperature is negligible. Recently, it has been
suggested that the situation can be changed if the atom in the
double potential well has the ground-state below the potential
barrier, but the first excited state is above it.\cite{Borda2003}
Then this objection could be avoided, but the value of the
splitting raises many questions. That theory for an intermediate
heavy atom gives too small $T_K$, not more than $0.1-0.2$\,K and
that cannot explain the resonances of typical width of $10-20$\,K.
There is, however, the possibility of light atoms like
hydrogen,\cite{Zawadowski2004} as the dominating energy scale is
the kinetic energy of the atom in the potential well. Therefore,
it would be very important in the future to study hydrogen defect
either in atomic states or water contamination, where the hydrogen
could still change its position.

Summarizing, the origin of slow TLS and the dubious resonances
deserve extended further studies.

\section{Acknowledgements}

We acknowledge to all the authors laying down the principles of
slow TLSs in point contacts. We are especially thankful to I.K.
Yanson and Sz. Csonka for the careful reading of the manuscript,
and also to K.E. Nagaev, J. von Delft, J.M. van Ruitenbeek, O.P.
Balkashin, A.N. Omelyanchouk, I.O. Kulik, G. Zar\'and, D. C.
Ralph, G. Mih\'aly, R. A. Buhrman and H. van Kempen for their
useful remarks. Part of this work was supported by the Hungarian
Research Funds OTKA T026327, TS040878, T037451, T038162, T034243.

\end{document}